%% file: stodile.tex
\documentclass[runningheads,fleqn]{cl2emult}
\usepackage{amssymb}

\usepackage{amsmath}

\usepackage{makeidx}
\usepackage{graphicx}
\usepackage{subeqnar}
\usepackage{multicol}
\usepackage{cropmark}
\usepackage{phys}

\makeindex

\begin{document}

\title*{Developing the MTO Formalism}
\toctitle{Developing the MTO Formalism}
\titlerunning{Developing the MTO Formalism}
\author{O. K. Andersen \and T. Saha-Dasgupta \and R. W. Tank \and C. Arcangeli \and %
O. Jepsen \and G. Krier}
\authorrunning{O.K. Andersen et al.} 
\institute{Max-Planck-Institut FKF,
D-70569 Stuttgart, FRG,\\andersen@and.mpi-stuttgart.mpg.de}
\date{\today}
\maketitle

\begin{abstract}
The TB-LMTO-ASA method is reviewed and generalized to an accurate and robust
TB-NMTO minimal-basis method, which solves Schr\"{o}dinger's equation to $N$%
th order in the energy expansion for an overlapping MT-potential, and which
may include any degree of downfolding. For $N=1,$ the simple TB-LMTO-ASA
formalism is preserved. For a discrete energy mesh, the NMTO basis set may
be given as: $\chi ^{\left( N\right) }\left( \mathbf{r}\right) =\sum_{n}\phi
\left( \varepsilon _{n},\mathbf{r}\right) L_{n}^{\left( N\right) }$ in terms
of \emph{kinked} partial waves, $\phi \left( \varepsilon ,\mathbf{r}\right) ,
$ evaluated on the mesh, $\varepsilon _{0},...,\varepsilon _{N}.$ This basis
solves Schr\"{o}dinger's equation for the MT-potential to within an error $%
\propto \left( \varepsilon -\varepsilon _{0}\right) ...\left( \varepsilon
-\varepsilon _{N}\right) .$ The Lagrange matrix-coefficients, $L_{n}^{\left(
N\right) },$ as well as the Hamiltonian and overlap matrices for the NMTO
set, have simple expressions in terms of energy derivatives on the mesh of
the Green matrix, defined as the inverse of the screened KKR matrix. The
variationally determined single-electron energies have errors $\propto $ $%
\left( \varepsilon -\varepsilon _{0}\right) ^{2}...\left( \varepsilon
-\varepsilon _{N}\right) ^{2}.$ A method for obtaining orthonormal NMTO sets
is given and several applications are presented.
\end{abstract}

\section{Overview}

\label{over}Muffin-tin orbitals (MTOs) have been used for a long time in 
\textit{ab initio }calculations of the electronic structure of condensed
matter. Over the years, several MTO-based methods have been devised and
further developed. The ultimate aim is to find a generally applicable
electronic-structure method which is \emph{accurate} and \emph{robust, }as
well as \emph{intelligible.}

In order to be \emph{intelligible, }such a method must employ a small,
single-electron basis of atom-centered, short-ranged orbitals. Moreover, the
single-electron Hamiltonian must have a simple, analytical form, which
relates to a two-center, orthogonal, tight-binding (TB) Hamiltonian.

In this sense, the conventional linear muffin-tin-orbitals method in the 
\emph{atomic-spheres approximation} (LMTO-ASA) \cite{AnJep,Kanpur} is
intelligible, because the orbital may be expressed as: 
\begin{equation}
\chi _{RL}\left( \mathbf{r}_{R}\right) \;=\;\phi _{RL}\left( \mathbf{r}%
_{R}\right) +\sum_{R^{\prime }L^{\prime }}\dot{\phi}_{R^{\prime }L^{\prime
}}\left( \mathbf{r}_{R^{\prime }}\right) \left( H_{R^{\prime }L^{\prime
},RL}-\varepsilon _{\nu }\delta _{R^{\prime }R}\delta _{L^{\prime }L}\right)
.  \label{a1}
\end{equation}
Here, $\phi _{RL}\left( \mathbf{r}_{R}\right) $ is the solution, $\varphi
_{Rl}\left( \varepsilon _{\nu },r_{R}\right) Y_{lm}\left( \mathbf{\hat{r}}%
_{R}\right) ,$ at a chosen energy, $\varepsilon _{\nu },$ of
Schr\"{o}dinger's differential equation inside the atomic sphere at site $R$
for the single-particle potential, $\sum_{R}v_{R}\left( r_{R}\right) \,,$
assumed to be spherically symmetric inside that sphere$.$ Moreover, $\mathbf{%
r}_{R}\equiv \mathbf{r-R}$ and $L\equiv lm.\;$The function $\varphi
_{Rl}\left( \varepsilon ,r\right) $ thus satisfies the one-dimensional,
radial Schr\"{o}dinger equation 
\begin{equation}
\frac{\partial ^{2}}{\partial r^{2}}r\varphi _{Rl}\left( \varepsilon
,r\right) =-\left[ \varepsilon -v_{R}\left( r\right) -\frac{l\left(
l+1\right) }{r^{2}}\right] r\varphi _{Rl}\left( \varepsilon ,r\right) .
\label{a16}
\end{equation}
In (\ref{a1}), $\dot{\phi}_{RL}\left( \mathbf{r}\right) $ are the
energy-derivative functions, $\left. \partial \varphi _{Rl}\left(
\varepsilon ,r\right) /\partial \varepsilon \right| _{\varepsilon _{\nu
}}Y_{lm}\left( \mathbf{\hat{r}}\right) .$ The radial functions, $\varphi $
and $\dot{\varphi},$ and also the potential, $v,$ are truncated outside
their own atomic sphere of radius $s,$ and the matrix, $H,$ is constructed
in such a way that the LMTO is continuous and differentiable in all space.
Equation (\ref{a1}) therefore expresses the LMTO at site $R$ and (pseudo)
angular momentum $L$ as the solution of Schr\"{o}dinger's equation at that
site, with that angular momentum, and at the chosen energy, plus a
'smoothing cloud' of energy-derivative functions, centered mainly at the
neighboring sites, and having around these, all possible angular momenta.

That a set of energy-\emph{in}dependent orbitals must have the form (\ref{a1}%
) in order to constitute a basis for the solutions $\Psi _{i}\left( \mathbf{r%
}\right) $ --with energies $\varepsilon _{i}$ in the neighborhood of $%
\varepsilon _{\nu }$-- of Schr\"{o}dinger's equation for the \emph{entire}
system, is intuitively obvious, because the corresponding linear
combinations, $\sum_{RL}\chi _{RL}\left( \mathbf{r}_{R}\right) \,c_{RL,i},$
will be those which locally, inside each atomic sphere and for each angular
momentum, have the right amount of $\dot{\varphi}$ --provided mainly by the
tails of the neighboring orbitals-- added onto the central orbital's $%
\varphi .$ Since by construction each $\varphi _{Rl}\left( \varepsilon
,r\right) $ is the correct solution, this right amount is of course $%
\varepsilon _{i}-\varepsilon _{\nu }.$ In math: since definitions can be
made such that the expansion matrix $H_{R^{\prime }L^{\prime },RL}$ is \emph{%
Hermitian,} its \emph{eigenvectors} are the coefficients of the proper
linear combinations, and its \emph{eigenvalues} are the energies: 
\begin{eqnarray}
\sum_{RL}\chi _{RL}\left( \mathbf{r}_{R}\right) \,c_{RL,i}\; &=&\;\sum_{RL}%
\left[ \phi _{RL}\left( \mathbf{r}_{R}\right) +\left( \varepsilon
_{i}-\varepsilon _{\nu }\right) \dot{\phi}_{RL}\left( \mathbf{r}_{R}\right) %
\right] c_{RL,i}  \notag \\
&\approx &\;\sum_{RL}\phi _{RL}\left( \varepsilon _{i},\mathbf{r}_{R}\right)
c_{RL,i}\;=\;\Psi _{i}\left( \mathbf{r}\right) .  \label{a2}
\end{eqnarray}
Hence, $H$ is a 1st-order \emph{Hamiltonian,} delivering energies and wave
functions with errors proportional to $\left( \varepsilon _{i}-\varepsilon
_{\nu }\right) ^{2},$ to leading order.

First-order energies seldom suffice, and in the conventional LMTO-ASA method
use is made of the variational principle for the Hamiltonian, 
\begin{equation}
\mathcal{H\;}\equiv \;-\nabla ^{2}\;+\;\sum\nolimits_{R}v_{R}\left(
r_{R}\right) \,,  \label{a20}
\end{equation}
so that errors of order $\left( \varepsilon _{i}-\varepsilon _{\nu }\right)
^{2}$ in the basis set merely give rise to errors of order $\left(
\varepsilon _{i}-\varepsilon _{\nu }\right) ^{4}$ in the energies. With that
approach, the energies and eigenvectors are obtained as solutions of the
generalized eigenvalue problem: 
\begin{equation}
\sum_{RL}\left[ \left\langle \chi _{R^{\prime }L^{\prime }}\left| \mathcal{H}%
-\varepsilon _{\nu }\right| \chi _{RL}\right\rangle -\left( \varepsilon
_{i}-\varepsilon _{\nu }\right) \left\langle \chi _{R^{\prime }L^{\prime
}}\mid \chi _{RL}\right\rangle \right] c_{RL,i}=0,  \label{a3}
\end{equation}
for all $R^{\prime }L^{\prime }.$ If we now insert (\ref{a1}) in (\ref{a3}),
we see that the Hamiltonian and overlap matrices are expressed in terms of
the 1st-order Hamiltonian, $H,$ plus two \emph{diagonal} matrices with the
respective elements 
\begin{equation*}
\left\langle \phi _{RL}\mid \dot{\phi}_{RL}\right\rangle
=\textstyle{\int_{0}^{s}}
\varphi _{Rl}\left( r\right) \dot{\varphi}_{Rl}\left( r\right)
r^{2}dr,\;\;\;\left\langle \dot{\phi}_{RL}\mid \dot{\phi}%
_{RL}\right\rangle =\int_{0}^{s}\dot{\varphi}_{Rl}\left( r\right)
^{2}r^{2}dr.
\end{equation*}
These matrices are diagonal by virtue of the ASA, which approximates
integrals over space by the sum of integrals over atomic spheres. If each
partial wave is normalized to unity in its sphere: $\int_{0}^{s}\varphi
_{Rl}\left( r\right) ^{2}r^{2}dr=1,$ then $\left\langle \phi \mid \phi
\right\rangle $ is the unit matrix in the ASA, and the Hamiltonian and
overlap matrices entering (\ref{a3}) take the simple forms: 
\begin{eqnarray}
\left\langle \chi \left| \mathcal{H}-\varepsilon _{\nu }\right| \chi
\right\rangle \; &=&\;\left( H-\varepsilon _{\nu }\right) \left[
1+\left\langle \phi \mid \dot{\phi}\right\rangle \left( H-\varepsilon _{\nu
}\right) \right]  \label{a4} \\[0.15cm]
\left\langle \chi \mid \chi \right\rangle \; &=&\;\left[ 1+\left(
H-\varepsilon _{\nu }\right) \left\langle \dot{\phi}\mid \phi \right\rangle %
\right] \left[ 1+\left\langle \phi \mid \dot{\phi}\right\rangle \left(
H-\varepsilon _{\nu }\right) \right]  \notag \\
&&\quad \quad \quad +\;\left( H-\varepsilon _{\nu }\right) \left[
\left\langle \dot{\phi}\mid \dot{\phi}\right\rangle -\left\langle \phi \mid 
\dot{\phi}\right\rangle ^{2}\right] \left( H-\varepsilon _{\nu }\right) . 
\notag
\end{eqnarray}
Here and in the following we use a vector-matrix notation according to
which, for example $\chi _{RL}\left( \mathbf{r}_{R}\right) $ and $\chi
_{RL}\left( \mathbf{r}_{R}\right) ^{\ast }$ are considered components of
respectively a row-vector, $\chi \left( \mathbf{r}\right) ,$ and a
column-vector, $\chi \left( \mathbf{r}\right) ^{\dagger }.$ The eigenvector, 
$c_{i},$ is a column vector with components $c_{RL,i}.$ Moreover, $1$ is the
unit matrix, $\varepsilon _{\nu }$ is a diagonal matrix, and $H$ is a
Hermitian matrix. Vectors and diagonal matrices are denoted by lower-case
Latin and Greek characters, and matrices by upper-case Latin characters.
Exceptions to this rule are: $Y\left( \mathbf{\hat{r}}\right) ,$ the vector
of spherical harmonics, the site and angular-momentum indices (subscripts) $%
R,\,L,\,I,$ and $A,$ and the orders (superscripts) $L,\,M,$ and $N.$
Operators are given in calligraphic, like $\mathcal{H},$ and an omitted
energy argument means that $\varepsilon =\varepsilon _{\nu }.$

With the $\phi \left( \mathbf{r}\right) $'s being orthonormal in the ASA,
the LMTO overlap matrix in (\ref{a4}) is seen to factorize to 1st order, and
it is therefore simple to transform to a set of \emph{nearly orthonormal }%
LMTOs: 
\begin{eqnarray}
\hat{\chi}\left( \mathbf{r}\right) \; &=&\;\chi \left( \mathbf{r}\right) %
\left[ 1+\left\langle \phi \mid \dot{\phi}\right\rangle \left( H-\varepsilon
_{\nu }\right) \right] ^{-1}  \label{a5} \\[0.15cm]
\left\langle \hat{\chi}\left| \mathcal{H}-\varepsilon _{\nu }\right| \hat{%
\chi}\right\rangle \; &\equiv &\;\hat{H}-\varepsilon _{\nu }\;=\;\left[
1+\left( H-\varepsilon _{\nu }\right) \left\langle \dot{\phi}\mid \phi
\right\rangle \right] ^{-1}\left( H-\varepsilon _{\nu }\right)  \notag \\
&=&\;H-\varepsilon _{\nu }\;-\;\left( H-\varepsilon _{\nu }\right)
\left\langle \dot{\phi}\mid \phi \right\rangle \left( H-\varepsilon _{\nu
}\right) \;+\;...  \notag \\[0.15cm]
\left\langle \hat{\chi}\mid \hat{\chi}\right\rangle \; &=&\;1\;+\;\left( 
\hat{H}-\varepsilon _{\nu }\right) \left\langle \overset{.}{\hat{\phi}}\mid 
\overset{.}{\hat{\phi}}\right\rangle \left( \hat{H}-\varepsilon _{\nu
}\right)  \notag \\[0.15cm]
\overset{.}{\hat{\phi}}\left( \mathbf{r}\right) \; &\equiv &\;\dot{\phi}%
\left( \mathbf{r}\right) \;-\;\phi \left( \mathbf{r}\right) \left\langle
\phi \mid \dot{\phi}\right\rangle .  \notag
\end{eqnarray}
Here, the energy-derivative function, $\overset{.}{\hat{\phi}}\left( \mathbf{%
r}\right) ,$ equals $\dot{\phi}\left( \mathbf{r}\right) ,$ orthogonalized to 
$\phi \left( \mathbf{r}\right) .$ Finally, we may transform to a set of 
\emph{orthonormal }LMTOs: 
\begin{eqnarray}
&&\check{\chi}\left( \mathbf{r}\right) \;=\;\hat{\chi}\left( \mathbf{r}%
\right) \left[ 1+\left( \hat{H}-\varepsilon _{\nu }\right) \left\langle 
\overset{.}{\hat{\phi}}\mid \overset{.}{\hat{\phi}}\right\rangle \left( \hat{%
H}-\varepsilon _{\nu }\right) \right] ^{-1/2}\;=  \label{a6} \\
&&\quad \quad \quad \quad \quad \quad \quad \quad \quad \quad \hat{\chi}%
\left( \mathbf{r}\right) \left[ 1-\frac{1}{2}\left( \hat{H}-\varepsilon
_{\nu }\right) \left\langle \overset{.}{\hat{\phi}}\mid \overset{.}{\hat{\phi%
}}\right\rangle \left( \hat{H}-\varepsilon _{\nu }\right) +..\right]  \notag
\\[0.15cm]
&&\left\langle \check{\chi}\left| \mathcal{H}-\varepsilon _{\nu }\right| 
\check{\chi}\right\rangle \;\equiv \;\check{H}-\varepsilon _{\nu }\;=\;\hat{H%
}-\varepsilon _{\nu }-  \notag \\
&&\frac{1}{2}\left( \hat{H}-\varepsilon _{\nu }\right) \left\langle \overset{%
.}{\hat{\phi}}\mid \overset{.}{\hat{\phi}}\right\rangle \left( \hat{H}%
-\varepsilon _{\nu }\right) ^{2}-\frac{1}{2}\left( \hat{H}-\varepsilon _{\nu
}\right) ^{2}\left\langle \overset{.}{\hat{\phi}}\mid \overset{.}{\hat{\phi}}%
\right\rangle \left( \hat{H}-\varepsilon _{\nu }\right) +..  \notag
\end{eqnarray}
We thus realize that of the Hamiltonians considered, $H$ is of 1st, $\hat{H}$
is of 2nd, and $\check{H}$ is of 3rd order. As the order increases, and the
energy window --inside which the eigenvalues of the Hamiltonian are useful
as single-electron energies-- widens, the real-space \emph{range} of the
Hamiltonian increases. For real-space calculations \cite
{Pessoa,Nowak,Bose,Vargas,Mook}, it is therefore important to be able to
express a higher-order Hamiltonian as a power series in a lower-order
Hamiltonian like in (\ref{a5}) and (\ref{a6}), because such a series may be
truncated when the energy window is sufficiently wide.

The energy-derivative of the radial function $\varphi \left( \varepsilon
,r\right) $ depends on the \emph{energy derivative} of its \emph{%
normalization.} If we choose to normalize according to: $\int_{0}^{s}\hat{%
\varphi}\left( \varepsilon ,r\right) ^{2}r^{2}dr=1,$ then it follows that $%
\int_{0}^{s}\hat{\varphi}\left( r\right) \overset{.}{\hat{\varphi}}\left(
r\right) r^{2}dr=0.$ Choosing another energy-dependent normalization: $%
\varphi \left( \varepsilon ,r\right) \equiv \hat{\varphi}\left( \varepsilon
,r\right) \left[ 1+\left( \varepsilon -\varepsilon _{\nu }\right) o\right] ,$
specified by a constant $o,$ then we see that: $\dot{\varphi}\left( r\right)
=\overset{.}{\hat{\varphi}}\left( r\right) +\varphi \left( r\right) o.$
Changing the energy derivative of the normalization thus adds some $\varphi
\left( r\right) $ to $\overset{.}{\hat{\varphi}}\left( r\right) $ and
thereby changes the shape of the 'tail function' $\dot{\varphi}\left(
r\right) .$ Since all LMTOs (\ref{a1}) should remain smooth upon this
change, also $H$ must change, and so must all LMTOs in the set. The diagonal
matrix $\left\langle \phi \mid \dot{\phi}\right\rangle ,$ whose elements are
the radial overlap integrals: $o=\int_{0}^{s}\varphi \left( r\right) \dot{%
\varphi}\left( r\right) r^{2}dr,$ thus determines the LMTO \emph{%
representation, }and the first and the last equations (\ref{a5}) specify the
linear transformation between representations. Values of the diagonal matrix 
$\left\langle \phi \mid \dot{\phi}\right\rangle $ exist, which yield \emph{%
short range} for the 1st-order Hamiltonian $H$ and, hence, for the LMTO set (%
\ref{a1}). Such an $H$ is therefore a \emph{two-center TB Hamiltonian} and
such an LMTO set is a \emph{first-principles TB basis.}

In order to obtain an explicit expression for $H,$ one needs to find the
spherical-harmonics expansions about the various site for a set of \emph{%
smooth} MTO \emph{envelope} functions. For a MT-potential, which is flat in
the interstitial, the envelope functions are wave-equation solutions with
pure spherical-harmonics character near the sites. Consistent with the idea
behind the ASA --to use 'space-filling spheres'-- is the use of envelope
functions with \emph{fixed} energy, specifically \emph{zero, }which is a
reasonable approximation for the kinetic energy between the atoms for a
valence state. The envelope functions in the ASA are thus \emph{screened
multipole potentials,} with the screening specified by a \emph{diagonal
matrix} of \emph{screening constants,} $\alpha _{Rl},$ related to the radial
overlaps $o_{Rl}$. The expansion of a \emph{bare} multipole potential at
site $R$ about a different site $R^{\prime }$ is well known: 
\begin{equation*}
\frac{Y_{L}\left( \mathbf{\hat{r}}_{R}\right) }{r_{R}^{l+1}}\sim
\sum_{R^{\prime }L^{\prime }}r_{R^{\prime }}^{l^{\prime }}Y_{L^{\prime
}}\left( \mathbf{\hat{r}}_{R^{\prime }}\right) \,\frac{Y_{l^{\prime \prime
}m^{\prime \prime }}\left( \widehat{\mathbf{R}^{\prime }\mathbf{-R}}\right) 
}{\left| \mathbf{R}^{\prime }\mathbf{-R}\right| ^{l^{\prime \prime }+1}}\sim
\sum_{R^{\prime }L^{\prime }}\,r_{R^{\prime }}^{l^{\prime }}Y_{L^{\prime
}}\left( \mathbf{\hat{r}}_{R^{\prime }}\right) \,\mathsf{S}_{R^{\prime
}L^{\prime },RL}^{0}\,.
\end{equation*}
Here, $l^{\prime \prime }\equiv l^{\prime }+l$ and $m^{\prime \prime }\equiv
m^{\prime }-m.$ With suitable normalizations, the \emph{bare structure
matrix,} $\mathsf{S}^{0},$ can be made Hermitian. The \emph{screened}
structure matrix is now related to the bare one through a Dyson equation: 
\begin{equation}
\left( \mathsf{S}^{\alpha }\right) ^{-1}\;\mathsf{=\;}\left( \mathsf{S}%
^{0}\right) ^{-1}-\,\alpha ,  \label{a12}
\end{equation}
which may be solved by inversion of the matrix $\mathsf{S}^{0}-\alpha ^{-1}.$
This inversion may be performed in \emph{real space,} that is in $\mathbf{R}$%
- rather than in $\mathbf{k}$-representation, provided that the screening
constants take values known from experience to give a short-ranged $\mathsf{S%
}^{\alpha }.$

In the end, it turns out that \emph{all} ingredients to the LMTO Hamiltonian
and overlap integrals$,$ $H,$ $\left\langle \phi \mid \dot{\phi}%
\right\rangle ,$ and $\left\langle \dot{\phi}\mid \dot{\phi}\right\rangle ,$
may be obtained from the screened Korringa-Kohn-Rostoker (KKR) matrix in the
ASA: 
\begin{equation}
\mathsf{K}_{R^{\prime }L^{\prime },RL}^{\alpha }\left( \varepsilon \right)
\equiv \mathsf{p}_{Rl}^{\alpha }\left( \varepsilon \right) \delta
_{R^{\prime }R}\delta _{L^{\prime }L}-\mathsf{S}_{R^{\prime }L^{\prime
},RL}^{\alpha }.  \label{a8}
\end{equation}
Here, $\mathsf{p}^{0}\left( \varepsilon \right) $ is a diagonal matrix of
potential functions obtained from the radial logarithmic derivative
functions, $\partial \left\{ \varphi \left( \varepsilon ,s\right) \right\}
\equiv \partial \ln \left| \varphi \left( \varepsilon ,r\right) \right|
\left/ \partial \ln r\right| _{s},$ evaluated at the MT-radius, and $\mathsf{%
p}^{\alpha }\left( \varepsilon \right) $ is related to $\mathsf{p}^{0}\left(
\varepsilon \right) $ via the diagonal version of Equation (\ref{a12})$.$
The results are: 
\begin{eqnarray}
&&H\;=\;\varepsilon _{\nu }-K\;=\;\varepsilon _{\nu }-\mathsf{p\dot{p}}^{-1}+%
\mathsf{\dot{p}}^{-\frac{1}{2}}\mathsf{S}\,\mathsf{\dot{p}}^{-\frac{1}{2}%
}\;\equiv \;\mathsf{c}+\mathsf{d}^{\frac{1}{2}}\mathsf{S}\,\mathsf{d}^{\frac{%
1}{2}},  \notag \\[0.1cm]
&&\left\langle \phi \mid \dot{\phi}\right\rangle =\frac{\ddot{K}}{2!}=\frac{1%
}{2!}\frac{\mathsf{\ddot{p}}}{\mathsf{\dot{p}}}\,,\quad \quad \left\langle 
\dot{\phi}\mid \dot{\phi}\right\rangle =\frac{\dddot{K}}{3!}=\frac{1}{3!}%
\frac{\mathsf{\dddot{p}}}{\mathsf{\dot{p}}}\,,  \label{a10}
\end{eqnarray}
expressed in terms of the KKR matrix, renormalized to have $\dot{K}=1:$%
\begin{equation}
K\left( \varepsilon \right) \;\equiv \;\mathsf{\dot{K}}^{-\frac{1}{2}}\,%
\mathsf{K}\left( \varepsilon \right) \mathsf{\dot{K}}^{-\frac{1}{2}}\;=\;%
\mathsf{p}\left( \varepsilon \right) \mathsf{\dot{p}}^{-1}-\mathsf{\dot{p}}%
^{-\frac{1}{2}}\mathsf{S}\,\mathsf{\dot{p}}^{-\frac{1}{2}}.  \label{a9}
\end{equation}
This corresponds to the partial-wave normalization: $\int_{0}^{s}\varphi
\left( r\right) ^{2}r^{2}dr=1,$ and $K\left( \varepsilon \right) $ is what
in the 2nd-generation method \cite{AnJep,Kanpur} is denoted $-h\left(
\varepsilon \right) ,$ but since the current notation identifies matrices by
capitals, we have changed. The LMTO Hamiltonian and overlap matrices are
thus expressed solely in terms of the structure matrix $\mathsf{S}$ and the
potential functions $\mathsf{p}\left( \varepsilon \right) ,$ specifically
the diagonal matrices $\mathsf{p},$ $\mathsf{\dot{p}},\;\mathsf{\ddot{p}},%
\mathrm{\;}$and$\;\mathsf{\dddot{p}}$. It may be realized that the
nearly-orthonormal representation is generated if the diagonal screening
matrix in (\ref{a12}) is set to the value $\gamma ,$ which makes $\mathsf{%
\ddot{p}}^{\gamma }$ vanish.

For calculations \cite{Abrikosov,Ruban,Turek} which employ the \emph{%
coherent-potential approximation} (CPA) to treat substitutional disorder, it
is important to be able to perform screening transformations of the \emph{%
Green matrix:} 
\begin{equation}
\mathsf{G}^{\alpha }\left( z\right) \equiv \mathsf{K}^{\alpha }\left(
z\right) ^{-1}=\left[ \mathsf{p}^{\alpha }\left( z\right) -\mathsf{S}%
^{\alpha }\right] ^{-1},  \label{a19}
\end{equation}
also called the resolvent, or the scattering path operator in multiple
scattering theory \cite{Weinberger}. In the 2nd generation MTO formalism, $%
\mathsf{G}^{a}\left( \varepsilon \right) $ was denoted $g^{\alpha }\left(
\varepsilon \right) .$ This screening transformation is: 
\begin{equation}
\mathsf{G}^{\beta }\left( z\right) =\left( \beta -\alpha \right) \frac{%
\mathsf{p}^{\alpha }\left( z\right) }{\mathsf{p}^{\beta }\left( z\right) }+%
\frac{\mathsf{p}^{\alpha }\left( z\right) }{\mathsf{p}^{\beta }\left(
z\right) }\mathsf{G}^{\alpha }\left( z\right) \frac{\mathsf{p}^{\alpha
}\left( z\right) }{\mathsf{p}^{\beta }\left( z\right) },  \label{a11}
\end{equation}
and is seen to involve no matrix multiplications, but merely
energy-dependent rescaling of matrix elements. As a transformation between
the nearly orthonormal, $\beta \mathrm{=}\gamma ,$ and the short-ranged
TB-representation, Eq. (\ref{a11}) has been useful also in \emph{%
Green-function} calculations for extended defects, surfaces, and interfaces 
\cite{Abrikosov,Turek,LamG,Green,Schilf}. However, calculations which start
out from the unperturbed Green matrices most natural for the problem
--namely those obtained from LMTO band-structure calculations in the nearly
orthonormal representation for the \emph{bulk} systems-- have usually been
limited to \emph{2nd-order} in $z-\varepsilon _{\nu }$, because $\mathsf{p}%
^{\gamma }\left( z\right) $ is linear to this order, and because attempts to
use 3rd-order expressions for $\mathsf{p}^{\gamma }\left( z\right) $
employing the potential parameter $\mathsf{\dddot{p}}^{\gamma }=3!\,\mathsf{%
\dot{p}}^{\gamma }\left\langle \overset{.}{\hat{\phi}}\mid \overset{.}{\hat{%
\phi}}\right\rangle ,$ induced false poles in the Green matrix.

What is \emph{not }intelligible in the TB-LMTO-ASA method is that the LMTO 
\emph{expansion} (\ref{a1}) must include \emph{all} $L^{\prime }$'s until
convergence is reached throughout each sphere, and \emph{all} $R^{\prime }$%
's until space is covered with spheres. This means that the LMTO-ASA basis
is \emph{minimal --}at most--\emph{\ }for elemental, closely packed
transition metals, the case for which it was in fact invented \cite{An}. The
supreme computational efficiency of the method soon made \emph{%
self-consistent} density-functional \cite{HKS} calculations possible, and
not only for elemental transition metals, but also for compounds. In order
to treat open structures such as diamond, \emph{empty} spheres were
introduced as a device for describing the repulsive potentials in the
interstices \cite{emp}. All of this then, led to misinterpretations of the
wave-function related output of such calculations in terms of the components
of the one-center expansions (\ref{a1}), typically the numbers of $s,$ $p,$
and $d$ electrons on the various atoms (including in the empty spheres!) and
the charge transfers between them. Absurd statements to the effect that CsCl
is basically a neutral compound with the Cs electron having a bit of $s$-,
more $p$-, quite some $d$-, and a bit of $f$-character were not uncommon.
Many practitioners of the ASA method did not realize that the role of the
MT-spheres is to describe the \emph{input potential, }rather than the output
wave-functions. For the latter, the one-center expansions truncated outside
the spheres constitute merely a decomposition which is used in the code for
selfconsistent calculations. The strange Cs electron is therefore little
more than the expansion about the Cs site of the tails of the neighboring Cl 
$p$ electrons spilling into the Cs sphere. That latter MT-sphere must of
course be chosen to have about the same size as that of Cl, because only
then is the shape of the Cs$^{+}$Cl$^{-}$ \emph{potential} in the
bi-partitioned structure well described.

Now, the so-called \emph{high }partial waves --they are those which are
shaped like $r^{l}$ in the outer part of the sphere where the potential
flattens out-- do enter the LMTO expansion (\ref{a1}), but \emph{not} the
eigenvalue problem (\ref{a3}) or the equivalent KKR equation: 
\begin{equation}
K\left( \varepsilon _{i}\right) c_{i}=0,  \label{a13}
\end{equation}
because they are part of the MTO envelope functions. This property of having
the high-$l$ limit correct is a strength of the MTO method, not shared by
for instance Gaussian orbitals, which are solutions of (\ref{a16}) for a 
\emph{parabolic} potential. There are, however, also other partial waves
--like the Cs $s$-waves, $d$-waves in non-transition metal atoms, $f$-waves
in transition-metal atoms, $s$-waves in oxygen and fluorine, and in positive
alkaline ions, and all partial waves in empty spheres-- which for the
problem at hand are judged to be \emph{inactive} and should therefore not
have corresponding LMTOs in the basis. In order to get rid of such inactive
LMTOs, one must first --by means of (\ref{a12}) or (\ref{a11})-- transform
to a representation in which the inactive partial waves appear only in the
'tails' (second term of (\ref{a1})) of the remaining LMTOs; only thereafter,
the inactive LMTOs can be deleted. This \emph{down-folding} procedure works
for the LMTO-ASA method, but it messes up the connection between the LMTO
Hamiltonian (\ref{a4})-(\ref{a9}) and the KKR Green-function formalisms (\ref
{a10})-(\ref{a13}), and it is not as efficient as one would have liked it to
be \cite{Kanpur}.\textit{\ }E.g.\textit{, }the Si valence band cannot be
described with an $sp$ LMTO basis set derived by down-folding of the Si $d$-
as well as all empty-sphere partial waves \cite{Lam}.

The basic reason for these failures is that the ASA envelopes are chosen to
be independent of energy --in order to avoid energy dependence of the
structure matrix-- because this is what forces us to carry out \emph{%
explicitly }the integrals involving all partial waves in all spheres
throughout space. What should be done is to include all inactive waves, $%
\varphi _{I}\left( \varepsilon ,r\right) ,$ in energy-\emph{de}pendent MTO-%
\emph{envelopes}, and \emph{then} to linearize these MTOs to form LMTOs.
This has been achieved with the development of the LMTO method of the \emph{%
3rd-generation }\cite{Trieste,MRS}, and will be dealt with in the present
paper. The reason why energy linearization still works in a window of useful
width, now that the energy dependence is kept throughout space, is due to
the \emph{screening} of the wave-equation solutions used as envelope
functions \cite{Boston}.

As an extreme example, it was demonstrated in Fig. 7 of Ref. \cite{MRS}
--and we shall present further results in Fig. \ref{Fig10} below-- how with
this method one may pick the orbital of \emph{one} band, with a particular
local symmetry and energy range, out of a complex of overlapping bands. This
goes beyond the construction of a Wannier function and has relevance for the
treatment of correlated electrons in narrow bands \cite{Tan,Ladder}. Another
example to be treated in the present paper is the valence and low-lying
conduction-band structure of GaAs calculated with the minimal Ga $spd$ As $sp
$ basis \cite{PolyT}. Other examples, not treated in this paper, concern the
calculation of \emph{chemical indicators,} such as the
crystal-orbital-overlap-projected densities of states (COOPs) \cite{COOP}
for describing chemical pair bonding. These indicators were originally
developed for the empirical H\"{u}ckel method where all parameters have been
standardized. When one tries to take this over to an \textit{ab initio}
method, one immediately gets confronted with the problems of \emph{%
representation.} For instance, COOPs will vanish in a basis of orthonormal
orbitals. Therefore, the COOPs first had to be substituted by COHPs, which
are Hamiltonian- rather than overlap projections, but still, the LMTO-ASA
method often gave strange results --for the above mentioned reasons \cite
{Dronsk}. What one has to do is --through downfolding-- to chose the \emph{%
chemically-correct} LMTO Hilbert space and --through screening-- choose the
chemically correct axes (orbitals) in this space. Only with such orbitals,
does it make sense to compute indicators \cite{Felser,Florent}.

A current criterion for an electronic-structure method to be \emph{accurate}
and \emph{robust }is that it can be used in \textit{ab initio }%
density-functional molecular-dynamics (DF-MD) calculations \cite{CarPar}.
According to this criterion, \emph{hardly }any existing LMTO method --and
the LMTO-ASA least of all-- is accurate and robust.

Most LMTO calculations include non-ASA corrections to the Hamiltonian and
overlap matrices, such as the \emph{combined correction} for the neglected
integrals over the interstitial region and the neglected high partial waves.
This brings in the first energy derivative of the structure matrix, $\mathsf{%
\dot{S}},$ in a way which makes the formalism clumsy \cite{Kanpur}. The code 
\cite{LMTO47} for the 2nd-generation LMTO method is useful \cite{Z} and
quite accurate for calculating energy bands, because it includes downfolding
in addition to the combined correction, as well as an automatic way of
dividing space into MT-spheres, but the underlying formalism is complicated.

There certainly \emph{are} LMTO methods sufficiently accurate to provide
structural energies and forces within density-functional theory \cite
{Wills,Savrasov,MethFP,Meth,Spring,Weyrich,Herman,GHJ,Woo}, but their basis
functions are defined with respect to MT-potentials which do not overlap. As
a consequence, in order to describe adequately the correspondingly large
interstitial region, these LMTO sets must include \emph{extra} degrees of
freedom, such as LMTOs centered at interstitial sites and LMTOs with more
than one radial quantum number. The latter include LMTOs with tails of
different kinetic energies (multiple kappa$\,$-sets) and LMTOs for semi-core
states. Moreover, these methods usually do not employ short-ranged
representations. Finally, since a non-overlapping MT potential is a poor
approximation to the self-consistent potential, these methods are forced to
include the matrix elements of the \emph{full potential.} Existing
full-potential methods are thus set up to provide final, numerical results
at relatively low cost, but since they are complicated, they have sofar
lacked the robustness needed for DF-MD, and their \emph{formalisms} provide
little insight to the physics and chemistry of the problem.

One of the early full-potential MTO methods did fold down extra orbitals and
furthermore contained a scheme by which the matrix elements of the full
potential could be efficiently approximated by integrals in \emph{overlapping%
} spheres \cite{Herman}. The formalism however remained complicated, and the
method apparently never took off. A decade later, it was shown \cite
{Boston,MRS} that the MT-potential, which defines the MTOs --and to which
the Hamiltonian (\ref{a20}) refers-- \emph{may} in fact have some overlap:
If one solves the \emph{exact} KKR equations \cite{KKR} with phase shifts
calculated for MT-wells which overlap, then the resulting wave function is
the one for the \emph{superposition }of these MT-wells, plus an error of 
\emph{2nd order} in the potential-overlap. This proof will be repeated in
Eq. (\ref{ovl}) of the present paper, and in Figs. \ref{Fig12} and \ref
{Fig13} we shall supplement the demonstration in Ref. \cite{MRS} that this
may be exploited to make the kind of extra LMTOs mentioned above
superfluous, provided that the MTO-envelopes have the proper energy
dependence, that is, provided that 3rd generation LMTOs are used. Presently
we can handle MT-potentials with up to $\sim $60\% radial overlap $\left(
s_{R}+s_{R^{\prime }}<1.6\left| \mathbf{R-R}^{\prime }\right| \right) $, and
it seems as if such potentials, with the MT-wells centered exclusively on
the atoms, are sufficiently realistic that we only need the minimal LMTO set
defined therefrom \cite{MRS,Catia}. It may even be that such fat
MT-potentials, without full-potential corrections to the Hamiltonian matrix,
will yield output charge densities which, when used in connection with the
Hohenberg-Kohn variational principle for the total energy \cite{HKS}, will
yield good structural energies \cite{Catia2}. Hence, we are getting rid of
one of the major obstacles to LMTO DF-MD calculations, the empty spheres.

Soon after the development of the TB-LMTO-ASA method, it was realized \cite
{Paw} that the \emph{full charge density} produced with this method --for
cases where atomic and interstitial MT-spheres fill space well-- is so
accurate, that it should suffice for the calculation of total energies,
provided that this charge density is used in connection with a variational
principle. However, it took ten years before the first successful
implementation was published \cite{Vitos}. The problem is as follows: The
charge density, $\rho \left( \mathbf{r}\right) =\sum_{i}^{occ}\left| \Psi
_{i}\left( \mathbf{r}\right) \right| ,$ is most simply obtained in the form
of one-center expansions: 
\begin{equation}
\rho \left( \mathbf{r}\right) =\sum_{R}\sum_{LL^{\prime }}\int_{occ}\phi
_{RL}\left( z,\mathbf{r}_{R}\right) \,{\rm Im}G_{RL,RL^{\prime }}\left(
z\right) \,\phi _{RL^{\prime }}\left( z,\mathbf{r}_{R}\right) ^{\ast }\frac{%
dz}{\pi },  \label{a14}
\end{equation}
where $G\left( z\right) \equiv K\left( z\right) ^{-1},$ as can be seen from (%
\ref{a1}) and (\ref{a2}), but these expansions have terribly bad $L$%
-convergence in the region between the atoms and cannot even be used to plot
the charge-density in that region. That was made possible by the
transformation to a short-ranged representation, because one could now use: 
\begin{equation}
\rho \left( \mathbf{r}\right) =\sum_{RL}\sum_{R^{\prime }L^{\prime }}\chi
_{RL}\left( \mathbf{r}_{R}\right) \,\left[ \int_{occ}{\rm Im}%
G_{RL,R^{\prime }L^{\prime }}\left( z\right) \,\frac{dz}{\pi }\right] \,\chi
_{R^{\prime }L^{\prime }}\left( \mathbf{r}_{R^{\prime }}\right) ^{\ast },
\label{a15}
\end{equation}
where the $L$-sums only run over active values, and where the double-sum
over sites converges fast. Nevertheless, to compute a value of $\chi
_{RL}\left( \mathbf{r}\right) $ with $\mathbf{r}$ far away from a site, one
must evaluate the LMTO envelope function, which is a superposition of the
bare ones, $Y_{L}\left( \mathbf{\hat{r}}_{R}\right) /r_{R}^{l+1},$ and this
means that (\ref{a15}) actually contains a 4-double summation over sites. At
that time, this appeared to make the evaluation of $\rho \left( \mathbf{r}%
\right) $ at a sufficient number of interstitial points too time-consuming
for DF-MD, although the full charge density from (\ref{a15}) was used
routinely for plotting the charge-density, the electron-localization
function \cite{Savin}, a.s.o. In order to evaluate the total energy, the
full charge density must also be expressed in a form practical for solving
the Poisson equation. If one insists on a real-space method, then fast
Fourier transformation is not an option. In Fig. \ref{Fig11} of the present
paper, we shall present results of a real-space scheme \cite{Ras,Ras2} used
in connection with 3rd-generation LMTOs for the phase diagram of Si \cite
{SiPhase}. This scheme is presently not\emph{\ }a full-potential, but a full
charge-density scheme, and the calculation of inter-atomic forces has still
not been implemented.

With 3rd generation LMTOs \cite{Trieste,MRS}, the simple ASA expressions (%
\ref{a1})-(\ref{a15}) still hold, provided that $\phi \left( \varepsilon ,%
\mathbf{r}\right) $ is suitably redefined, and that $K\left( \varepsilon
\right) $ is substituted by the proper screened KKR matrix whose \emph{%
structure matrix depends on energy.} The LMTO Hamiltonian and overlap
matrices are given in terms of $K,$ and its first three energy derivatives, $%
\dot{K},\;\ddot{K},\;$and $\overset{...}{K},$ which are not diagonal.
Downfolding, the interstitial region, and potential-overlap to first order
are now all included in this simple ASA-like formalism \cite{AnJep}. In due
course, we thus hope to be able to perform DF-MD calculations with an
electronic Hamiltonian which is little more complicated than (\ref{a4}), (%
\ref{a5}), or (\ref{a6}).

A final problem with the LMTO basis is that even with the conventional $spd$%
-basis and space-filling spheres, the LMTO set is insufficient for cases
where semi-core states and excited states must be described by \emph{one}
minimal basis set, and in \emph{one} energy panel. This problem becomes even
more acute in the 3rd-generation method where, due to the proper treatment
of the interstitial region, the expansion energy $\varepsilon _{\nu }$ must
be \emph{global,} that is, $\varepsilon _{\nu }$ is now the \emph{unit}
matrix times $\varepsilon _{\nu },$ rather than a \emph{diagonal} matrix
with elements $\varepsilon _{\nu Rl}\delta _{RR^{\prime }}\delta
_{LL^{\prime }}.$ The same problem was met when attempting to apply the
formally elegant relativistic, spin-polarized LMTO method of Ref. \cite{Soly}
to narrow, spin-orbit split $f$-bands. Finally, as MT-spheres get larger,
and as more partial waves are being folded into the MTO envelopes, the
energy window inside which the LMTO basis gives accurate results shrinks.
This means, that the 3rd-generation LMTO method described in \cite{MRS} may
not be sufficiently \emph{robust.}

The idea emerging from the LMTO construction (\ref{a1}) seems to be: Divide
space into local regions inside which Schr\"{o}dinger's equation separates
due to spherical symmetry and which are so small that the energy dependence
of the radial functions is weak over the energy range of interest. Then
expand this energy dependence in a Taylor series to first order around the
energy $\varepsilon _{\nu }$ at the center of interest: $\phi \left(
\varepsilon \mathbf{,r}\right) \approx \phi \left( \mathbf{r}\right) +\left(
\varepsilon -\varepsilon _{\nu }\right) \dot{\phi}\left( \mathbf{r}\right) .$
Finally, substitute the energy by a Hamiltonian to obtain the energy-\emph{in%
}dependent LMTO. The question therefore arises (Fig. \ref{Fig1}): Can we
develop a more general, \emph{polynomial }MTO scheme of degree $N,$ which
allows us to use an $N$th-order Taylor series or --more generally-- allows
us to use a mesh of $N+1$ \emph{discrete} energy points, and thereby obtain
good results over a wider energy range, \emph{without increasing the size of
the basis set\thinspace }? Such an NMTO scheme has recently been developed 
\cite{PolyO} and shown to be very powerful \cite{PolyT}. We shall preview it
in the present paper.

\input{fig1.tex}

Most aspects of the 3rd-generation LMTO method have been dealt with in a set
of lecture notes \cite{Trieste} and a recent review \cite{MRS}. Here, we
shall try to avoid repetition but, nevertheless, give a self-contained
description of two selected aspects of the new method: the \emph{basic
concepts} and the new \emph{polynomial NMTO scheme}, to be presented here
for the first time\emph{.}

We first explain (Sect. \ref{KPW}) what the functions $\phi \left(
\varepsilon ,\mathbf{r}\right) $ actually are in the 3rd-generation
formalism. This we do using conventional notation in terms of spherical
Bessel functions and phase shifts --like in Ref. \cite{Boston}-- and only
later, we renormalize to the notation used in Refs. \cite{Trieste} and \cite
{MRS}. It turns out that the \emph{bare} $\phi $'s are the energy-\emph{de}%
pendent MTOs of the 1st generation \cite{1gen}. The \emph{screened} $\phi $%
's are the screened, energy-dependent MTOs of the 2nd generation \cite
{Boston}, with the proviso that $\kappa ^{2}\equiv \varepsilon .$ This
proviso --together with truncations of the screening divergencies at the
sites, inside the so-called \emph{screening spheres}-- is what makes the
screened $\phi $'s equal to the so-called \emph{unitary }\cite{Trieste}\emph{%
\ }or \emph{kinked } \cite{MRS} \emph{partial waves }in the formalism of the
3rd generation. We then derive the screened KKR equations and repeat the
proof from Refs. \cite{Boston} and \cite{MRS} that overlapping MT-potentials
are treated correctly to leading (1st) order in the potential overlap.
Towards the end of this first section, we introduce the so-called \emph{%
contracted }Green function $\phi \left( \varepsilon ,\mathbf{r}\right)
G\left( \varepsilon \right) ,$ which will play a crucial role in the
development of the polynomial NMTO scheme, and we derive the 3rd-generation
version of the scaling relation (\ref{a11}) for screening the Green function.

In Sect. \ref{poly} we show how to get rid of the energy dependence of the
kinked-partial wave set: First, we introduce a set of energy-dependent
NMTOs, $\chi ^{\left( N\right) }\left( \varepsilon ,\mathbf{r}\right) ,$
which --like the $\phi \left( \varepsilon ,\mathbf{r}\right) $ set-- spans
the solutions of Schr\"{o}dinger's equation for the chosen MT-potential, and
whose contracted Green function, $\chi ^{\left( N\right) }\left( \varepsilon
,\mathbf{r}\right) G\left( \varepsilon \right) ,$ differs from $\phi \left(
\varepsilon ,\mathbf{r}\right) G\left( \varepsilon \right) $ by a function
which is \emph{analytical in energy.} Like in classical polynomial
approximations, we choose a mesh of arbitrarily spaced energies, $%
\varepsilon _{0},...,\varepsilon _{N},$ and subsequently adjust the
analytical function in such a way that, $\chi ^{\left( N\right) }\left(
\varepsilon _{0},\mathbf{r}\right) =...=\chi ^{\left( N\right) }\left(
\varepsilon _{N},\mathbf{r}\right) .$ The latter then, constitutes the set
of energy-\emph{in}dependent NMTOs. The 0th-order set, $\chi ^{\left(
0\right) }\left( \mathbf{r}\right) ,$ is seen to be the set of kinked
partial waves$,$ $\phi \left( \varepsilon _{0},\mathbf{r}\right) ,$ at the
energy $\varepsilon _{0},$ and the 1st-order set, $\chi ^{\left( 1\right)
}\left( \mathbf{r}\right) ,$ to be the set of tangent or chord-LMTOs
--depending on whether the mesh is condensed or discrete. For the case of a
condensed mesh --which is the simplest-- the matrices, which substitute for
the energies in the Taylor series (\ref{a1}) --generalized to $N$th order--
turn out to be: 
\begin{equation}
E^{\left( M\right) }-\varepsilon _{\nu }\;=\;\frac{\overset{\left(
M-1\right) }{G}}{\left( M-1\right) !}\left( \frac{\overset{\left( M\right) }{%
G}}{M!}\right) ^{-1},\quad \quad \mathrm{for\;}1\leq M\leq N,  \label{a24}
\end{equation}
in terms of the $M$th and the $\left( M-1\right) $st energy derivatives of
the Green matrix. Moreover, the expressions for the Hamiltonian and overlap
matrices are: 
\begin{eqnarray}
\left\langle \chi ^{\left( N\right) }\left| \mathcal{H}-\varepsilon _{\nu
}\right| \chi ^{\left( N\right) }\right\rangle \; &=&\;-\left( \frac{%
\overset{\left( N\right) }{G}}{N!}\right) ^{-1}\frac{\overset{\left(
2N\right) }{G}}{\left( 2N\right) !}\,\left( \frac{\overset{\left( N\right) }{%
G}}{N!}\right) ^{-1},  \label{a23} \\[0.15cm]
\left\langle \chi ^{\left( N\right) }\mid \chi ^{\left( N\right)
}\right\rangle \; &=&\;-\left( \frac{\overset{\left( N\right) }{G}}{N!}%
\right) ^{-1}\frac{\overset{\left( 2N+1\right) }{G}}{\left( 2N+1\right) !}%
\,\left( \frac{\overset{\left( N\right) }{G}}{N!}\right) ^{-1},  \notag
\end{eqnarray}
which, for $N=1,$ are easily seen to reduce to (\ref{a4}) upon insertion of (%
\ref{a10}). In retrospect, it is convenient that these basic NMTO results
are expressed in terms of energy derivatives of the Green matrix $G\left(
\varepsilon \right) $ --rather than in terms of those of its inverse ,the
KKR matrix, as we are used to from the LMTO-ASA method (\ref{a10})-- because
if we imagine generalizing (\ref{a1}) to $N$th order and using it to form
the Hamiltonian and overlap matrices like in (\ref{a4}), then each matrix
will consist of $N^{2}$ terms, among which a number of relations can be
shown to exist. We also realize, that the problem mentioned above about
using Green matrices beyond 2nd order in $z-\varepsilon _{\nu },$ is solved
by using --instead of $G\left( z\right) $-- the NMTO Green function: 
\begin{equation}
\left\langle \chi ^{\left( N\right) }\left| z-\mathcal{H}\right| \chi
^{\left( N\right) }\right\rangle ^{-1}=\frac{\overset{\left( N\right) }{G}}{%
N!}\left[ \frac{\overset{\left( 2N\right) }{G}}{\left( 2N\right) !}-\left(
z-\varepsilon _{\nu }\right) \frac{\overset{\left( 2N+1\right) }{G}}{\left(
2N+1\right) }\right] ^{-1}\frac{\overset{\left( N\right) }{G}}{N!},
\label{a25}
\end{equation}
which equals $G\left( z\right) $ to $\left( 2N+1\right) $st order. This
Green function has the additional advantage of allowing for a simple
treatment of non-MT perturbations. We admit that this route to
energy-independent MTO basis sets has little in common with the twisted path
we cut the first time, but once found, it is easy to accept and understand
the results --which are simple.

In practice, it is cumbersome to differentiate a KKR matrix --not to speak
of a Green matrix-- many times with respect to energy. Hence, one uses a
discrete energy mesh. With that, the derivatives in (\ref{a24}) and the pre-
and post factors in (\ref{a23}) and (\ref{a25}) turn out to be \emph{divided
differences,} while those at the centers of (\ref{a23}) turn out to be the
highest derivative of that approximating polynomial which is fitted not only
to the values of $G\left( \varepsilon \right) $ at the mesh points, but also
to its slopes. Hence, they are related to classical \emph{Hermite
interpolation }\cite{Hamming}.

In both Sections \ref{KPW} and \ref{poly}, special attention is paid to the
so-called \emph{triple-valuedness,} because this was not previously
explained in any detail, but has turned out to be crucial for the further
developments and will be even more so when we come to evaluate the
inter-atomic forces. A related aspect is the fact that a \emph{screening
transformation} in the formalism of the 3rd-generation is \emph{linear} as
regards the envelope functions, but \emph{non-linear} as regards the NMTOs.
This means, that changing the screening, changes the NMTO Hilbert space.
This was not the case for 2nd-generation LMTOs. This is the reason why we
took care to denote the nearly-orthonormal and orthonormal LMTO sets arrived
at by the \emph{linear} transformations (\ref{a5}) and (\ref{a6}) by
respectively $\hat{\chi}$ and $\check{\chi},$ rather than by $\chi ^{\gamma }
$ and $\chi ^{\perp },$ as in the 2nd-generation LMTO scheme, where
screening transformations were linear and denoted by superscripts. Screening
transformations like (\ref{a12}) and (\ref{a11}) still hold for the
3rd-generation structure- and Green-matrices, but the \emph{partial waves}
providing the spatial factors of the Green function (see(\ref{a14})) are 
\emph{different}: they have tails extending into the interstitial region. A
tail is attached continuously, but with a kink, at the screening sphere,
which is concentric with, but smaller than, its own MT-sphere, and the
resulting kinked partial wave,\emph{\ }or 0th-order energy-dependent MTO, is
--for the purpose of evaluating its properties in a simple, approximate
way-- triple-valued in the shell between these two spheres. The radii, $%
a_{RL},$ define the screening and determine the shape of the MTO envelopes.
Now, for a superposition of kinked partial waves given by a solution of the
KKR equations (\ref{a13}), the kinks and the triple-valuedness cancel, but
for a \emph{single} NMTO, a triple-valuedness of order $\left( r-a\right)
^{2N+1}\left( \varepsilon _{i}-\varepsilon _{0}\right) ...\left( \varepsilon
_{i}-\varepsilon _{N}\right) $ --which is the same as the error caused by
the energy interpolation-- remains. For this reason: The smaller the
screening radii --i.e.\textit{\ }the weaker the screening-- the smaller the
energy window inside which an energy-\emph{in}dependent NMTO set gives good
results. The extreme case is the \emph{bare }$\left( a\rightarrow 0\right) $ 
$N=0$ set, which is the set of 1st-generation MTOs \cite{1gen}, but defined 
\emph{without }freezing the energy dependence outside the central MT-sphere.
The tail-cancellation condition for this set leads to the original KKR
equations \cite{KKR}, which --we know-- must be solved energy-by-energy,
that is, the energy window can be very narrow, depending on the application.
Specifically, for free electrons the width is zero.

At the end of Sect. \ref{poly}, we demonstrate the power of the new NMTO
methods by applying the differential and discrete LMTO, QMTO, and CMTO
variational methods to the valence and conduction-band structure of GaAs
using a minimal Ga $spd$ As $sp$ basis, and to the conduction band of CaCuO$%
_{2}$ using only \emph{one} orbital, all others being removed by massive
downfolding \cite{PolyT}. We also give simple expressions for the charge
density and show the total energy as a function of volume for the various
crystalline phases of Si calculated with the full-charge, differential LMTO
method \cite{Ras,Ras2,SiPhase}. Finally, numerical results are presented for
the error of the valence-band energy of diamond-structured Si --as a
function of the potential overlap-- obtained from LMTOs constructed for a
potential whose MT-wells are centered exclusively on the atoms. In addition,
results of a scheme which corrects for the error of 2nd order in the overlap
will be presented \cite{Catia}.

In Sect. \ref{linear} we show that energy-dependent, linear transformations
of the set of kinked partial waves --such as a normalization-- merely leads
to similarity transformations among the NMTO basis functions and, hence,
does not change the Hilbert space spanned by the NMTO set.

This is exploited in Sect. \ref{ortho} to generate nearly orthonormal basis
sets, $\hat{\chi}^{\left( N\right) }\left( \mathbf{r}\right) ,$ for which
the energy matrices defined in (\ref{a24}) become Hermitian, Hamiltonian
matrices, $\hat{H}^{\left( M\right) }.$ We also show how to generate \emph{%
orthonormal} sets, $\check{\chi}^{\left( N\right) }\left( \mathbf{r}\right) ,
$ of general order, and we demonstrate by the example of the minimal MTO set
for GaAs that this technique works numerically efficiently --at least up to
and including $N=3.$ This development of orthonormal basis sets should be
important e.g. for the construction of correlated, multi-orbital
Hamiltonians for real materials \cite{Ladder,LDAU}.

In the last Sect. \ref{ASA} we show explicitly how --for $N=1$ and a
condensed mesh-- the general, nearly-orthonormal NMTO formalism reduces to
the simple ASA formalism of the present Overview.

In the Appendix we have derived those parts of the classical formalism for
polynomial approximation --Lagrange, Newton, and Hermite interpolation--
needed for the development of the NMTO method for discrete meshes \cite
{Hamming}.

\section{Kinked partial waves}

\label{KPW}In this section we shall define 0th-order energy-dependent MTOs
and show that linear combinations can be formed which solve Schr\"{o}%
dinger's equation for the MT-potential used to construct the MTOs. The
coefficients of these linear combinations are the solutions of the
(screened) KKR equations. By renormalization and truncation of the irregular
parts of the screened MTOs inside appropriately defined screening spheres,
these 0th-order energy-dependent MTOs become the kinked partial waves of the
3rd generation.

If we continue the regular solution $\varphi _{Rl}\left( \varepsilon
,r\right) $ of the radial Schr\"{o}dinger equation (\ref{a16}) for the
single potential well, $v_{R}\left( r\right) ,$ smoothly outside that well,
it becomes: 
\begin{equation}
\varphi _{Rl}\left( \varepsilon ,r\right) =n_{l}\left( \kappa r\right)
-j_{l}\left( \kappa r\right) \cot \eta _{Rl}\left( \varepsilon \right)
\equiv \varphi _{Rl}^{\circ }\left( \varepsilon ,r\right) ,\quad \mathrm{for}%
\;r>s_{R},  \label{b1}
\end{equation}
in terms of the spherical Bessel and Neumann functions, $j_{l}\left( \kappa
r\right) $ and $n_{l}\left( \kappa r\right) ,$ which are regular
respectively at the origin and at infinity, and a phase shift defined by: 
\begin{equation*}
\cot \eta \left( \varepsilon \right) =\frac{n\left( \kappa s\right) }{%
j\left( \kappa s\right) }\frac{\partial \ln \left| \varphi \left(
\varepsilon ,r\right) \right| \left/ \partial \ln r\right| _{s}-\partial \ln
\left| n\left( \kappa r\right) \right| \left/ \partial \ln r\right| _{s}}{%
\partial \ln \left| \varphi \left( \varepsilon ,r\right) \right| \left/
\partial \ln r\right| _{s}-\partial \ln \left| j\left( \kappa r\right)
\right| \left/ \partial \ln r\right| _{s}}.
\end{equation*}
In the latter expression, we have dropped the subscripts. Note that we no
longer distinguish between 'inside' and 'outside' kinetic energies, $%
\varepsilon -v\left( r\right) $ and $\kappa ^{2}\equiv \varepsilon -V_{mtz},$
and that we have returned to the common practice of setting $V_{mtz}\equiv 0.
$ If the energy is negative, $n_{l}\left( \kappa r\right) $ denotes a
spherical, exponentially decreasing Hankel function. Note also that --unlike
in the ASA-- the radial function is not truncated outside its MT-sphere, and
is not normalized to unity inside. In fact, we shall meet three different
normalizations throughout the bulk of this paper, and (\ref{b1}) is the
first.

\input{fig2.tex}

\subsection{Bare MTOs}

The bare, energy-dependent muffin-tin orbital (MTO) remains the one of the
1st generation \cite{1gen}: 
\begin{eqnarray}
\phi _{RL}\left( \varepsilon ,\mathbf{r}\right) &\equiv &Y_{L}\left( 
\widehat{\mathbf{r}}\right) \left[ \varphi _{Rl}\left( \varepsilon ,r\right)
+j_{l}\left( \kappa r\right) \cot \eta _{Rl}\left( \varepsilon \right) %
\right]  \notag \\[0.15cm]
&=&Y_{L}\left( \widehat{\mathbf{r}}\right) \left\{ 
\begin{array}{c}
\varphi _{Rl}\left( \varepsilon ,r\right) +j_{l}\left( \kappa r\right) \cot
\eta _{Rl}\left( \varepsilon \right) \quad \mathrm{for\quad }r\leq s_{R} \\ 
n_{l}\left( \kappa r\right) \quad \quad \quad \quad \quad \quad \quad \quad
\quad \quad \;\mathrm{for\quad }r>s_{R}
\end{array}
\right.  \notag \\[0.15cm]
&=&Y_{L}\left( \widehat{\mathbf{r}}\right) \left[ \varphi _{Rl}\left(
\varepsilon ,r\right) -\varphi _{Rl}^{\circ }\left( \varepsilon ,r\right)
+n_{l}\left( \kappa r\right) \right] ,  \label{b2}
\end{eqnarray}
and is seen to have pure angular momentum and to be regular in all space.
The reason for denoting this 0th-order MTO $\phi \left( \varepsilon ,\mathbf{%
r}\right) ,$ rather than $\chi ^{\left( N=0\right) }\left( \varepsilon ,%
\mathbf{r}\right) ,$ should become clear later.

In Fig. \ref{Fig2} we show the radial part of this MTO for a Si $p$-orbital,
a MT-sphere which is so large that it reaches 3/4 the distance to the next
site in the diamond lattice, and an energy in the valence-band, which --in
this case of a large MT-sphere-- is slightly negative (see Fig. 11 in Ref. 
\cite{MRS}). The full line shows the MTO as defined in (\ref{b2}), while the
various broken lines show it 'the 3-fold way': The radial Schr\"{o}dinger
equation for the potential $v\left( r\right) $ is integrated outwards, from
the origin to the MT radius, $s,$ yielding the regular solution, $\varphi
\left( \varepsilon ,r\right) ,$ shown by the dot-dashed curve. At $s,$ the
integration is continued with reversed direction and with the potential
substituted by the flat potential, whose value is defined as the zero of
energy. This inwards integration results in the radial function 'seen from
the outside of the atom', $\varphi ^{\circ }\left( \varepsilon ,r\right) ,$
shown by the dotted curve. The inwards integration is continued to the
origin, where $\varphi ^{\circ }\left( \varepsilon ,r\right) $ joins the
'outgoing' solution for the flat potential, that is the one which is regular
at infinity: $n\left( \kappa r\right) .$ The latter is the envelope function
for the bare MTO.

As usual, the envelope-function for the MTO centered at $\mathbf{R}$ may be
expanded in spherical-harmonics about another site $\mathbf{R}^{\prime }$ $%
\left( \neq \mathbf{R}\right) $: 
\begin{equation*}
\kappa n_{l}\left( \kappa r_{R}\right) Y_{L}\left( \mathbf{\hat{r}}%
_{R}\right) =\sum_{L^{\prime }}j_{l^{\prime }}\left( \kappa r_{R^{\prime
}}\right) Y_{L^{\prime }}\left( \mathbf{\hat{r}}_{R^{\prime }}\right)
B_{R^{\prime }L^{\prime },RL}\left( \varepsilon \right) ,
\end{equation*}
where the expansion coefficients form the Hermitian KKR structure matrix: 
\begin{equation}
B_{R^{\prime }L^{\prime },RL}\left( \varepsilon \right) \equiv \sum_{l"}4\pi
\,i^{-l+l^{\prime }-l^{\prime \prime }}C_{LL^{\prime }l^{\prime \prime
}}\,\kappa n_{l^{\prime \prime }}\left( \kappa \left| \mathbf{R-R}^{\prime
}\right| \right) Y_{l^{\prime \prime },m^{\prime \prime }}^{\ast }\left( 
\widehat{\mathbf{R-R}^{\prime }}\right)  \label{b38}
\end{equation}
as conventionally \cite{KKR} defined, albeit in $R$-space. The spherical
harmonics are as defined by Condon and Shortley, $m^{\prime \prime }\equiv
m^{\prime }-m,$ the summation runs over $l^{\prime \prime }=\left| l^{\prime
}-l\right| ,\;\left| l^{\prime }-l\right| +2,\,...,\;l^{\prime }+l,$ and $%
i^{-l+l^{\prime }-l^{\prime \prime }}$ is real, because $C_{LL^{\prime
}L^{\prime \prime }}\equiv \int Y_{L}(\hat{r})Y_{L^{\prime }}^{\ast }(\hat{r}%
)Y_{L^{\prime \prime }}(\hat{r})d\hat{r}$.

If for the on-site elements of $B\left( \varepsilon \right) ,$ we define: $%
B_{RL,RL^{\prime }}\left( \varepsilon \right) \equiv 0,$ and use the
notation: $f_{L}\left( \varepsilon ,\mathbf{r}_{R}\right) \equiv f_{l}\left(
\kappa r_{R}\right) Y_{L}\left( \mathbf{\hat{r}}_{R}\right) ,$ as well as
the vector-matrix notation introduced in connection with (\ref{a4}), we may
express the spherical-harmonics expansion of the bare envelope about any
site as: 
\begin{equation}
\kappa n\left( \varepsilon ,\mathbf{r}\right) =j\left( \varepsilon ,\mathbf{r%
}\right) B\left( \varepsilon \right) +\kappa n\left( \varepsilon ,\mathbf{r}%
\right) .  \label{b5}
\end{equation}
When we now form a linear combination, $\sum_{RL}\phi _{RL}\left(
\varepsilon ,\mathbf{r}_{R}\right) c_{RL},$ of energy-dependent MTOs (\ref
{b2}), and require that it be a solution of Schr\"{o}dinger's equation, then
the condition is that, inside any MT-sphere $\left( R^{\prime }\right) $ and
for any angular momentum $\left( L^{\prime }\right) ,$ the contributions
from the tails should cancel the $j_{l^{\prime }}\left( \kappa r\right) \cot
\eta _{R^{\prime }l^{\prime }}\left( \varepsilon \right) $-term from their
own MTO, $\phi _{R^{\prime }L^{\prime }}\left( \varepsilon ,\mathbf{r}%
_{R^{\prime }}\right) $, thus leaving behind the term $\varphi _{R^{\prime
}l^{\prime }}\left( \varepsilon ,r\right) ,$ which is a solution by
construction. This gives rise to the original KKR equations \cite{KKR}: 
\begin{eqnarray}
&&\sum_{RL}\left[ B_{R^{\prime }L^{\prime },RL}\left( \varepsilon
_{i}\right) +\kappa \cot \eta _{Rl}\left( \varepsilon _{i}\right) \delta
_{R^{\prime }R}\delta _{L^{\prime }L}\right] c_{RL,i}  \notag \\
&\equiv &\;\sum_{RL}K_{R^{\prime }L^{\prime },RL}\left( \varepsilon
_{i}\right) c_{RL,i}\;=\;0,  \label{b6}
\end{eqnarray}
which have non-zero solutions, $c_{RL,i},$ for those energies, $\varepsilon
_{i},$ where the determinant of the KKR matrix vanishes.

With those equations satisfied, the wave function is 
\begin{eqnarray}
\sum_{RL}\phi _{RL}\left( \varepsilon _{i},\mathbf{r}_{R}\right) c_{RL,i}
&=&\sum_{l^{\prime }=0}^{\infty }\sum_{m^{\prime }=-l^{\prime }}^{l^{\prime
}}\varphi _{R^{\prime }l^{\prime }}\left( \varepsilon _{i},r_{R^{\prime
}}\right) Y_{L^{\prime }}\left( \mathbf{\hat{r}}_{R^{\prime }}\right)
c_{R^{\prime }L^{\prime },i}+  \label{b7} \\
&&\sum_{R\neq R^{\prime }}\sum_{L}\left[ \varphi _{Rl}\left( \varepsilon
_{i},r_{R}\right) -\varphi _{Rl}^{\circ }\left( \varepsilon
_{i},r_{R}\right) \right] Y_{L}\left( \mathbf{\hat{r}}_{R}\right) c_{RL,i} 
\notag
\end{eqnarray}
near site $R^{\prime }$. Since according to (\ref{b1}) the function $\varphi
-\varphi ^{\circ }$ vanishes outside its own MT-sphere, the terms in the
second line vanish for a non-overlapping MT-potential so that, in this case,
(\ref{b7}) solves Schr\"{o}dinger's equation exactly. If the potential from
a neighboring site $\left( R\right) $ \emph{overlaps} the central site $%
\left( R^{\prime }\right) $, then $\varphi _{RL}-\varphi _{RL}^{\circ }$ 
\emph{tongues} stick into the MT-sphere at $R^{\prime }.$ The radial part of
such a tongue is $\frac{1}{2}\left( s_{R}-r_{R}\right) ^{2}v_{R}\left(
s_{R}\right) \varphi _{RL}\left( s_{R}\right) ,$ to lowest order in $%
s_{R}-r_{R},$ as may be seen from the radial Schr\"{o}dinger equation (\ref
{a16}). Let us now operate on the smooth function $\Psi _{i}\left( \mathbf{r}%
\right) \equiv \sum_{RL}\phi _{RL}\left( \varepsilon _{i},\mathbf{r}%
_{R}\right) c_{RL,i}\,,$ of which (\ref{b7}) is the expansion around site $%
R^{\prime },$ with $\mathcal{H}-\varepsilon _{i}$ as given by (\ref{a20}) to
find the error: 
\begin{eqnarray}
&&\left( \mathcal{H}-\varepsilon _{i}\right) \Psi _{i}\left( \mathbf{r}%
\right) =  \notag \\[0.15cm]
&&\sum_{R^{\prime }}v_{R^{\prime }}\left( r_{R^{\prime }}\right) \sum_{R\neq
R^{\prime }}\sum_{L}\left[ \varphi _{Rl}\left( \varepsilon _{i},r_{R}\right)
-\varphi _{Rl}^{\circ }\left( \varepsilon _{i},r_{R}\right) \right]
Y_{L}\left( \mathbf{\hat{r}}_{R}\right) c_{RL,i}  \label{ovl} \\[0.15cm]
&\sim &\frac{1}{2}\sum_{RR^{\prime }}^{pairs}v_{R^{\prime }}\left(
s_{R^{\prime }}\right) \left[ \left( s_{R^{\prime }}-r_{R^{\prime }}\right)
^{2}+\left( s_{R}-r_{R}\right) ^{2}\right] v_{R}\left( s_{R}\right) \Psi
_{i}\left( \mathbf{r}\right) .  \notag
\end{eqnarray}
This shows that the wave function (\ref{b7}) solves Schr\"{o}dinger's
equation for the \emph{superposition} of MT-wells to within an error, which
is of \emph{second }order in the potential overlap \cite{Boston,MRS}.

\subsection{Screened MTOs}

Screening is the characteristic of 2nd-generation MTOs and was first
discovered as the transformation (\ref{a5}) to a \emph{nearly-orthonormal}
representation, in which the Hamiltonian is of \emph{second} order \cite
{Gent,Gunnar}. Shortly thereafter it was realized that there exists a whole
set of screening transformations which may be used to make the orbitals 
\emph{short ranged}, so that the structure matrix may be generated in real
space. It was also realized that the screening transformation could be used
to downfold inactive channels and, hence, to produce \emph{minimal} basis
sets \cite{AnJep,Lam,Paw}. These applications were all for the ASA with $%
\kappa ^{2}$=0. Only long time after \cite{Boston}, did it become clear that
screening would work for \emph{positive }energies as well, and at that time
a fourth virtue of screening became clear, namely, that sceening the range
of the orbitals, simultaneously reduces their energy dependence \emph{to the
extent} that the full energy dependence may be kept in the interstitial
region, thus making the $\kappa ^{2}$=0-part of the ASA superfluous. Most of
this was shown in the last paper on the 2nd-generation formalism \cite
{Boston}. Nevertheless, this paper was unable to devise a generally useful
recipe for choosing the energy-dependent screening constants, it failed to
realize that screening allows the return to: $\kappa ^{2}\mathrm{=}%
\varepsilon ,$ and for those reasons it missed the elegant
energy-linearization of the MTOs achieved by the 3rd generation.

The \emph{screened envelopes} of the 2nd-generation method are linear
superpositions, 
\begin{equation}
n^{\alpha }\left( \varepsilon ,\mathbf{r}\right) \equiv n\left( \varepsilon ,%
\mathbf{r}\right) S^{\alpha }\left( \varepsilon \right) ,  \label{b8a}
\end{equation}
of the envelope functions, $n\left( \varepsilon ,\mathbf{r}\right) ,$ with
the property that the spherical-harmonics expansions of the set of screened
envelopes be: 
\begin{equation}
\kappa n\left( \varepsilon ,\mathbf{r}\right) S^{\alpha }\left( \varepsilon
\right) \equiv \kappa n^{\alpha }\left( \varepsilon ,\mathbf{r}\right)
=j^{\alpha }\left( \varepsilon ,\mathbf{r}\right) B^{\alpha }\left(
\varepsilon \right) +\kappa n\left( \varepsilon ,\mathbf{r}\right) ,
\label{b8}
\end{equation}
which are (\ref{b5}) with the substitutions: 
\begin{equation}
j_{l}\left( \kappa r\right) \quad \rightarrow \quad j_{Rlm}^{\alpha }\left(
\varepsilon ,r\right) \equiv j_{l}\left( \kappa r\right) -n_{l}\left( \kappa
r\right) \tan \alpha _{Rlm}\left( \varepsilon \right) ,  \label{b9}
\end{equation}
and: $B\left( \varepsilon \right) \rightarrow B^{\alpha }\left( \varepsilon
\right) ,$ which will be determined below. In contrast to its bare
counterpart, a screened envelope does \emph{not} have pure angular momentum,
i.e., cannot be factorized as a radial function times a spherical harmonics,
and it depends \emph{explicitly} on its surroundings. The \emph{background
phase shifts} $\alpha \left( \varepsilon \right) $ --which may even depend
on $m$ (see for instance Fig. \ref{Fig10})-- specify the \emph{shapes} of
the screened \emph{envelopes.} Whereas the bare envelopes are regular in all
space --except at their own site where they diverge like $Y_{lm}\left( 
\mathbf{\hat{r}}\right) /r^{l+1}$-- the screened envelopes \emph{diverge} at
any site where there is a finite background phase shift in at least one $L$%
-channel.

Note that only in the Overview did we use ASA $\kappa ^{2}$=0-notation with
Greek letters denoting screening constants and $\mathsf{S}^{\alpha }\mathsf{%
\,}$the structure matrix. In the bulk of the present paper, we use Greek
letters to denote background phase shifts, and $B^{\alpha }$ and $S^{\alpha }
$ to denote respectively the structure matrix and the screening
transformation.

We now find the screened structure matrix and the transformation matrix by
expanding also the bare envelope on the left hand side of (\ref{b8}) by
means of (\ref{b5}). Comparisons of the coefficients to $\kappa n_{L^{\prime
}}\left( \varepsilon ,\mathbf{r}_{R^{\prime }}\right) $ and $j_{L^{\prime
}}\left( \varepsilon ,\mathbf{r}_{R^{\prime }}\right) $ yield respectively: 
\begin{equation}
S^{\alpha }\left( \varepsilon \right) =1-\frac{\tan \alpha \left(
\varepsilon \right) }{\kappa }B^{\alpha }\left( \varepsilon \right) ,\quad 
\mathrm{and:\quad }B^{\alpha }\left( \varepsilon \right) =B\left(
\varepsilon \right) S^{\alpha }\left( \varepsilon \right)   \label{b10}
\end{equation}
with the quantities regarded as matrices, e.g. $\kappa ^{-1}\tan \alpha $ is
considered a diagonal matrix with elements $\kappa ^{-1}\tan \alpha
_{RL}\,\delta _{RR^{\prime }}\delta _{LL^{\prime }}.$ As a result of (\ref
{b10}): 
\begin{equation}
B^{\alpha }\left( \varepsilon \right) ^{-1}=B\left( \varepsilon \right)
^{-1}+\frac{\tan \alpha \left( \varepsilon \right) }{\kappa },  \label{b11}
\end{equation}
which shows that, like the bare structure matrix, also the screened one is
Hermitian. In contrast to the bare structure matrix, the screened one has
non-vanishing on-site elements. For background phase shifts known to give a
short-ranged $B^{\alpha }\left( \varepsilon \right) ,$ the inversion of the
matrix $B\left( \varepsilon \right) +\kappa \cot \alpha \left( \varepsilon
\right) ,$ implied by (\ref{b11}), may be performed in real space, although
the \emph{bare} structure matrix is long-ranged. Eq. (\ref{b11}) is the $%
\kappa ^{2}\mathrm{=}\varepsilon $ equivalent of the ASA 'Dyson equation' (%
\ref{a12}).

For the \emph{inactive }channels $\left( RL\equiv I\right) ,$ we choose the
background phase shifts to be equal to the \emph{real} phase shifts: 
\begin{equation}
\alpha _{I}\left( \varepsilon \right) \equiv \eta _{I}\left( \varepsilon
\right)   \label{b12}
\end{equation}
so that for these channels, 
\begin{equation*}
j_{I}^{\alpha }\left( \varepsilon ,r\right) \;=\;j_{I}\left( \kappa r\right)
-n_{I}\left( \kappa r\right) \tan \eta _{I}\left( \varepsilon \right)
\;=\;-\varphi _{I}^{\circ }\left( \varepsilon ,r\right) \tan \eta _{I}\left(
\varepsilon \right) .
\end{equation*}
That is, we shape the set of screened envelope functions in such a way that,
for the inactive channels, the radial functions, $\varphi _{I}^{\circ
}\left( \varepsilon ,r\right) ,$ may be \emph{substituted smoothly} by the
regular solutions, $\varphi _{I}\left( \varepsilon ,r\right) ,$ of the
radial \emph{Schr\"{o}dinger} equation$.$ This is what we call \emph{%
downfolding.} This substitution makes the screened envelopes become the
so-called \emph{screened spherical waves, }$\psi ,$ of the 3rd-generation
method. Only the screened spherical waves corresponding to the remaining,
so-called \emph{active} channels $\left( RL=A\right) $ will be used to
construct the MTO; they are: 
\begin{eqnarray}
&&\psi _{RL}^{\alpha }\left( \varepsilon ,\mathbf{r}_{R}\right) \;\equiv
\;n_{RL}^{\alpha }\left( \varepsilon ,\mathbf{r}_{R}\right) \;+  \label{b14}
\\[0.15cm]
&&\sum_{I}\left[ \varphi _{I}^{\circ }\left( \varepsilon ,r_{R^{\prime
}}\right) -\varphi _{I}\left( \varepsilon ,r_{R^{\prime }}\right) \right] 
\frac{\tan \eta _{I}\left( \varepsilon \right) }{\kappa }Y_{I}\left( \mathbf{%
\hat{r}}_{R^{\prime }}\right) B_{I,RL}^{\alpha }\left( \varepsilon \right) ,
\notag
\end{eqnarray}
which --in contrast to $n_{RL}^{\alpha }\left( \varepsilon ,\mathbf{r}%
_{R}\right) $-- are \emph{regular} in all inactive channels, albeit \emph{%
irregular} in the active channels. In (\ref{b14}), $I\equiv R^{\prime
}L^{\prime }.$ Below, we shall choose to truncate the active channels inside
their screening spheres. Due to the augmentation (substitution), the
screened spherical waves do \emph{not} transform linearly like (\ref{b8a}).

For the partial waves of high $l,$ the phase shifts vanish due to the
dominance of the centrifugal term over the potential term in the radial
Schr\"{o}dinger equation (\ref{a16}). As a consequence, the matrices
involved in the Dyson equation (\ref{b11}) --whose indices run over all
active as well as inactive channels-- truncate above a certain $l$ of about
3--4.

Before specifying our choice of background phase shifts for the \emph{active}
channels, let us define the energy-dependent, \emph{screened} MTO analogous
to the third equation (\ref{b2}) as the (augmented) envelope function, plus
a term proportional to the function $\varphi -\varphi ^{\circ },$ which
vanishes (quadratically) outside the central MT-sphere and has pure
angular-momentum character. That is: 
\begin{eqnarray}
\phi _{RL}^{\alpha }\left( \varepsilon ,\mathbf{r}_{R}\right)  &\equiv
&Y_{L}\left( \mathbf{\hat{r}}_{R}\right) \left[ \varphi _{Rl}\left(
\varepsilon ,r_{R}\right) -\varphi _{Rl}^{\circ }\left( \varepsilon
,r_{R}\right) \right] \frac{\tan \eta _{Rl}\left( \varepsilon \right) }{\tan
\eta _{RL}^{\alpha }\left( \varepsilon \right) }+\psi _{RL}^{\alpha }\left(
\varepsilon ,\mathbf{r}_{R}\right)   \notag \\[0.2cm]
&\equiv &Y_{L}\left( \mathbf{\hat{r}}_{R}\right) \left[ \varphi
_{Rl}^{\alpha }\left( \varepsilon ,r_{R}\right) -\varphi _{Rl}^{\circ
\,\alpha }\left( \varepsilon ,r_{R}\right) \right] +\psi _{RL}^{\alpha
}\left( \varepsilon ,\mathbf{r}_{R}\right)   \label{b15}
\end{eqnarray}
and $RL\in A.$ Here, the coefficient to $\varphi -\varphi ^{\circ }$ has
been chosen in such a way that, in its own channel and outside any other
MT-sphere, the screened MTO is $\varphi ^{\alpha }+j^{\alpha }\cot \eta
^{\alpha }$ plus a term from the diagonal element of the screened structure
matrix.

To check this, we project onto the 'eigen-channel,' making use of (\ref{b14}%
), (\ref{b8}), (\ref{b1}), and (\ref{b9}), and neglecting any contribution
from $\varphi _{I}\left( \varepsilon ,r_{R^{\prime }}\right) $'s from
overlapping neighboring MT-spheres: 
\begin{eqnarray}
\mathcal{P}_{RL}\phi _{RL}^{\alpha }\left( \varepsilon ,\mathbf{r}%
_{R}\right) &=&\varphi _{Rl}^{\alpha }\left( \varepsilon ,r_{R}\right)
-\varphi _{Rl}^{\circ \,\alpha }\left( \varepsilon ,r_{R}\right) +\mathcal{P}%
_{RL}\psi _{RL}^{\alpha }\left( \varepsilon ,\mathbf{r}_{R}\right)  \notag \\
&=&\left[ \varphi -n+\left( j^{\alpha }+n\tan \alpha \right) \cot \eta %
\right] \frac{\tan \eta }{\tan \eta ^{\alpha }}+n+j^{\alpha }\frac{B^{\alpha
}}{\kappa }  \notag \\
&=&\varphi ^{\alpha }+j^{\alpha }\cot \eta ^{\alpha }-n\frac{\tan \eta -\tan
\alpha }{\tan \eta ^{\alpha }}+n+j^{\alpha }\frac{B^{\alpha }}{\kappa } 
\notag \\
&=&\varphi ^{\alpha }+j^{\alpha }\cot \eta ^{\alpha }+j^{\alpha }\frac{%
B^{\alpha }}{\kappa }  \label{b18}
\end{eqnarray}
For simplicity, we have dropped all arguments and indices in the last three
lines. We see that the new phase shift, $\eta ^{\alpha },$ is given by: 
\begin{equation}
\tan \eta _{RL}^{\alpha }\left( \varepsilon \right) \equiv \tan \eta
_{Rl}\left( \varepsilon \right) -\tan \alpha _{RL}\left( \varepsilon \right)
,  \label{b16}
\end{equation}
as expected for the phase shift on the background of $\alpha .$ This is the
same transformation as the one obtained from (\ref{b11}) for $-B^{\alpha
}\left( \varepsilon \right) ^{-1}.$ The definition of the renormalized free
radial solution given in (\ref{b15}) may be written as: 
\begin{eqnarray}
\varphi _{RL}^{\circ \,\alpha }\left( \varepsilon ,r\right) &\equiv
&n_{l}\left( \kappa r\right) -j_{RL}^{\alpha }\left( \varepsilon ,r\right)
\cot \eta _{RL}^{\alpha }\left( \varepsilon \right)  \label{b17} \\
&=&\left[ n_{l}\left( \kappa r\right) \tan \eta _{Rl}\left( \varepsilon
\right) -j_{l}\left( \kappa r\right) \right] \cot \eta _{RL}^{\alpha }\left(
\varepsilon \right) ,  \notag
\end{eqnarray}
and $\varphi _{Rl}^{\alpha }\left( \varepsilon ,r_{R}\right) $ is the
solution of the radial Schr\"{o}dinger equation, normalized in such a way
that it matches onto $\varphi _{RL}^{\circ \,\alpha }\left( \varepsilon
,r\right) $ at the MT radius, $s_{R}.$ The definition (\ref{b17}) reduces to
(\ref{b1}) when $\alpha =0.$

The \emph{set} of screened MTOs now consists of the screened MTOs (\ref{b15}%
) of all active channels. Since the $\varphi -\varphi ^{\circ }$ function
has pure angular-momentum character, the mixed character of the screened MTO
stems solely from the $\psi $-function. The result of projecting the
screened MTO onto an active channel $R^{\prime }L^{\prime }$ different from
its own is seen from (\ref{b8}) to be: 
\begin{equation}
\mathcal{P}_{R^{\prime }L^{\prime }}\phi _{RL}^{\alpha }\left( \varepsilon ,%
\mathbf{r}_{R}\right) =\mathcal{P}_{R^{\prime }L^{\prime }}\psi
_{RL}^{\alpha }\left( \varepsilon ,\mathbf{r}_{R}\right) =j_{R^{\prime
}L^{\prime }}^{\alpha }\left( \varepsilon ,r_{R^{\prime }}\right) \frac{%
B_{R^{\prime }L^{\prime },RL}^{\alpha }\left( \varepsilon \right) }{\kappa },
\label{b19}
\end{equation}
when $r_{R^{\prime }}$ is so small that $\mathbf{r}$ lies inside only \emph{%
one} MT-sphere, the one centered at $R^{\prime }.$ From (\ref{b19}) and (\ref
{b18}) it is then obvious that, in order to get a \emph{smooth} linear
combination $\sum_{A}\phi _{A}^{\alpha }\left( \varepsilon ,\mathbf{r}%
_{A}\right) c_{A}^{\alpha }$ of screened MTOs, \emph{all }$j^{\alpha }$\emph{%
-functions must cancel.} This leads to the condition that the energy must be
such that the coefficients can satisfy 
\begin{equation}
\sum_{A}\left[ B_{A^{\prime }A}^{\alpha }\left( \varepsilon _{i}\right)
+\kappa \cot \eta _{A}^{\alpha }\left( \varepsilon _{i}\right) \delta
_{A^{\prime }A}\right] c_{A,i}^{\alpha }\;\equiv \;\sum_{A}K_{A^{\prime
}A}^{\alpha }\left( \varepsilon _{i}\right) c_{A,i}^{\alpha }=0,  \label{b20}
\end{equation}
for all active $R^{\prime }L^{\prime }\equiv A^{\prime }.$ These are the 
\emph{screened KKR equations, }and $K^{\alpha }\left( \varepsilon \right) $
is the screened KKR matrix. If these equations are satisfied, the linear
combination of screened MTOs is: 
\begin{eqnarray}
\sum_{A}\phi _{A}^{\alpha }\left( \varepsilon _{i},\mathbf{r}_{R}\right)
c_{A,i}^{\alpha } &=&\sum_{l^{\prime }=0}^{\infty }\sum_{m^{\prime }}\varphi
_{R^{\prime }L^{\prime }}^{\alpha }\left( \varepsilon _{i},r_{R^{\prime
}}\right) Y_{L^{\prime }}\left( \mathbf{\hat{r}}_{R^{\prime }}\right)
c_{R^{\prime }L^{\prime },i}^{\alpha }+  \label{b21} \\
&&\sum_{R\neq R^{\prime }}\sum_{L}\left[ \varphi _{RL}^{\alpha }\left(
\varepsilon _{i},r_{R}\right) -\varphi _{RL}^{\circ \,\alpha }\left(
\varepsilon _{i},r_{R}\right) \right] Y_{L}\left( \mathbf{\hat{r}}%
_{R}\right) c_{RL,i}^{\alpha }  \notag
\end{eqnarray}
near site $R^{\prime }.$ As long as the MT-spheres do not overlap, this is a
solution of Schr\"{o}dinger's equation for the MT-potential and, if the
potentials overlap, then the $\varphi -\varphi ^{\circ }$ tongues from the
neighboring sites in the second line of (\ref{b21}) make the wave function
correct to first order in the overlap \cite{MRS}. This is exactly as in (\ref
{b7}). The summation over spherical-harmonics around the central site
includes the contributions $-\varphi _{I}\left( \varepsilon ,\mathbf{r}%
_{R^{\prime }}\right) \kappa ^{-1}\tan \eta _{I}\left( \varepsilon \right)
\sum_{A}B_{I,A}^{\alpha }\left( \varepsilon \right) c_{A,i}^{\alpha }$
provided by the screened-spherical-wave part of the MTO (see (\ref{b15}) and
(\ref{b14})).

Although energy-dependent MTO sets with different screenings are not
linearly related, they all solve Schr\"{o}dinger's equation for the
MT-potential used for their construction via the corresponding KKR equation.
E.g. had one chosen a representation in which a channel making a significant
contribution to a wave function $\Psi _{i}\left( \mathbf{r}\right) $ with
energy $\varepsilon _{i}=\varepsilon $ were downfolded, then the
corresponding solution of the KKR equation (\ref{b20}) would arise from $%
B^{\alpha }\left( \varepsilon \right) $ being long ranged and, as a function
of $\varepsilon ,$ going through a zero-pole pair near $\varepsilon _{i}.$
If the energy were now \emph{fixed }at some energy $\varepsilon _{\nu },$
and the energy-independent set $\phi ^{\alpha }\left( \varepsilon _{\nu },%
\mathbf{r}\right) $ were used as the 0th-order MTO basis in a variational
calculation, then a useful result could in principle be obtained, but only
if $\varepsilon _{\nu }$ were chosen very close to $\varepsilon _{i}.$

\input{fig3.tex}

\subsection{Hard-sphere interpretation and redefinitions}

\label{hard}We now wish to choose the background phase shifts for the active
channels in a way which reduces the spatial range and the energy dependence
of the MTO envelopes. It is obvious, that for the orbitals to be localized,
they must have energies \emph{below the bottom of the continuum of the
background} --defined as the system which has the same structure as the real
system, but has all phase shifts equal to those of the background. Hence,
the active $\alpha \left( \varepsilon \right) $'s should be defined in such
a way that the energy band defined by: $\left| B^{0}\left( \varepsilon
\right) +\kappa \cot \alpha \left( \varepsilon \right) \right| =0,$ lie as
high as possible.

The discovery of a useful way of determining this background, turned out to
be the unplanned birth of the 3rd MTO generation \cite{Trieste,MRS}.
Realizing that the weakest point of the ASA was its solution of Poisson's
--and not Schr\"{o}dinger's-- equation, and unhappy with the complexities of
existing full-potential schemes, we \cite{unitary} were looking for those
linear combinations of Hankel functions --like (\ref{b8a})-- which would fit
the charge density continuously at spheres. With Methfessel's formulation 
\cite{Meth}: What we wanted was those solutions of the wave equation which
are $Y_{L}\left( \mathbf{\hat{r}}_{R}\right) $ at their own sphere and for
their own angular momentum, and zero at all other spheres and for all other
angular momenta. This set was therefore named \emph{unitary} spherical
waves. The solution to this boundary-value problem is of course a particular
screening transformation (\ref{b11}).

Our way of defining the background was thus in terms of hard \emph{%
screening-spheres} for the active channels; the larger the screening
spheres, the larger the excluded volume and the higher the bottom of the
continuum. The screening spheres are not allowed to overlap --at least not
if all $l$-channels were active, because then a unitary spherical wave would
be asked to take both values, 1 \emph{and} 0, on the circle common to the
central and an overlapping sphere. As a consequence, in order to reduce the
range and the energy dependence of the MTO envelope functions, the screening
spheres should in general be \emph{nearly touching.} Now, since the
screening radii, $,$ control the shapes of the envelopes, the \emph{relative}
sizes of the screening spheres should be determined by \emph{chemical}
considerations, i.e.\textit{\ the }$a$'s may be covalent- or ionic radii in
order that results obtained from an electronic-structure calculation be
interpretable in terms of covalency, ionicity etc. Referring to the
discussion in the Overview, one could say: The MT-spheres $\left( s\right) $
are potential-spheres and the screening-spheres $\left( a\right) $ are
charge-spheres.

Inspired by Ref. \cite{Boston}, practitioners of multiple-scattering theory
--who traditionally take the Kohn-Rostoker \cite{KKR} Green-function point
of view-- found another useful way of determining the background phase
shifts, namely in terms of \emph{repulsive potentials} \cite{Zeller}.

For a given active channel $\left( RL=A\right) $, the radial positions, $r=$ 
$a_{A}\left( \varepsilon \right) ,$ of the \emph{nodes} of the background
functions $j^{\alpha }$ given by (\ref{b9}) are the solutions of the
equation: 
\begin{equation*}
0=j_{A}^{\alpha }\left( \varepsilon ,a_{A}\left( \varepsilon \right) \right)
=j_{l}\left( \kappa a_{A}\left( \varepsilon \right) \right) -n_{l}\left(
\kappa a_{A}\left( \varepsilon \right) \right) \tan \alpha _{A}\left(
\varepsilon \right) .
\end{equation*}
Whereas \emph{attractive} potentials usually do not give positive radii
--for an example, see the dotted curve in Fig. \ref{Fig2}-- repulsive
potentials do, as may be seen from the radial Schr\"{o}dinger equation (\ref
{a16}). For a \emph{hard} repulsive potential, the position of the node is 
\emph{independent} of energy and of $l.$ What we shall use for the active
channels are therefore screening-sphere radii, $a_{A},$ which are
independent of energy and which usually depend little on $L$ among the
active channels. In terms of such a screening radius, the corresponding
background phase shift is given by: 
\begin{equation}
\tan \alpha _{A}\left( \varepsilon \right) =j_{l}\left( \kappa a_{A}\right)
\left/ n_{l}\left( \kappa a_{A}\right) \right. .  \label{b22}
\end{equation}

Now, instead of having screened spherical waves (\ref{b14}) and MTOs (\ref
{b15}) whose active channels are irregular at the origin --the
irregularities of the \emph{in}active channels were already gotten rid of by
downfolding, followed by $\varphi _{I}^{\circ }\left( \varepsilon ,r\right)
\rightarrow \varphi _{I}\left( \varepsilon ,r\right) $ substitutions-- we
prefer that the active channels have merely \emph{kinks.} This is achieved
by \emph{truncating} all \emph{active} $j^{\alpha }$-functions \emph{inside}
their \emph{screening spheres,} that is, we perform the substitution: 
\begin{equation}
j_{A}^{\alpha }\left( \varepsilon ,r\right) \rightarrow \left\{ 
\begin{array}{l}
0\quad \quad \quad \quad \quad \quad \quad \quad \quad \quad \quad \quad
\quad \quad \quad \quad \mathrm{for\quad }r<a_{A} \\ 
j_{l}\left( \kappa r\right) -n_{l}\left( \kappa r\right) j_{l}\left( \kappa
a_{A}\right) \left/ n_{l}\left( \kappa a_{A}\right) \right. \quad \;\mathrm{%
for\quad }r\geq a_{A}
\end{array}
\right. ,  \label{b23}
\end{equation}
which is continuous but not differentiable, for the screened spherical waves
and for its own $j^{\alpha }$-function of the MTO --that is the second term
on the last two lines of (\ref{b18}). With that substitution, a screened
spherical wave, $\psi _{RL}^{\alpha }\left( \varepsilon ,\mathbf{r}%
_{R}\right) ,$ vanishes inside all screening spheres of the active channels
--except inside its own, where it equals $n_{l}\left( \kappa r_{R}\right)
Y_{L}\left( \mathbf{\hat{r}}_{R}\right) .$ This may be seen from (\ref{b19})
and the two first lines of (\ref{b18}). Finally, if we renormalize according
to: 
\begin{equation}
\psi _{RL}^{a}\left( \varepsilon ,\mathbf{r}_{R}\right) \equiv \psi
_{RL}^{\alpha }\left( \varepsilon ,\mathbf{r}_{R}\right) \left/ n_{l}\left(
\kappa a_{RL}\right) \right.   \label{b24}
\end{equation}
--note the difference between the superscripts $a$ and $\alpha $-- we
finally arrive at the screened (unitary) spherical wave as defined in Refs. 
\cite{Trieste,MRS}.

$\psi _{RL}^{a}\left( \varepsilon ,\mathbf{r}_{R}\right) $ is that solution
of the wave equation which is $Y_{L}\left( \mathbf{\hat{r}}_{R}\right) $ on
its own screening sphere, has vanishing $Y_{L^{\prime }}\left( \mathbf{\hat{r%
}}_{R^{\prime }}\right) $-average on the screening spheres of the other
active channels, and joins smoothly onto the regular solutions of the radial
Schr\"{o}dinger equations of the inactive channels. In those, the regular
Schr\"{o}dinger solutions are, in fact, substituted for the wave-equation
solutions.

It is now obvious, that overlap of screening spheres will cause complicated,
and hence long-ranged spatial behavior of the screened spherical waves, and
the worse, the more spherical harmonics are active.

With the normalization (\ref{b24}), there is apparently no need for
functions, like spherical Bessel and Neumann or Hankel functions, which have
a branch-cut at zero energy, and this was the point of view taken in the
first accounts \cite{Trieste,MRS} of the 3rd-generation method. However, the
normalization (\ref{b24}) is not appropriate for $a\mathrm{=}0,$ and
expressing the screened structure matrix in terms of the bare one (\ref{b38}%
) --which is the only one computable in terms of elementary functions-- was
slightly painful in Ref. \cite{Trieste}; moreover, in that paper downfolding
was not presented in its full generality. In these respects, the present,
conventional derivation is simpler, but it takes more equations.

With the $\alpha \rightarrow a$ redefinitions (\ref{b23})-(\ref{b24}), the
MTO remains as defined by (\ref{b15}), but with the screened spherical waves
and its own $j^{\alpha }$-function truncated as described above. We may also
renormalize the MTO like in (\ref{b24}): 
\begin{equation}
\phi _{RL}^{a}\left( \varepsilon ,\mathbf{r}_{R}\right) \equiv \phi
_{RL}^{\alpha }\left( \varepsilon ,\mathbf{r}_{R}\right) \left/ n_{l}\left(
\kappa a_{RL}\right) \right. ,  \label{b25}
\end{equation}
whereby these energy-dependent 0th-order MTOs become identical with the 
\emph{kinked partial waves} of Refs. \cite{Trieste,MRS}. This normalization
corresponds to: 
\begin{equation}
\varphi _{Rl}^{\circ \,a}\left( \varepsilon ,a_{RL}\right) \equiv 1.
\label{b26}
\end{equation}
Note that this will cause the normalization of the radial Schr\"{o}%
dinger-equation solution, $\varphi ^{a}\left( \varepsilon ,r\right) ,$ to
depend on $m$ in case the corresponding screening radius is chosen to do so.

In Fig. \ref{Fig3} we show the screened counterpart of the bare Si $p$
orbital in Fig. \ref{Fig2}. Since only the two first terms of (\ref{b15})
--but not the screened spherical wave-- has pure angular momentum, we cannot
plot just the radial wave function like in Fig. \ref{Fig2}. Rather, we show
the MTO together with its three parts along the [111]-line between the
central atom and one of its four nearest neighbors in the diamond structure.
The positions of the central and the nearest-neighbor atoms are indicated on
the axis (Si), and so is the intersection with the \emph{central} MT-sphere $%
\left( s\right) $. The $p$ orbital chosen is the one pointing along this
[111] direction. The Si $spd$ channels were taken as active, and to have one
and the same screening radius, $a=0.75t,$ where $t$ is half the
nearest-neighbor distance, i.e.\textit{,} the touching-sphere radius. The
places where the central and the nearest-neighbor screening spheres
intersect the [111]-line are indicated by '$\leftarrow a$' and '$%
a\rightarrow $' with the arrow pointing towards the respective center. We
see that the central MT-sphere is so large, that it overlaps the screening
sphere of the neighboring atom. Like in Fig. \ref{Fig2}, the full curve
shows the MTO $\left( \phi ^{a}\right) $, and the dot-dashed $\left( \varphi
^{a}Y\right) $, the dotted $\left( \varphi ^{\circ \,a}Y\right) $, and the
dashed $\left( \psi ^{a}\right) $ curves show the three terms in the
renormalized version of equation (\ref{b15}). The dot-dashed and the dotted
curves are identical with those in Fig. \ref{Fig2}, except for the
normalization; they are the outwards-integrated solution $\left( \varphi
^{a}Y\right) $ of the radial Schr\"{o}dinger equation, continued by the
inwards-integrated solution $\left( \varphi ^{\circ \,a}Y\right) $ for the
flat potential$.$ These two curves have been deleted outside the central
MT-sphere where their contribution to the MTO (\ref{b15}) cancels. The
inwards integration ends at the screening sphere, inside which $\varphi
^{\circ \,a}$ --with $j^{a}$ truncated-- cancels its own-part, $n_{l}\left(
\kappa r\right) /n_{l}\left( \kappa a\right) ,$ of the screened spherical
wave, $\psi ,$ shown by the dashed curve (see Eqs. (\ref{b18}) and (\ref{b23}%
)). Neither of these cancelling parts are shown in the figure, and the
dashed curve inside the central screening sphere therefore merely shows the
contribution to the screened spherical wave from the inactive channels $%
\left( l\geq 3\right) $. Due to the $j^{a}$-truncations, the screened
spherical wave has kinks at \emph{all} screening spheres and, inside these
spheres, only the contribution from the inactive partial waves --which are
regular solutions of the radial Schr\"{o}dinger equations-- remain. The full
curve is the MTO, which is identical with the screened spherical wave
outside its own MT-sphere. At its own screening sphere, its kink differs
from that of the screened spherical wave due to the truncation of the $j^{a}$%
-contribution to $\varphi ^{\circ \,a}.$ Compared with the bare MTO in Fig. 
\ref{Fig2}, the screened MTO in Fig. \ref{Fig3} is considerably more
localized, even though a negative energy was chosen.

\input{fig4.tex}

If one demands that the valence band --as well as the lower part of the
conduction band-- of Si be described from first principles using merely the
minimal 4 orbitals per atom, one cannot use a set with $p$ orbitals such as
those shown in Figs. \ref{Fig2} and \ref{Fig3}; the $d$-MTOs must be folded
into the envelopes of the remaining $sp$ set by use of the appropriate
structure matrix obtained from Eq. (\ref{b11}) with the choice (\ref{b12})
for the Si $d$-channels. The corresponding Si $p_{111}$-MTO is shown in Fig. 
\ref{Fig4}. Little is changed inside the central screening sphere, but the
tail extending into the nearest-neighbor atom has attained a lot of $d$%
-character around that site, and the MTO is correspondingly more delocalized.

The Si $p_{111}$-MTO for use in an $sp$ MTO basis constructed from the
conventional Si+E potential --for which the diamond structure is packed bcc
with equally large space-filling spheres-- is obtained by down-folding of
the Si $d$ and all empty-sphere channels. It turns out to be so similar to
the one obtained from the fat Si-centered potential shown in Fig. \ref{Fig4}%
, that we will not take the space to show it.

Whereas the bare MTO in Fig. \ref{Fig2} is what has always been called a
bare MTO, the screened ones in Figs. \ref{Fig3} and \ref{Fig4} look more
like a partial wave, $\varphi Y,$ with a tail attached at its own screening
sphere --and with kinks at all screening spheres. Hence the name 'kinked
partial wave' given in Ref. \cite{Trieste}. In this original derivation,
kinked partial waves with $a=s\leq t$ were considered first, and only later,
the limiting case $a\rightarrow 0$ gave rise to a painful exercise. The
kinked partial waves have in common with Slater's original Augmented Plane
Waves (APWs) \cite{APW}, that they are partial waves, $\varphi \left(
\varepsilon ,r\right) Y,$ of the proper energy inside non-overlapping
spheres, which are joined continuously --but with kinks-- to wave-equation
solutions in the interstitial. In that region, the APW is a wave-equation
solution with a given \emph{wave-vector,} whereas the MTO is a solution with
the same \emph{energy.} Moreover, whereas the APW method uses identical
potential and augmentation spheres, this is not the case for MTOs.

If --for the third time in this section-- we make a linear combination of
MTOs --this time defined with kinks-- and demand that it solves Schr\"{o}%
dinger's equation, then the condition is, that the kinks --rather than the $%
j^{\alpha }$-functions-- from the tails should cancel the ones in the head.
This condition is of course equivalent with the one for $j^{\alpha }$%
-cancellation. Nevertheless, let us express the KKR equations in this
language because it will turn out to have three further advantages: The
artificial dependence on $\kappa \equiv \sqrt{\varepsilon }$ and the
associated change between Neumann and decaying Hankel functions will
disappear, there will be a simple expression for the integral of the product
of two MTOs, and we will be led to a contracted Green function of great use
in the following section.

Since the kinks arise because the $j^{\alpha }$-functions are truncated
inside their screening spheres, the kink in a certain active channel of an
MTO is proportional to the slope of the corresponding $j^{\alpha }$-function
at $a_{+}.$ An expression for this slope is most easily found from the
Wronskian, which in general is defined as: $r^{2}\left[ f\left( r\right)
g^{\prime }\left( r\right) -g\left( r\right) f^{\prime }\left( r\right) %
\right] \equiv \left\{ f,g\right\} _{r},$ and is independent of $r$ when the
two functions considered are solutions of the same linear, second-order
differential equation. As a consequence, $\left\{ n,j^{\alpha }\right\}
=\left\{ n,j-n\tan \alpha \right\} =\left\{ n,j\right\} =-\kappa ^{-1},$ and
therefore: 
\begin{equation}
\partial j^{\alpha }\left( \varepsilon ,r\right) \left/ \partial r\right|
_{a_{+}}=-\left[ a^{2}\kappa n\left( \kappa a\right) \right] ^{-1}.
\label{b27}
\end{equation}

We now define the elements $K_{R^{\prime }L^{\prime },RL}^{a}\left(
\varepsilon \right) $ --where $R^{\prime }L^{\prime }$ and $RL$ both refer
to active channels-- of a \emph{kink matrix }\cite{Trieste,MRS} as $%
a_{R^{\prime }L^{\prime }}^{2}$ times the kink in the $R^{\prime }L^{\prime }
$-channel of $\phi _{RL}^{a}\left( \varepsilon ,\mathbf{r}_{R}\right) .$
From the expression for $\partial j^{\alpha }\left/ \partial r\right|
_{a_{+}}$, the last forms of the spherical-harmonics expansions (\ref{b18})
and (\ref{b19}), the definition (\ref{b20}) of the screened KKR matrix, and
the renormalization (\ref{b25}), this is seen to be: 
\begin{equation}
K_{R^{\prime }L^{\prime },RL}^{a}\left( \varepsilon \right) \;=\;\frac{%
-K_{R^{\prime }L^{\prime },RL}^{\alpha }\left( \varepsilon \right) }{\kappa
n_{l^{\prime }}\left( \kappa a_{R^{\prime }L^{\prime }}\right) \,\kappa
n_{l}\left( \kappa a_{RL}\right) }.  \label{b34}
\end{equation}
Note that this is the kink matrix as defined in Ref. \cite{MRS}, whereas the
one defined in Ref. \cite{Trieste} has the opposite sign. As presently
defined, the energy derivative of the kink matrix is positive definite, as
we shall se in the next section.

Screening and the definition (\ref{b34}) have removed the spurious energy
dependencies of $K^{\alpha =0}\left( \varepsilon \right) $. To see this more
clearly, let us use the first --rather than the last-- forms of the
spherical-harmonics expansions (\ref{b18}) and (\ref{b19}), which are also
more closely related to the definition (\ref{b15}) of the MTO, and to 
Figs.
\ref{Fig3} and \ref{Fig4}: The kink matrix for $\psi _{A}^{a}\left(
\varepsilon ,\mathbf{r}_{R}\right) $ is $-\left[ \kappa n_{l^{\prime
}}\left( \kappa a_{A^{\prime }}\right) \right] ^{-1}B_{A^{\prime }A}^{\alpha
}\left( \varepsilon \right) \left[ \,\kappa n_{l}\left( \kappa a_{A}\right) %
\right] ^{-1}$. Moreover, $\psi _{A}^{a}\left( \varepsilon ,\mathbf{r}%
_{R}\right) $ contains the diverging term $n\left( \kappa r\right) /n\left(
\kappa a\right) $ in its own channel, which in the MTO is being cancelled by
a term from $\varphi ^{\circ \,a}$ (see the third equation (\ref{b18}) and (%
\ref{b16})). The kink matrix for the MTO set is now seen to equal the one
for the set of screened spherical waves, plus --in the diagonal-- the kink
in the function $\varphi ^{a}-\varphi ^{\circ \,a}+n\left( \kappa r\right)
/n\left( \kappa a\right) .$ Since $\varphi -\varphi ^{\circ }$ is smooth,
this kink is the one between the radial functions $\varphi ^{\circ
\,a}\left( \varepsilon ,r\right) $ and $n\left( \kappa r\right) /n\left(
\kappa a\right) .$ We thus arrive at the expression: 
\begin{eqnarray}
K_{R^{\prime }L^{\prime },RL}^{a}\left( \varepsilon \right) \; &=&\;-\frac{%
B_{R^{\prime }L^{\prime },RL}^{\alpha }\left( \varepsilon \right) }{\kappa
n_{l^{\prime }}\left( \kappa a_{R^{\prime }L^{\prime }}\right) \,\kappa
n_{l}\left( \kappa a_{RL}\right) }  \label{b36} \\[0.15cm]
&&\quad \quad \quad \quad \quad +a_{RL}\left[ \partial \left\{ n_{l}\left(
\varepsilon ,a\right) \right\} -\partial \left\{ \varphi _{l}^{\circ }\left(
\varepsilon ,a\right) \right\} \right] \delta _{R^{\prime }R}\delta
_{L^{\prime }L}  \notag \\[0.15cm]
&=&a_{R^{\prime }L^{\prime }}^{2}\left. \frac{\partial }{\partial r}\mathcal{%
P}_{R^{\prime }L^{\prime }}\psi _{RL}^{a}\left( \varepsilon ,\mathbf{r}%
_{R}\right) \right| _{a}-a_{A}\partial \left\{ \varphi _{l}^{\circ }\left(
\varepsilon ,a\right) \right\} \delta _{R^{\prime }R}\delta _{L^{\prime }L} 
\notag \\[0.15cm]
&\equiv &B_{R^{\prime }L^{\prime },RL}^{a}\left( \varepsilon \right)
\;-\;a_{RL}\,\partial \left\{ \varphi _{l}^{\circ }\left( \varepsilon
,a\right) \right\} \,\delta _{R^{\prime }R}\delta _{L^{\prime }L},
\label{b37}
\end{eqnarray}
in terms of the \emph{logarithmic-derivative function} at the screening
sphere of the inwards-integrated radial function, $\partial \left\{ \varphi
_{l}^{\circ }\left( \varepsilon ,a\right) \right\} \equiv \partial \ln
\left| \varphi _{l}^{\circ }\left( \varepsilon ,r\right) \right| \left/
\partial \ln r\right| _{a}$. Remember that $RL$ and $R^{\prime }L^{\prime }$
refer to active channels.

In the third line of (\ref{b36}) we have pointed to the fact that the first, 
\emph{potential-independent} part of the kink matrix is $a_{A^{\prime }}^{2}$
times the outwards slope of the screened spherical wave and in (\ref{b37})
we have denoted this \emph{slope matrix} $B_{R^{\prime }L^{\prime
},RL}^{a}\left( \varepsilon \right) .$ Note that, as presently defined, this
slope matrix is Hermitian and equals $a_{R^{\prime }L^{\prime }}$ times the
non-Hermitian slope matrix defined in Refs. \cite{Trieste,MRS}; moreover,
the transformation from $B^{\alpha }$ to $B^{a}$ is not quite (\ref{b34}),
but differs from it by the term $a\partial \ln \left| n_{l}\left( \kappa
r\right) \right| \left/ \partial \ln r\right| _{a}$. We may switch from
Neumann to Bessel functions, using again that $j_{l}\left( \kappa a\right)
=n_{l}\left( \kappa a\right) \tan \alpha ,$ and that $\left\{ j,n\right\}
=1/\kappa .$ We get: 
\begin{eqnarray}
B^{a}\left( \varepsilon \right)  &=&-\frac{\tan \alpha \left( \varepsilon
\right) }{\kappa j\left( \kappa a\right) }\left[ B^{\alpha }\left(
\varepsilon \right) -\kappa \cot \alpha \left( \varepsilon \right) \right] 
\frac{\tan \alpha \left( \varepsilon \right) }{\kappa j\left( \kappa
a\right) }+a\partial \left\{ j\left( \kappa a\right) \right\}   \notag \\%
[0.15cm]
&=&\frac{1}{j\left( \kappa a\right) }\left[ B\left( \varepsilon \right)
+\kappa \cot \alpha \left( \varepsilon \right) \right] ^{-1}\frac{1}{j\left(
\kappa a\right) }+a\partial \left\{ j\left( \kappa a\right) \right\} ,
\label{b33}
\end{eqnarray}
where the last equation has been obtained with the help of (\ref{b11}), and
where $B\left( \varepsilon \right) \equiv B^{\alpha =0}\left( \varepsilon
\right) $ is the bare KKR structure matrix (\ref{b38}). The matrix $B\left(
\varepsilon \right) +\kappa \cot \alpha \left( \varepsilon \right) $ is the
bare KKR matrix for the background-potential and has dimension $\left(
A+I\right) ^{2};$ it only truncates when $\alpha _{I}\left( \varepsilon
\right) \equiv \eta _{I}\left( \varepsilon \right) =0,$ as it happens for
high $l.$

\subsubsection{Computational Procedure.}

The recipe for a computation could be: Solve the radial Schr\"{o}dinger
equations outwards, and then inwards to $a\sim 0.8t,$ for all channels up $%
l\lesssim 3.$ Then, compute the Green matrix of the background, $G^{\alpha
=0}\left( \varepsilon \right) \equiv \left[ B^{\alpha =0}\left( \varepsilon
\right) +\kappa \cot \alpha \left( \varepsilon \right) \right] ^{-1},$ by
inversion in real space, choosing the strong screening just mentioned, i.e.%
\textit{\ }nearly touching screening spheres for all $spd\left( f\right) $
channels. This gives the strongly screened structure matrix, $B^{\alpha
}\left( \varepsilon \right) $ or $B^{a}\left( \varepsilon \right) ,$
according to (\ref{b33}), and the KKR matrix, $K^{\alpha }\left( \varepsilon
\right) $ or $K^{a}\left( \varepsilon \right) ,$ for the real potential in
the strongly screened representation according to (\ref{b20}) or (\ref{b37}%
). For a crystal, Bloch-sum the KKR matrix. Now, invert this matrix in real
space to obtain the Green matrix, $G^{\alpha }\left( \varepsilon \right)
\equiv K^{\alpha }\left( \varepsilon \right) ^{-1}$ or $G^{a}\left(
\varepsilon \right) \equiv K^{a}\left( \varepsilon \right) ^{-1}.$ Next,
choose the physically and chemically motivated screening $\left( \beta
\right) $ and rescreen the Green matrix to the downfolded representation, $%
G^{\beta }\left( \varepsilon \right) $ or $G^{b}\left( \varepsilon \right) ,$
using the scaling relations (\ref{b32}) or (\ref{b44}) derived below. As
will be explained in the following Sect. \ref{poly}, this should be done for
a number of energies. In addition, one will need the first energy
derivatives $\dot{G}^{b}\left( \varepsilon \right) .$ The latter may be
obtained from $\dot{K}^{a}\left( \varepsilon \right) $ via numerical
differentiation of the weakly energy dependent structure matrix, $%
B^{a}\left( \varepsilon \right) ,$ and calculation of $\int_{0}^{s}\varphi
^{a}\left( \varepsilon ,r\right) ^{2}r^{2}dr-\int_{a}^{s}\varphi
_{RL}^{\circ \,a}\left( \varepsilon ,r\right) ^{2}r^{2}dr$ for the energy
derivative of the logarithmic derivative function in (\ref{b37}), as will be
shown in (\ref{b52})-(\ref{b55}) below. With this $\dot{K}^{a}\left(
\varepsilon \right) ,$ compute $\dot{G}^{a}\left( \varepsilon \right) $ from
(\ref{b55}) and, finally, rescreen to $\dot{G}^{b}\left( \varepsilon \right) 
$ using the energy derivative of (\ref{b44}) given below.

In order to evaluate the wave function (\ref{b21}), one needs in addition to 
$B_{A^{\prime }A}^{b}\left( \varepsilon \right) ,$ the block $%
B_{IA}^{b}\left( \varepsilon \right) ,$ and this may be obtained from (\ref
{b33}).\bigskip

The relation of the screening constants, the structure matrix, and the KKR
matrix to those --see (\ref{a12}) and (\ref{a8})-- of the conventional ASA
is simple, but not as straightforward as the $\alpha $-to-$a$
transformations of the present section, so for this topic we refer to Refs. 
\cite{Trieste,MRS}.

This completes our \emph{exact} transformation of the original KKR matrix (%
\ref{b6}) which has long range and strong energy dependence --$B^{0}\left(
\varepsilon ,\mathbf{k}\right) $ has poles at the free-electron parabola: $%
\sum_{G}\left| \mathbf{k+G}\right| ^{2}\mathrm{=}\varepsilon $--\quad to a
screened and renormalized KKR matrix which --depending on the screening--
may be short ranged and weakly energy dependent. The kink matrix is
expressed in terms of a slope matrix, which only depends on the energy and
the structure of the background, and the logarithmic derivatives of the
active radial functions extrapolated inwards to the appropriate screening
radius.

\subsection{Re-screening the Green matrix}

In the ASA, it is simpler to re-screen the Green matrix (\ref{a11}) than the
structure matrix (\ref{a12}), because the former involves additions to the
diagonal and energy-dependent rescaling of rows and columns, but no matrix
inversions. The same holds for the fully energy-dependent matrices of the
3rd-generation, as may be seen from (\ref{b11}) or (\ref{b33}) for the
structure matrix. For the Green matrix (\ref{b20}), we get with the help of (%
\ref{b33}) and a bit of algebra: 
\begin{eqnarray*}
G^{\alpha }\left( \varepsilon \right)  &\equiv &K^{\alpha }\left(
\varepsilon \right) ^{-1}=\kappa ^{-1}\tan \alpha \left( \varepsilon \right) 
\left[ 1-\tan \alpha \left( \varepsilon \right) \cot \eta \left( \varepsilon
\right) \right]  \\
&&+\left[ 1-\tan \alpha \left( \varepsilon \right) \cot \eta \left(
\varepsilon \right) \right] G^{\alpha =0}\left( \varepsilon \right) \left[
1-\tan \alpha \left( \varepsilon \right) \cot \eta \left( \varepsilon
\right) \right] ,
\end{eqnarray*}
which has the form (\ref{a11}). Solving for $G^{\alpha =0}\left( \varepsilon
\right) $ and setting the result equal to $G^{\beta }\left( \varepsilon
\right) $ yields the following relation for re-screening of the Green
matrix: 
\begin{equation}
G^{\beta }\left( \varepsilon \right) =\frac{\tan \eta ^{\beta }\left(
\varepsilon \right) }{\tan \eta ^{\alpha }\left( \varepsilon \right) }%
G^{\alpha }\left( \varepsilon \right) \frac{\tan \eta ^{\beta }\left(
\varepsilon \right) }{\tan \eta ^{\alpha }\left( \varepsilon \right) }-\frac{%
\tan \alpha \left( \varepsilon \right) -\tan \beta \left( \varepsilon
\right) }{\kappa }\frac{\tan \eta ^{\beta }\left( \varepsilon \right) }{\tan
\eta ^{\alpha }\left( \varepsilon \right) }. \label{b32}
\end{equation}
In $a$-language, where according to (\ref{b34}): $G^{a}\left( \varepsilon
\right) =-\kappa n\left( \kappa a\right) G^{\alpha }\left( \varepsilon
\right) \kappa n\left( \kappa a\right) ,$ the diagonal matrices in (\ref{b32}%
) become $\left[ n\left( \kappa b\right) /n\left( \kappa a\right) \right] %
\left[ \tan \eta ^{\beta }\left( \varepsilon \right) /\tan \eta ^{\alpha
}\left( \varepsilon \right) \right] $ and $\kappa n\left( \kappa a\right)
n\left( \kappa b\right) \left[ \tan \alpha \left( \varepsilon \right) -\tan
\beta \left( \varepsilon \right) \right] $ and may, in fact, be expressed
more simply in terms of the inwards-integrated radial wave function,
renormalized according to (\ref{b26}). In order to see this, we first use
the form (\ref{b17}): 
\begin{equation*}
\varphi ^{\circ \,a}\left( \varepsilon ,r\right) =\frac{n\left( \kappa
r\right) \tan \eta ^{\alpha }\left( \varepsilon \right) -j^{\alpha }\left(
\varepsilon ,r\right) }{n\left( \kappa a\right) \tan \eta ^{\alpha }\left(
\varepsilon \right) },
\end{equation*}
and then evaluate this at the screening-radius $b:$ 
\begin{equation*}
\varphi ^{\circ \,a}\left( \varepsilon ,b\right) =\frac{n\left( \kappa
b\right) \tan \eta ^{\alpha }\left( \varepsilon \right) -j^{\alpha }\left(
\varepsilon ,b\right) }{n\left( \kappa a\right) \tan \eta ^{\alpha }\left(
\varepsilon \right) }=\frac{n\left( \kappa b\right) \tan \eta ^{\beta
}\left( \varepsilon \right) }{n\left( \kappa a\right) \tan \eta ^{\alpha
}\left( \varepsilon \right) }.
\end{equation*}
To obtain this result, we have also used: 
\begin{equation*}
j^{\alpha }\left( \varepsilon ,b\right) =j\left( \kappa b\right) -n\left(
\kappa b\right) \tan \alpha \left( \varepsilon \right) =n\left( \kappa
b\right) \left[ \tan \beta \left( \varepsilon \right) -\tan \alpha \left(
\varepsilon \right) \right] ,
\end{equation*}
from (\ref{b9}) and (\ref{b22}). The second, readily computable function is
that solution of the radial wave equation which vanishes at $a$ with slope $%
1/a^{2}:$%
\begin{equation}
j^{a}\left( \varepsilon ,r\right) \equiv \frac{j^{\alpha }\left( \varepsilon
,r\right) }{a^{2}\partial j^{\alpha }\left( \varepsilon ,r\right) \left/
\partial r\right| _{a}}=-\kappa n\left( \kappa a\right) \,j^{\alpha }\left(
\varepsilon ,r\right) .  \label{b43}
\end{equation}
Evaluation at $r=b$ yields: 
\begin{equation*}
j^{a}\left( \varepsilon ,b\right) =-\kappa n\left( \kappa a\right)
\,j^{\alpha }\left( \varepsilon ,b\right) =\kappa n\left( \kappa a\right)
n\left( \kappa b\right) \left[ \tan \alpha \left( \varepsilon \right) -\tan
\beta \left( \varepsilon \right) \right] ,
\end{equation*}
which is the second function needed. Hence, we have found the following simple
and practical scaling relation for re-screening of the Green matrix: 
\begin{equation}
G^{b}\left( \varepsilon \right) \;=\;\varphi ^{\circ \,a}\left( \varepsilon
,b\right) \,G^{a}\left( \varepsilon \right) \,\varphi ^{\circ \,a}\left(
\varepsilon ,b\right) \;+\;j^{a}\left( \varepsilon ,b\right) \,\varphi
^{\circ \,a}\left( \varepsilon ,b\right) .  \label{b44}
\end{equation}

\subsection{Green functions, matrix elements, and charge density}

The kinked partial wave is the solution of the inhomogeneous Schr\"{o}dinger
equation: 
\begin{equation}
\left( \mathcal{H}-\varepsilon \right) \phi _{R^{\prime }L^{\prime
}}^{a}\left( \varepsilon ,\mathbf{r}\right) =-\sum_{RL}\delta \left(
r_{R}-a_{RL}\right) Y_{L}\left( \mathbf{\hat{r}}_{R}\right) K_{RL,R^{\prime
}L^{\prime }}^{a}\left( \varepsilon \right) ,  \label{e2}
\end{equation}
provided that we define the MTO (\ref{b15}) the 3-fold way indicated in
Figs. \ref{Fig2}--\ref{Fig4}, and therefore --for the MT-Hamiltonian $%
\mathcal{H}$ (\ref{a20})-- use the radial Schr\"{o}dinger equation (\ref{a16}%
) channel-wise.

The \emph{kinks} of the MTO are given correctly by (\ref{e2}), but the
proper MTO does not solve Schr\"{o}dinger's differential equation in the
shells between the screening and the MT-spheres; here we need the 3-fold
way. This way \emph{must} not be an approximation: For instance, 
when applied to
those \emph{linear combinations} of MTOs which solve the KKR equations --and
hence Schr\"{o}dinger's equation-- equation (\ref{e2}) is correct (and
yields zero), because for each active channel, $A^{\prime },$ the two
solutions, $\mathcal{P}_{A^{\prime }}\sum_{A}\psi _{A}^{a}\left( \varepsilon
,\mathbf{r}_{R}\right) c_{A}^{a}$ and $\varphi _{A^{\prime }}^{\circ
\,a}\left( \varepsilon ,r_{R^{\prime }}\right) c_{A^{\prime }}^{a},$ of the
radial wave equation match in value and slope at $a_{R^{\prime }L^{\prime }},
$ and therefore cancel \emph{throughout} the shell $s_{R^{\prime
}}-a_{R^{\prime }L^{\prime }}.$ Expressed in another way: For
energy-dependent MTOs, kink-cancellation leads to cancellation of the
triple-valuedness. For the energy-independent NMTOs to be derived in the
next section, special considerations will be necessary.

Solving (\ref{e2}) for $\delta \left( r_{R}-a_{RL}\right) Y_{L}\left( 
\mathbf{\hat{r}}_{R}\right) ,$ leads to: 
\begin{equation}
\left( \mathcal{H}-\varepsilon \right) \sum_{R^{\prime }L^{\prime }}\phi
_{R^{\prime }L^{\prime }}^{a}\left( \varepsilon ,\mathbf{r}\right)
G_{R^{\prime }L^{\prime },RL}^{a}\left( \varepsilon \right) =-\delta \left(
r_{R}-a_{RL}\right) Y_{L}\left( \mathbf{\hat{r}}_{R}\right)  \label{e0}
\end{equation}
which shows that the linear combinations 
\begin{equation}
\gamma _{RL}^{a}\left( \varepsilon ,\mathbf{r}\right) =\sum_{R^{\prime
}L^{\prime }}\phi _{R^{\prime }L^{\prime }}^{a}\left( \varepsilon ,\mathbf{r}%
\right) G_{R^{\prime }L^{\prime },RL}^{a}\left( \varepsilon \right) ,
\label{e6}
\end{equation}
of MTOs --all with the same energy and screening-- is a \emph{contraction}
of $\mathbf{r}^{\prime }$ onto the screening spheres $\left( \mathbf{r}%
^{\prime }\rightarrow a_{RL},RL\right) $ \emph{of the Green function}
defined by:

\begin{equation*}
\left( \mathcal{H}_{\mathbf{r}}-\varepsilon \right) G\left( \varepsilon ;%
\mathbf{r,r}^{\prime }\right) =-\delta \left( \mathbf{r-r}^{\prime }\right) .
\end{equation*}
The \emph{contracted Green function} $\gamma _{RL}^{a}\left( \varepsilon ,%
\mathbf{r}\right) $ has kink 1 in its own channel and kink 0 in all other
active channels $\left( \neq RL\right) .$ This function is therefore a
solution of the Schr\"{o}dinger equation (defined the 3-fold way) which is
smooth everywhere except at its own screening sphere. $\gamma
_{RL}^{a}\left( \varepsilon ,\mathbf{r}\right) $ is usually \emph{de}%
localized, and when the energy, $\varepsilon ,$ coincides with a pole, $%
\varepsilon _{j},$ of the Green matrix, $\gamma _{RL}^{a}\left( \varepsilon ,%
\mathbf{r}\right) $ diverges everywhere in space. This means, that when $%
\varepsilon =\varepsilon _{j}$, then the \emph{renormalized} function is
smooth also at its own sphere, and it therefore solves Schr\"{o}dinger's
equation.\emph{\ }In vector-matrix notation, equations (\ref{e2}) and (\ref
{e0}) become: 
\begin{eqnarray*}
\left( \mathcal{H}-\varepsilon \right) \phi ^{a}\left( \varepsilon ,\mathbf{r%
}\right) &=&-\delta ^{a}\left( \mathbf{r}\right) K^{a}\left( \varepsilon
\right) , \\
\left( \mathcal{H}-\varepsilon \right) \phi ^{a}\left( \varepsilon ,\mathbf{r%
}\right) G^{a}\left( \varepsilon \right) &\equiv &\left( \mathcal{H}%
-\varepsilon \right) \gamma ^{a}\left( \varepsilon ,\mathbf{r}\right)
=-\delta ^{a}\left( \mathbf{r}\right) ,
\end{eqnarray*}
where we have defined a \emph{set} of spherical harmonics on the $a$-shells
with the following members$:$ 
\begin{equation}
\delta _{RL}^{a}\left( \mathbf{r}_{R}\right) \equiv \delta \left(
r_{R}-a_{RL}\right) Y_{L}\left( \mathbf{\hat{r}}_{R}\right) .  \label{b51}
\end{equation}

If expressed in real space, our Green matrix, $G^{a}\left( \varepsilon
\right) ,$ is what in multiple-scattering theory \cite{Weinberger} is
usually called the scattering path operator and denoted $\tau \left(
\varepsilon \right) $. In the 2nd-generation LMTO formalism, it was denoted $%
g\left( \varepsilon \right) ,$ but in the present paper we denote matrices
by capitals.

Since in the 3-fold way, an MTO takes the value one at its own screening
sphere and zero at all other screening spheres, expression (\ref{e2}) yields
for the matrix element of $\mathcal{H}-\varepsilon $ with another, or the
same, MTO in the set: 
\begin{equation}
\left\langle \phi _{R^{\prime }L^{\prime }}^{a}\left( \varepsilon \right)
\left| \mathcal{H}-\varepsilon \right| \phi _{RL}^{a}\left( \varepsilon
\right) \right\rangle =-K_{R^{\prime }L^{\prime },RL}^{a}\left( \varepsilon
\right) \equiv -G_{R^{\prime }L^{\prime },RL}^{a}\left( \varepsilon \right)
^{-1},  \label{b60}
\end{equation}
which says that the negative of the kink matrix is the \emph{Hamiltonian}
matrix, minus the energy, in the basis of energy-dependent 0th-order MTOs.

For the \emph{overlap integral} between screened spherical waves, with
possibly different energies and in the interstitial between the screening
spheres, defined channel-by-channel, we obtain the simple expression \cite
{Trieste}: 
\begin{eqnarray}
\left\langle \psi _{R^{\prime }L^{\prime }}^{a}\left( \varepsilon ^{\prime
}\right) \mid \psi _{RL}^{a}\left( \varepsilon \right) \right\rangle \; &=&\;%
\frac{B_{R^{\prime }L^{\prime },RL}^{a}\left( \varepsilon ^{\prime }\right)
-B_{R^{\prime }L^{\prime },RL}^{a}\left( \varepsilon \right) }{\varepsilon
^{\prime }-\varepsilon }  \label{b52} \\[0.15cm]
&\longrightarrow &\;\dot{B}_{R^{\prime }L^{\prime },RL}^{a}\left(
\varepsilon \right) \quad \mathrm{if\quad }\varepsilon ^{\prime }\rightarrow
\varepsilon   \notag
\end{eqnarray}
by use of Green's second theorem, together with expression (\ref{b37}) for
the surface integrals. Note that, \emph{neither} active channels different
from the eigen-channels, $R^{\prime }L^{\prime }$ and $RL,$ \emph{nor} the
inactive channels contribute to the surface integrals. The reasons are that $%
\psi _{R^{\prime }L^{\prime }}^{a}\left( \varepsilon ^{\prime },\mathbf{r}%
\right) $ and $\psi _{RL}^{a}\left( \varepsilon ,\mathbf{r}\right) $ vanish
on all 'other' screening spheres, and that they are regular in the inactive
channels. The latter means that, in the inactive channels, the
'screening-sphere interstitial' extends all the way to the sites $\left(
a_{I}\rightarrow 0\right) $. For the overlap integral between kinked partial
waves, the 3-fold way yields: 
\begin{eqnarray}
&&\left\langle \phi _{R^{\prime }L^{\prime }}^{a}\left( \varepsilon ^{\prime
}\right) \mid \phi _{RL}^{a}\left( \varepsilon \right) \right\rangle
\;\equiv \;\left\langle \psi _{R^{\prime }L^{\prime }}^{a}\left( \varepsilon
^{\prime }\right) \mid \psi _{RL}^{a}\left( \varepsilon \right)
\right\rangle \;+\;\delta _{R^{\prime }R}\delta _{L^{\prime }L}\times  
\notag \\[0.15cm]
&&\;\left( \int_{0}^{s_{R}}\varphi _{RL}^{a}\left( \varepsilon ^{\prime
},r\right) \varphi _{RL}^{a}\left( \varepsilon ,r\right)
r^{2}dr-\int_{a_{RL}}^{s_{R}}\varphi _{RL}^{\circ \,a}\left( \varepsilon
^{\prime },r\right) \varphi _{RL}^{\circ \,a}\left( \varepsilon ,r\right)
r^{2}dr\right)   \notag \\[0.2cm]
&=&\;\frac{K_{R^{\prime }L^{\prime },RL}^{a}\left( \varepsilon ^{\prime
}\right) -K_{R^{\prime }L^{\prime },RL}^{a}\left( \varepsilon \right) }{%
\varepsilon ^{\prime }-\varepsilon }\quad \longrightarrow \quad \dot{K}%
_{R^{\prime }L^{\prime },RL}^{a}\left( \varepsilon \right) \;\mathrm{if\;}%
\varepsilon ^{\prime }\rightarrow \varepsilon .  \label{b54}
\end{eqnarray}
For the overlap matrix for the set of contracted Green functions, this
gives: 
\begin{eqnarray}
\left\langle \gamma ^{a}\left( \varepsilon ^{\prime }\right) \mid \gamma
^{a}\left( \varepsilon \right) \right\rangle \; &=&\;-\frac{G^{a}\left(
\varepsilon ^{\prime }\right) -G^{a}\left( \varepsilon \right) }{\varepsilon
^{\prime }-\varepsilon }  \label{b55} \\[0.15cm]
&\rightarrow &\;-\dot{G}^{a}\left( \varepsilon \right) \;=\;G^{a}\left(
\varepsilon \right) \dot{K}^{a}\left( \varepsilon \right) G^{a}\left(
\varepsilon \right) \quad \mathrm{if\;}\varepsilon ^{\prime }\rightarrow
\varepsilon .  \notag
\end{eqnarray}
We see that $\dot{B}^{a}\left( \varepsilon \right) ,$ $\dot{K}^{a}\left(
\varepsilon \right) ,$ and $\dot{G}^{a}\left( \varepsilon \right) $ are
Hermitian, just like $B^{a}\left( \varepsilon \right) ,\;K^{a}\left(
\varepsilon \right) ,$ and $G^{a}\left( \varepsilon \right) $. Whereas $\dot{%
B}^{a}\left( \varepsilon \right) $ and $\dot{K}^{a}\left( \varepsilon
\right) $ are positive definite matrices, that is, their eigenvalues are
positive or zero, $\dot{G}^{a}\left( \varepsilon \right) $ is negative
definite. For well-screened MTOs, the logarithmic derivative functions in
the diagonal of the kink matrix (\ref{b37}) depend more strongly on energy
than the slope matrix. The way to compute the energy derivative $\dot{K}%
^{a}\left( \varepsilon \right) $ is therefore to compute $\dot{B}^{a}\left(
\varepsilon \right) $ by numerical differentiation, and the remaining terms
by integration as in (\ref{b54}).

In the following we shall stay with the normalization (\ref{b24})-(\ref{b26}%
) denoted by Latin --rather than Greek-- superscripts and shall rarely
change the screening. We therefore usually drop the superscript $a$
altogether. Some well-screened representation\ is usually what we have in
mind, but also heavily down-folded --and therefore long-ranged--
representations will be considered. In those cases, some parts of the
computation must of course be performed in the Bloch --or $\mathbf{k}$%
-space-- representation.

The wave function is $\Psi _{i}\left( \mathbf{r}\right) =\phi \left(
\varepsilon _{i},\mathbf{r}\right) c_{i}\,,$ where the eigen(column)vector $%
c_{i}$ solves the KKR equations, $K\left( \varepsilon _{i}\right) c_{i}=0,$
and is normalized according to: $1=c_{i}^{\dagger }\dot{K}\left( \varepsilon
_{i}\right) c_{i},$ in order that $\left\langle \Psi _{i}\mid \Psi
_{i}\right\rangle =1.$ From the definition (\ref{b15}) of the MTO, we see
that an accurate approximation for the \emph{charge density,} which is
consistent with the 3-fold way and, hence, with the normalization, has the
simple form: 
\begin{equation}
\rho \left( \mathbf{r}\right) =\rho ^{\psi }\left( \mathbf{r}\right)
+\sum_{R}\left[ \rho _{R}^{\varphi }\left( \mathbf{r}_{R}\right) -\rho
_{R}^{\varphi ^{\circ }}\left( \mathbf{r}_{R}\right) \right]   \label{b56}
\end{equation}
where the global contribution is: 
\begin{equation}
\rho ^{\psi }\left( \mathbf{r}\right) \equiv \sum_{RR^{\prime
}}\sum_{LL^{\prime }}\int^{\varepsilon _{F}}\psi _{RL}\left( \varepsilon ,%
\mathbf{r}_{R}\right) \,\Gamma _{RL,R^{\prime }L^{\prime }}\left(
\varepsilon \right) \,\psi _{R^{\prime }L^{\prime }}\left( \varepsilon ,%
\mathbf{r}_{R^{\prime }}\right) ^{\ast }d\varepsilon   \label{b57}
\end{equation}
and the local contributions, $\rho _{R}^{\varphi }\left( \mathbf{r}%
_{R}\right) -\rho _{R}^{\varphi ^{\circ }}\left( \mathbf{r}_{R}\right) ,$
which vanish smoothly at their respective MT-sphere, are given by: 
\begin{eqnarray}
\rho _{R}^{\varphi }\left( \mathbf{r}\right)  &=&\sum_{LL^{\prime
}}Y_{L}\left( \mathbf{\hat{r}}\right) Y_{L^{\prime }}^{\ast }\left( \mathbf{%
\hat{r}}\right) \int^{\varepsilon _{F}}\varphi _{Rl}\left( \varepsilon
,r\right) \,\Gamma _{RL,RL^{\prime }}\left( \varepsilon \right) \,\varphi
_{Rl^{\prime }}\left( \varepsilon ,r\right) d\varepsilon   \notag \\[0.1cm]
\rho _{R}^{\varphi ^{\circ }}\left( \mathbf{r}\right)  &=&\sum_{LL^{\prime
}}Y_{L}\left( \mathbf{\hat{r}}\right) Y_{L^{\prime }}^{\ast }\left( \mathbf{%
\hat{r}}\right) \int^{\varepsilon _{F}}\varphi _{Rl}^{\circ }\left(
\varepsilon ,r\right) \,\Gamma _{RL,RL^{\prime }}\left( \varepsilon \right)
\,\varphi _{Rl^{\prime }}^{\circ }\left( \varepsilon ,r\right) d\varepsilon
\,.  \label{b58}
\end{eqnarray}
The common density-of-states matrix in these equations is: 
\begin{equation}
\Gamma _{RL,R^{\prime }L^{\prime }}\left( \varepsilon \right)
=\sum_{i}^{occ}c_{RL,i}\delta \left( \varepsilon -\varepsilon _{i}\right)
c_{R^{\prime }L^{\prime },i}^{\ast }=\frac{1}{\pi }{\rm Im}G_{RL,R^{\prime
}L^{\prime }}\left( \varepsilon +i\delta \right) .  \label{b59}
\end{equation}
The approximations inherent in (\ref{b56}) are that all cross-terms between
products of $\psi $-, $\varphi $-, and $\varphi ^{\circ }$-functions, and
between $\varphi $- or $\varphi ^{\circ }$-functions on different sites are
neglected.

\section{Polynomial MTO approximations}

\label{poly}In this section we shall show how energy-\emph{in}dependent
basis sets may be derived from the kinked partial waves, that is, how we get
rid of the energy dependence of the MTOs. Specifically, we shall preview the
generalization \cite{PolyO,PolyT} of the 3rd-generation LMTO method \cite
{Trieste,MRS} mentioned in connection with Fig. \ref{Fig1}. This
generalization is to an 'N'MTO method in which the basis set consists of
energy-\emph{in}dependent NMTOs,
\begin{eqnarray}
&&\chi _{RL}^{\left( N\right) }\left( \mathbf{r}\right)
=\sum_{n=0}^{N}\sum_{R^{\prime }L^{\prime }}\phi _{R^{\prime }L^{\prime
}}\left( \varepsilon _{n},\mathbf{r}\right) \,L_{R^{\prime }L^{\prime
},RL;n}^{\left( N\right) }\,,  \label{d6} \\[0.15cm]
&&\mathrm{where}\quad \sum_{n=0}^{N}L_{R^{\prime }L^{\prime },RL;n}^{\left(
N\right) }=\delta _{R^{\prime }R}\delta _{L^{\prime }L},  \notag
\end{eqnarray}
constructed as linear combinations of the kinked partial waves at a mesh of $%
N+1$ energies, in such a way that the NMTO basis can describe the solutions, 
$\Psi _{i}\left( \mathbf{r}\right) ,$ of Schr\"{o}dinger's equation
correctly to within an error proportional to $\left( \varepsilon
_{i}-\varepsilon _{0}\right) \left( \varepsilon _{i}-\varepsilon _{1}\right)
...\left( \varepsilon _{i}-\varepsilon _{N}\right) .$ Note the difference
between one-electron energies denoted $\varepsilon _{i}$ and $\varepsilon
_{j},$ and mesh points denoted $\varepsilon _{n}$ and $\varepsilon _{m},$
with $n$ and $m$ taking integer values. The set, $\chi ^{\left( N=0\right)
}\left( \mathbf{r}\right) ,$ is therefore simply $\phi \left( \varepsilon
_{0},\mathbf{r}\right) ,$ and this is the reason why, right at the beginning
of the previous section, $\phi \left( \varepsilon ,\mathbf{r}\right) $ was
named the set of 0th-order energy-\emph{de}pendent MTOs. For $N>0,$ the
NMTOs are smooth and their triple-valuedness decreases with increasing $N.$
For the mesh condensing to one energy, $\varepsilon _{\nu },$ the NMTO basis
is of course constructed as linear combinations of $\phi \left( \varepsilon
_{\nu },\mathbf{r}\right) $ and its first $N$ energy derivatives at $%
\varepsilon _{\nu }.$ For $N$=1, this is the well-known LMTO set.

The immediate practical use of this new development is to widen and sharpen
the energy window inside which the method gives good wave functions,\emph{\
without increasing the size of the basis set. }One may even \emph{decrease}
the size of the basis through downfolding, and still maintain an acceptable
energy window by increasing the \emph{order} of the basis set. The prize for
increasing $N$ is: More computation and increased range of the basis
functions.

\subsection{Energy-independent NMTOs}

What we have done in the previous sections --one might say-- is to factorize
out of the contracted Green function, $\gamma \left( \varepsilon ,\mathbf{r}%
\right) ,$ some spatial functions, $\phi _{RL}\left( \varepsilon ,\mathbf{r}%
\right) ,$ which are so localized that, for two energies inside the
energy-window of interest, the corresponding functions, $\phi _{RL}\left(
\varepsilon ,\mathbf{r}\right) $ and $\phi _{RL}\left( \varepsilon ^{\prime
},\mathbf{r}\right) ,$ \emph{cannot be orthogonal.} In other words: The
kinked partial waves are so well separated through localization and angular
symmetry that we need only \emph{one radial quantum number} for each
function.

Now, we want to get rid of the kinks and to \emph{reduce the
triple-valuedness and the energy dependence} of \emph{each} kinked partial
wave --retaining its $RL$-character-- to a point where the triple-valuedness
and the energy-dependence may both be neglected. This we do, first by
passing from the set $\phi \left( \varepsilon ,\mathbf{r}\right) $ to a set
of so-called $N$th-order energy-dependent MTOs, $\chi ^{\left( N\right)
}\left( \varepsilon ,\mathbf{r}\right) ,$ whose contracted Green function, 
\begin{equation}
\chi ^{\left( N\right) }\left( \varepsilon ,\mathbf{r}\right) G\left(
\varepsilon \right) \;\equiv \;\phi \left( \varepsilon ,\mathbf{r}\right)
G\left( \varepsilon \right) \;-\;\sum_{n=0}^{N}\phi \left( \varepsilon _{n},%
\mathbf{r}\right) G\left( \varepsilon _{n}\right) A_{n}^{\left( N\right)
}\left( \varepsilon \right) ,  \label{c1}
\end{equation}
differs from $\phi \left( \varepsilon ,\mathbf{r}\right) G\left( \varepsilon
\right) $ by a function which remains in the Hilbert space spanned by the
set $\phi \left( \varepsilon ,\mathbf{r}\right) $ with energies inside the
window of interest, and which is \emph{analytical in energy}. The two
contracted Green functions thus have the same poles, and both
energy-dependent basis sets, $\phi \left( \varepsilon ,\mathbf{r}\right) $
and $\chi ^{\left( N\right) }\left( \varepsilon ,\mathbf{r}\right) ,$ can
therefore yield the exact Schr\"{o}dinger-equation solutions. The analytical
functions of energy we wish to determine in such a way that $\chi ^{\left(
N\right) }\left( \varepsilon ,\mathbf{r}\right) $ takes the \emph{same }%
value, $\chi ^{\left( N\right) }\left( \mathbf{r}\right) ,$ at the $N+1$
points, $\varepsilon _{0},...,\varepsilon _{N}.$ With the set $\chi ^{\left(
N\right) }\left( \varepsilon ,\mathbf{r}\right) $ defined that way, we can
finally \emph{neglect} its energy dependence, and the resulting $\chi
^{\left( N\right) }\left( \mathbf{r}\right) $ is then the set of $N$th-order
energy-\emph{in}dependent MTOs.

Other choices for the analytical functions of energy, involving for instance
complex energies or Chebyshev polynomials, await their exploration.

One solution with the property that $\chi _{RL}^{\left( N\right) }\left(
\varepsilon ,\mathbf{r}\right) $ takes the same\emph{\ }value for $%
\varepsilon $ at any of the $N+1$ mesh points, is of course given by the 
\emph{polynomial:} 
\begin{equation*}
A_{n;R^{\prime }L^{\prime },RL}^{\left( N\right) }\left( \varepsilon \right)
=\delta _{R^{\prime }R}\delta _{L^{\prime }L}\prod_{m=0,\neq n}^{N}\frac{%
\varepsilon -\varepsilon _{m}}{\varepsilon _{n}-\varepsilon _{m}},
\end{equation*}
of $N$th degree. But this solution is useless, because it yields: $\chi
^{\left( N\right) }\left( \mathbf{r}\right) =0$. If, instead, we try a
polynomial of $(N-1)$st degree for the analytical function, then we can
write down the corresponding expression for the set $\chi ^{\left( N\right)
}\left( \mathbf{r}\right) $ without explicitly solving for the $\left(
N+1\right) ^{2}$ matrices $A_{n}^{\left( N\right) }\left( \varepsilon
_{m}\right) ,$ and then prove afterwards that each basis function has its
triple-valuedness reduced \emph{consistently }with the remaining error $%
\propto \left( \varepsilon _{i}-\varepsilon _{0}\right) \left( \varepsilon
_{i}-\varepsilon _{1}\right) ...\left( \varepsilon _{i}-\varepsilon
_{N}\right) $ of the set.

Since we want $\chi ^{\left( N\right) }\left( \varepsilon _{n},\mathbf{r}%
\right) $ to be independent of $n$ for $0\leq n\leq N,$ \emph{all} its \emph{%
divided differences} on the mesh --up to and including the divided
difference of order $N$-- \emph{vanish,} with the exception of the 0th
divided difference, which is $\chi ^{\left( N\right) }\left( \mathbf{r}%
\right) $. As a consequence, the $N$th divided difference of $\chi ^{\left(
N\right) }\left( \varepsilon ,\mathbf{r}\right) G\left( \varepsilon \right) $
on the left-hand side of (\ref{c1}) is $\chi ^{\left( N\right) }\left( 
\mathbf{r}\right) $ times the $N$th divided difference of the Green matrix.
Now, the $N$th divided difference of the last term on the right-hand side
vanishes, because it is a polynomial of order $N-1,$ and as a consequence, 
\begin{equation}
\chi ^{\left( N\right) }\left( \mathbf{r}\right) =\frac{\Delta ^{N}\phi
\left( \mathbf{r}\right) G}{\Delta \left[ 0...N\right] }\left( \frac{\Delta
^{N}G}{\Delta \left[ 0...N\right] }\right) ^{-1}.  \label{c2}
\end{equation}
This basically solves the problem of finding the energy-independent NMTOs!
What remains, is to factorize the divided difference of the product $\phi
\left( \varepsilon ,\mathbf{r}\right) G\left( \varepsilon \right) $ into
spatial functions, $\phi \left( \varepsilon _{n},\mathbf{r}\right) ,$ which
are vectors in $RL,$ and matrices, $G\left( \varepsilon _{n}\right) ,$ with $%
n=0,...,N.$ Equivalently, we could use a binomial divided-difference series
in terms of $\phi \left( \varepsilon _{0},\mathbf{r}\right) $ and its first $%
N$ divided differences on the mesh together with $G\left( \varepsilon
_{N}\right) $ and its corresponding divided differences.

For a \emph{condensed} energy mesh, defined by: $\varepsilon _{n}\rightarrow
\varepsilon _{\nu }$ for $0\leq n\leq N,$ the $N$th divided difference
becomes $\frac{1}{N!}$ times the $N$th derivative: 
\begin{equation}
\frac{\Delta ^{N}f}{\Delta \left[ 0...N\right] }\;\equiv \;f\left[ 0...N%
\right] \quad \rightarrow \quad \frac{1}{N!}\left. \frac{d^{N}f\left(
\varepsilon \right) }{d\varepsilon ^{N}}\right| _{\varepsilon _{\nu }}\,,
\label{c3}
\end{equation}
but since a discrete mesh with arbitrarily spaced points is much more
powerful in the present case where the time-consuming part of the
computation is the evaluation of the Green matrix (and its first energy
derivative for use in Eq. (\ref{b55})) at the energy points, we shall
proceed using the language appropriate for a discrete mesh. In (\ref{c3}) we
have introduced the form $f\left[ 0...N\right] $ because it may --more
easily than $\Delta ^{N}f/\Delta \left[ 0...N\right] $-- be modified to
include another kind of divided differences, the so-called Hermite divided
differences, which we shall meet later.

Readers interested in the details of the discrete formalism are referred to
the Appendix where we review relevant parts of the classical theory of
polynomial approximation, and derive formulae indispensable for the NMTO
formalism for discrete meshes. Readers merely interested in an overview, may
be satisfied with the formalism as applied to a \emph{condensed} mesh and
for this, they merely need the translation (\ref{c3}) together with the
divided-difference form of the NMTO to be described in the following.
Details about the Lagrange form may be ignored.

\input{fig4a.tex}

\subsubsection{Lagrange form.}

We first use the Lagrange form (\ref{aa6}) of the divided difference to
factorize the energy-independent NMTO (\ref{c2}) and obtain: 
\begin{equation}
\chi ^{\left( N\right) }\left( \mathbf{r}\right) \;=\;\sum_{n=0}^{N}\frac{%
\phi _{n}\left( \mathbf{r}\right) \,G_{n}}{\prod_{m=0,\neq n}^{N}\left(
\varepsilon _{n}-\varepsilon _{m}\right) }G\left[ 0..N\right] ^{-1},
\label{d0}
\end{equation}
Here and in the following, $\phi _{n}\left( \mathbf{r}\right) \equiv \phi
\left( \varepsilon _{n},\mathbf{r}\right) $ and $G_{n}\equiv G\left(
\varepsilon _{n}\right) .$ Eq. (\ref{d0}) has the form (\ref{d6}) and we
see, that the weight with which the MTO set at $\varepsilon _{n}$ enters the
NMTO set, is:
\begin{equation}
L_{n}^{\left( N\right) }=\frac{G_{n}}{\prod_{m=0,\neq n}^{N}\left(
\varepsilon _{n}-\varepsilon _{m}\right) }G\left[ 0..N\right] ^{-1}.
\label{d1}
\end{equation}
By application of (\ref{aa6}) to the Green matrix, we may verify that these
Lagrange weights sum up to the unit matrix. For this reason, the $RL$
characters of the NMTO basis functions will correspond to those of the
kinked partial waves.

As an example, for $N$=1 we get the so-called chord-LMTO: 
\begin{eqnarray}
\chi ^{\left( 1\right) }\left( \mathbf{r}\right)  &=&\phi _{0}\left( \mathbf{%
r}\right) G_{0}\left( G_{0}-G_{1}\right) ^{-1}+\phi _{1}\left( \mathbf{r}%
\right) G_{1}\left( G_{1}-G_{0}\right) ^{-1}  \notag \\
&=&\phi _{0}\left( \mathbf{r}\right) \left( K_{1}-K_{0}\right)
^{-1}K_{1}+\phi _{1}\left( \mathbf{r}\right) \left( K_{0}-K_{1}\right)
^{-1}K_{0}  \label{d15} \\
&=&\phi _{0}\left( \mathbf{r}\right) -\phi \left( \left[ 01\right] ,\mathbf{r%
}\right) K\left[ 01\right] ^{-1}K_{0}  \notag \\[0.15cm]
&\rightarrow &\phi \left( \mathbf{r}\right) -\dot{\phi}\left( \mathbf{r}%
\right) \dot{K}^{-1}K.  \notag
\end{eqnarray}
In this case, there is only one energy difference, $\varepsilon
_{0}-\varepsilon _{1},$ so it cancels out. In the 3rd line, we have
reordered the terms in such a way that the \emph{Newton form,} to be derived
for general $N$ in (\ref{d25}) and (\ref{d26}) below, is obtained. In the
4th line, we have condensed the mesh onto $\varepsilon _{\nu },$ whereby the
well-known tangent-LMTO \cite{Trieste,MRS} is obtained. The latter is shown
by the full curve in Fig.\ref{Fig4a} for the case of the Si $p_{111}$%
-orbital belonging to an $sp$ set. The dashed curve is the corresponding
kinked partial wave, $\phi \left( \mathbf{r}\right) ,$ shown by the full
curve in Fig. \ref{Fig4}. Compared to the latter, $\chi ^{\left( 1\right)
}\left( \mathbf{r}\right) $ is smooth, but has longer range. The strong
contributions to the tail of the LMTO from $\dot{\phi}\left( \mathbf{r}%
\right) $'s on the nearest neighbor are evident. It is also clear, that for
computations involving wave functions --e.g.\textit{\ }of the charge
density-- the building blocks will rarely be the NMTOs, but the kinked
partial waves, $\phi _{n}\left( \mathbf{r}\right) ,$ which are more compact.

One might fear that the discrete NMTO scheme would fail when one of the mesh
points is close to a one-electron energy, that is, to a pole of the Green
matrix, but that does not happen: If one of the $G_{n}$'s diverges, this
just means that the corresponding Lagrange weight is 1, and the others 0.
Hence, in this case the NMTO is just $\phi _{n}\left( \mathbf{r}\right) ,$
and this is the correct result. Moreover, the kink of this single $\phi
_{n}\left( \mathbf{r}\right) $ does not matter, because in this case where $%
G\left( \varepsilon \right) $ is at a pole, the determinant of its inverse
vanishes, so that the kink-cancellation equations, $K_{n}c_{n}=0,$ have a
non-zero solution, $c_{n},$ which yields a smooth linear combination, $\phi
_{n}\left( \mathbf{r}\right) c_{n},$ of NMTOs$.$

\subsubsection{Kinks and triple-valuedness.}

The energy-independent NMTOs have been defined through (\ref{c1}) and (\ref
{c2}) in such a way that $\chi ^{\left( N\right) }\left( \varepsilon ,%
\mathbf{r}\right) -\chi ^{\left( N\right) }\left( \mathbf{r}\right) \propto
\left( \varepsilon -\varepsilon _{0}\right) ...\left( \varepsilon
-\varepsilon _{N}\right) $. We now show, that also the
kink-and-triple-valuedness of $\chi ^{\left( N\right) }\left( \mathbf{r}%
\right) $ is of that order, and therefore negligible.

The result of projecting the energy-dependent MTO onto $Y_{L^{\prime
}}\left( \mathbf{\hat{r}}_{R^{\prime }}\right) $ for an active channel was
given in (\ref{b18}) for its own channel, and in (\ref{b19}) for any other
active channel. Together, these results may be expressed as: 
\begin{eqnarray*}
\mathcal{P}_{R^{\prime }L^{\prime }}\phi _{RL}^{\alpha }\left( \varepsilon ,%
\mathbf{r}_{R}\right) &=&\varphi _{Rl}^{\alpha }\left( \varepsilon
,r_{R}\right) \delta _{R^{\prime }R}\delta _{L^{\prime }L}+j_{R^{\prime
}L^{\prime }}^{\alpha }\left( \varepsilon ,r_{R^{\prime }}\right) \kappa
^{-1}\times \\
&&\quad \quad \quad \quad \quad \quad \left[ \kappa \cot \eta _{RL}^{\alpha
}\left( \varepsilon \right) \delta _{R^{\prime }R}\delta _{L^{\prime
}L}+B_{R^{\prime }L^{\prime },RL}^{\alpha }\left( \varepsilon \right) \right]
\end{eqnarray*}
or, in terms of the renormalized functions (\ref{b23}), (\ref{b25}), (\ref
{b26}), and (\ref{b43}), as well as the kink matrix defined in (\ref{b34}),
as: 
\begin{equation*}
\mathcal{P}_{R^{\prime }L^{\prime }}\phi _{RL}^{a}\left( \varepsilon ,%
\mathbf{r}_{R}\right) =\varphi _{Rl}^{a}\left( \varepsilon ,r_{R}\right)
\delta _{R^{\prime }R}\delta _{L^{\prime }L}+j_{R^{\prime }L^{\prime
}}^{a}\left( \varepsilon ,r_{R^{\prime }}\right) K_{R^{\prime }L^{\prime
},RL}^{a}\left( \varepsilon \right) .
\end{equation*}
Here, like in (\ref{b18}) and (\ref{b19}), contributions from MT-overlaps
--which are irrelevant for the present discussion-- have been neglected.
Without kinks and triple-valuedness, $\mathcal{P}_{R^{\prime }L^{\prime
}}\phi _{RL}^{a}\left( \varepsilon ,\mathbf{r}_{R}\right) $ would be given
by the first term, and the kinks and the triple-valuedness are therefore
given by the second term: 
\begin{equation}
\mathcal{T}_{R^{\prime }L^{\prime }}\phi _{RL}^{a}\left( \varepsilon ,%
\mathbf{r}_{R}\right) =j_{R^{\prime }L^{\prime }}^{a}\left( \varepsilon
,r_{R^{\prime }}\right) K_{R^{\prime }L^{\prime },RL}^{a}\left( \varepsilon
\right) .  \label{d2}
\end{equation}
This vanishes for those \emph{linear combinations} of MTOs which solve the
kink-cancellation conditions.

What now happens for the energy-independent approximation, $\chi ^{\left(
0\right) }\left( \mathbf{r}\right) \equiv \phi _{0}\left( \mathbf{r}\right) ,
$ to the 0th-order energy-dependent MTO, $\chi ^{\left( 0\right) }\left(
\varepsilon ,\mathbf{r}\right) \equiv \phi \left( \varepsilon ,\mathbf{r}%
\right) ,$ is that the former has kinks and triple-valuedness, but both are
proportional to $K\left( \varepsilon _{0}\right) $ which --according to (\ref
{e2})-- is proportional to $\mathcal{H}-\varepsilon _{0}$ and, hence, to $%
\varepsilon _{i}-\varepsilon _{0}.$ The kinks and triple-valuedness are thus
of the same order as the error of $\chi ^{\left( 0\right) }\left( \mathbf{r}%
\right) .$ Similarly, for $N>0,$ the fact that the $A_{n}^{\left( N\right)
}\left( \varepsilon \right) $'s are polynomials of $(N-1)$st degree,\emph{\ }%
reduces the triple-valuedness of $\chi ^{\left( N\right) }\left( \mathbf{r}%
\right) $ to being proportional to $\left( \varepsilon -\varepsilon
_{0}\right) ...\left( \varepsilon -\varepsilon _{N}\right) ,$ as we shall
now see: Multiplication of (\ref{d2}) with $G^{a}\left( \varepsilon \right) $
from the right yields: $\mathcal{T}\phi ^{a}\left( \varepsilon ,\mathbf{r}%
\right) G^{a}\left( \varepsilon \right) =j^{a}\left( \varepsilon ,\mathbf{r}%
\right) ,$ and for the kinks and the triple-valuedness of the contracted
Green function (\ref{c1}) we therefore get: 
\begin{equation*}
\mathcal{T}\chi ^{\left( N\right) }\left( \varepsilon ,\mathbf{r}\right)
G\left( \varepsilon \right) =j^{a}\left( \varepsilon ,r\right)
-\sum_{n=0}^{N}j^{a}\left( \varepsilon _{n},r\right) A_{n}^{\left( N\right)
}\left( \varepsilon \right) .
\end{equation*}
Taking again the $N$th divided difference for the mesh on which $\chi
^{\left( N\right) }\left( \varepsilon ,\mathbf{r}\right) $ is constant
yields: 
\begin{eqnarray}
\mathcal{T}\chi ^{\left( N\right) }\left( \mathbf{r}\right)  &=&j^{a}\left( %
\left[ 0...N\right] ,r\right) G^{a}\left[ 0...N\right] ^{-1}  \label{d3} \\%
[0.15cm]
&=&-j^{a}\left( \left[ 0...N\right] ,r\right) \left( E^{\left( 0\right)
}-\varepsilon _{0}\right) \left( E^{\left( 1\right) }-\varepsilon
_{1}\right) ...\left( E^{\left( N\right) }-\varepsilon _{N}\right) ,  \notag
\end{eqnarray}
for the kinks and the triple-valuedness of the energy-independent NMTO. In
the last line, we have used an expression --which will be proved in (\ref
{d21})-- for the inverse of the $N$th divided difference of the Green matrix
in terms of the product of energy matrices to be defined in (\ref{d17}). At
present, it suffices to note that differentiation of the Green function, 
\begin{equation}
\check{G}\left( \varepsilon \right) \equiv \sum_{j}\frac{1}{\varepsilon
-\varepsilon _{j}},  \label{d4}
\end{equation}
for a model with \emph{one, normalized orbital }yields: 
\begin{equation*}
\left[ \frac{1}{N!}\left. \frac{d^{N}\check{G}\left( \varepsilon \right) }{%
d\varepsilon ^{N}}\right| _{\varepsilon _{\nu }}\right] ^{-1}=-\left[
\sum_{j}\frac{1}{\left( \varepsilon _{j}-\varepsilon _{\nu }\right) ^{N+1}}%
\right] ^{-1}\approx -\left( \varepsilon _{i}-\varepsilon _{\nu }\right)
^{N+1},
\end{equation*}
where the last approximation holds when the mesh is closer to the
one-electron energy of interest, $\varepsilon _{i},$ than to any other
one-electron energy, $\varepsilon _{j}\neq \varepsilon _{i}$. Note that $j$
--and not $n$-- denotes the radial quantum number. Similarly, this model
Green function has a divided difference on a discrete mesh of $N$+1 points,
whose inverse is: 
\begin{equation}
\check{G}\left[ 0..N\right] ^{-1}=-\left[ \sum_{j}\frac{1}{%
\prod_{n=0}^{N}\left( \varepsilon _{j}-\varepsilon _{n}\right) }\right]
^{-1}\approx -\prod_{n=0}^{N}\left( \varepsilon _{i}-\varepsilon _{n}\right)
,  \label{d20}
\end{equation}
as proved in Eq. (\ref{h10}) of the Appendix. We have thus seen that the
triple-valuedness is of the \emph{same} order as the error present in $\chi
^{\left( N\right) }\left( \mathbf{r}\right) $ due to the neglect of the
energy-dependence of $\chi ^{\left( N\right) }\left( \varepsilon ,\mathbf{r}%
\right) .$

The radial function $j^{a}\left( \varepsilon ,r\right) $ in (\ref{d2})
vanishes for $r\leq a,$ where it has a kink of value $1/a^{2},$ and it
solves the radial wave equation for $r\geq a$. As shown in \cite{PolyO}, its
expansion in powers of $r-a\geq 0$ is: 
\begin{eqnarray*}
rj^{a}\left( \varepsilon ,r\right) \; &=&\;\frac{r-a}{a}+\frac{1}{3!}\left[
l\left( l+1\right) -\varepsilon a^{2}\right] \left( \frac{r-a}{a}\right)
^{3}\;-\;\frac{l\left( l+1\right) }{3!}\left( \frac{r-a}{a}\right) ^{4} \\
&&+\;\frac{1}{5!}\left[ 18l\left( l+1\right) +\left( l\left( l+1\right)
-\varepsilon a^{2}\right) ^{2}\right] \left( \frac{r-a}{a}\right)
^{5}+\;...\,.
\end{eqnarray*}
This means the $N$th divided-difference function entering (\ref{d3})
satisfies: 
\begin{equation*}
j^{a}\left( \left[ 0...N\right] ,r\right) \;\propto \;\left( r-a\right)
^{2N+1}.
\end{equation*}
The \emph{kink and triple-valuedness} (\ref{d3}) in the $s-a$ shell of $\chi
^{\left( N\right) }\left( \mathbf{r}\right) $ is thus proportional to $%
\left( r-a\right) ^{2N+1}\prod_{n=0}^{N}\left( \varepsilon _{i}-\varepsilon
_{n}\right) ,$ and for this reason the energy-window \emph{widens} as $s-a$
decreases, that is, as the screening \emph{increases.}

\subsubsection{Transfer matrices and correspondence with Lagrange
interpolation.}

We need to work out the effect of the Hamiltonian on the NMTO set. Since the
NMTOs with $N>0$ are smooth, the contributions from the delta-function on
the right-hand side of (\ref{e0}) for the contracted Green function will
cancel in the end. Operation on (\ref{c1}) therefore yields: 
\begin{eqnarray*}
\mathcal{H}\left[ \phi \left( \varepsilon ,\mathbf{r}\right) -\chi ^{\left(
N\right) }\left( \varepsilon ,\mathbf{r}\right) \right] G\left( \varepsilon
\right)  &=&\phi \left( \varepsilon ,\mathbf{r}\right) \,\varepsilon G\left(
\varepsilon \right) -\mathcal{H}\chi ^{\left( N\right) }\left( \varepsilon ,%
\mathbf{r}\right) G\left( \varepsilon \right)  \\
&=&\sum\nolimits_{n=0}^{N}\phi _{n}\left( \mathbf{r}\right) \varepsilon
_{n}G_{n}A_{n}^{\left( N\right) }\left( \varepsilon \right) 
\end{eqnarray*}
and by taking the $N$th divided difference for the mesh on which $\chi
^{\left( N\right) }\left( \varepsilon ,\mathbf{r}\right) $ is constant, we
obtain: 
\begin{eqnarray}
\mathcal{H}\gamma \left( \left[ 0...N\right] ,\mathbf{r}\right) \; &=&\;%
\mathcal{H}\chi ^{\left( N\right) }\left( \mathbf{r}\right) G\left[ 0...N%
\right] \;=\;\left( \phi \varepsilon G\right) \left( \left[ 0...N\right] ,%
\mathbf{r}\right)   \notag \\
&=&\;\gamma \left( \left[ 0..N-1\right] ,\mathbf{r}\right) \;+\;\varepsilon
_{N}\,\gamma \left( \left[ 0...N\right] ,\mathbf{r}\right) ,  \label{d16}
\end{eqnarray}
using (\ref{aa5}) with the choice of the last point on the mesh. Solving for
the NMTOs yields: 
\begin{equation}
\left( \mathcal{H}-\varepsilon _{N}\right) \chi ^{\left( N\right) }\left( 
\mathbf{r}\right) \;=\;\chi ^{\left( N-1\right) }\left( \mathbf{r}\right)
\,\left( E^{\left( N\right) }-\varepsilon _{N}\right)   \label{d18}
\end{equation}
where $\chi ^{\left( N-1\right) }\left( \mathbf{r}\right) \equiv \gamma
\left( \left[ 0..N-1\right] ,\mathbf{r}\right) \,G\left[ 0..N-1\right] ^{-1}$
is the energy-independent MTO of order $N-1,$ obtained by \emph{not} using
the last point. Moreover, 
\begin{eqnarray}
E^{\left( N\right) } &\equiv &\;\varepsilon _{N}\;+\;G\left[ 0..N-1\right] G%
\left[ 0...N\right] ^{-1}=\left( \varepsilon G\right) \left[ 0...N\right] \,G%
\left[ 0...N\right] ^{-1}  \notag \\[0.15cm]
&=&\;\sum_{n=0}^{N}\frac{\varepsilon _{n}G_{n}}{\prod_{m=0,\neq n}\left(
\varepsilon _{n}-\varepsilon _{m}\right) }G\left[ 0...N\right]
^{-1}\;=\;\sum_{n=0}^{N}\varepsilon _{n}L_{n}^{\left( N\right) },
\label{d17}
\end{eqnarray}
is the \emph{energy matrix }which --in contrast to $\chi ^{\left( N-1\right)
}\left( \mathbf{r}\right) $-- is independent of which point on the mesh is
omitted. The first equation (\ref{d17}) shows how to compute $E^{\left(
N\right) }$ and the last equation shows that $E^{\left( N\right) }$ is the
energy \emph{weighted} on the $0...N$-mesh by the Lagrange matrices (\ref{d1}%
). For a condensed mesh, the results is the simple one (\ref{a24}) quoted in
the Overview.

We now consider a sequence of energy meshes, starting with the single-point
mesh, $\varepsilon _{0},$ then adding $\varepsilon _{1}$ in order to obtain
the two-point mesh $\varepsilon _{0},\,\varepsilon _{1}$, then adding $%
\varepsilon _{2}$ obtaining the three-point mesh $\varepsilon
_{0},\,\varepsilon _{1},\,\varepsilon _{2},$ a.s.o. Associated with these
meshes we obtain a sequence of NMTO sets: the kinked-partial wave set, $\chi
^{\left( 0\right) }\left( \mathbf{r}\right) ,\,$the LMTO set, $\chi ^{\left(
1\right) }\left( \mathbf{r}\right) ,\,$the QMTO set, $\chi ^{\left( 2\right)
}\left( \mathbf{r}\right) ,$ a.s.o. Working \emph{downwards}, we thus always 
\emph{delete} the point with the \emph{highest} index. Equation (\ref{d18})
now shows that $\mathcal{H}-\varepsilon _{N}$ may be viewed as the \emph{%
step-down }operator and $E^{\left( N\right) }-\varepsilon _{N}$ as the
corresponding \emph{transfer matrix} with respect to the \emph{order} of the
NMTO set.

In this sequence we may include the case $N$=0, provided that we define: 
\begin{equation}
E^{\left( 0\right) }-\varepsilon _{0}\;\equiv \;-K\left( \varepsilon
_{0}\right) \quad \mathrm{and\quad }\chi ^{\left( -1\right) }\left( \mathbf{r%
}\right) \;\equiv \;\delta \left( \mathbf{r}\right) .  \label{d19}
\end{equation}
$N+1$ successive step-down operations on the NMTO set thus yield: 
\begin{equation*}
\left( \mathcal{H}-\varepsilon _{0}\right) ...\left( \mathcal{H}-\varepsilon
_{N}\right) \,\chi ^{\,\left( N\right) }\left( \mathbf{r}\right) \;=\;\delta
\left( \mathbf{r}\right) \left( E^{\left( 0\right) }-\varepsilon _{0}\right)
...\left( E^{\left( N\right) }-\varepsilon _{N}\right) 
\end{equation*}
which, first of all, tells us that one has to operate $N$ times with $\nabla
^{2}$ --that is, with $\nabla ^{2N}$-- before getting to the non-smoothness
of an NMTO. This is consistent with the conclusion about kinks and
triple-valuedness reached in the preceding sub-section. Secondly, it tells
us that the higher the $N$, the more spread out the NMTOs; if we let $%
r\left( M\right) $ denote the range of the $E^{\left( M\right) }$-matrix,
then the range of the NMTO is roughly $\sum_{M=0}^{N}$ $r\left( M\right) .$

The product of $E^{\left( 0\right) }-\varepsilon _{0}$ and all the transfer
matrices on the right-hand side of the above equation is seen from (\ref{d17}%
) and (\ref{d19}) to be simply: $-G\left[ 0...N\right] ^{-1}.$ Hence, we
have found the \emph{matrix equivalent} of the elementary relation (\ref{d20}%
): 
\begin{equation}
-G\left[ 0...N\right] ^{-1}\;=\;\left( E^{\left( 0\right) }-\varepsilon
_{0}\right) \left( E^{\left( 1\right) }-\varepsilon _{1}\right) ...\left(
E^{\left( N\right) }-\varepsilon _{N}\right) \,.  \label{d21}
\end{equation}
The other way around: Recursive use of (\ref{d21}) with increasing $N$, will
generate the transfer matrices and will lead to the first equation (\ref{d17}%
). Note that although the order of the arguments in the divided difference
on the left-hand side is irrelevant, the order of the factors on the
right-hand side is \emph{not, }since the transfer matrices\emph{\ do not
commute.} That $G\left[ 0...N\right] $ is Hermitian, is not so obvious from (%
\ref{d21}) either. Finally, we may note that $G\left[ 0..n-1,n+1..N\right] $
is \emph{not} defined by (\ref{d21}) but by (\ref{aa4}): 
\begin{equation}
G\left[ 0..n-1,n+1..N\right] \;\equiv \;G\left[ 0...N-1\right] \;+\;\left(
\varepsilon _{N}-\varepsilon _{n}\right) G\left[ 0....N\right] .  \notag
\end{equation}

Relation (\ref{d21}) now gives the following form for the Lagrange weights (%
\ref{d1}): 
\begin{equation}
L_{n}^{\left( N\right) }\;=\;\left( E^{\left( n\right) }-\varepsilon
_{n}\right) ^{-1}\frac{\left( E^{\left( 0\right) }-\varepsilon _{0}\right)
..\left( E^{\left( n\right) }-\varepsilon _{n}\right) ..\left( E^{\left(
N\right) }-\varepsilon _{N}\right) }{\left( \varepsilon _{n}-\varepsilon
_{0}\right) ..\left( \varepsilon _{n}-\varepsilon _{n-1}\right) \left(
\varepsilon _{n}-\varepsilon _{n+1}\right) ..\left( \varepsilon
_{n}-\varepsilon _{N}\right) },  \label{d22}
\end{equation}
and this is seen to pass over to the classical expression (\ref{e32}) for
the Lagrange coefficients if we substitute all energy matrices by the
energy: $E^{\left( M\right) }\rightarrow \varepsilon .$ This correspondence
between --on the one side-- the set $\phi \left( \varepsilon ,\mathbf{r}%
\right) $ and the Lagrange polynomial approximation (\ref{e32}) to its
energy dependence (Fig. \ref{Fig1}) and --on the other side-- the set $\chi
^{\left( N\right) }\left( \mathbf{r}\right) $ expressed by (\ref{d6}) with
the matrix form (\ref{d22}), is conceptually very pleasing. What is not so
obvious --but comforting-- is that the Hilbert space spanned by the NMTO set
is invariant under 
energy-dependent linear transformations, $\hat{\phi}\left(
\varepsilon ,\mathbf{r}\right) \equiv \phi \left( \varepsilon ,\mathbf{r}%
\right) T\left( \varepsilon \right) ,$ of the kinked partial waves. This
will be shown in a later section.

By taking matrix elements of (\ref{d18}), the transfer matrix may be
expressed as: 
\begin{equation}
E^{\left( N\right) }-\varepsilon _{N}\;=\;\left\langle \chi ^{\left(
N\right) }\mid \chi ^{\left( N-1\right) }\right\rangle ^{-1}\left\langle
\chi ^{\left( N\right) }\left| \mathcal{H}-\varepsilon _{N}\right| \chi
^{\left( N\right) }\right\rangle .  \label{d23}
\end{equation}
This holds also for $N$=0, provided that we take the value of $\chi ^{\left(
0\right) }\left( \mathbf{r}\right) $ at its screening sphere to be $\varphi
^{\circ \,a}\left( \varepsilon ,a\right) =1$ --as dictated by the 3-fold
way-- so that $\left\langle \chi ^{\left( 0\right) }\mid \chi ^{\left(
-1\right) }\right\rangle =1.$ The form (\ref{d23}) shows that the transfer
matrices with $N\geq 1$ are \emph{not} Hermitian, but short ranged, as one
may realize by recursion starting from $N$=0. Finally, it should be
remembered that the NMTOs considered sofar have particular normalizations,
which are \emph{not:} $\left\langle \chi ^{\left( N\right) }\mid \chi
^{\left( N\right) }\right\rangle =1,$ and so do the transfer matrices. We
shall return to this point.

\subsubsection{Newton form.}

Instead of using the Lagrange form (\ref{aa6}) to factorize the NMTO (\ref
{c2}), we may use the divided-difference expression (\ref{aa22}). With the
substitutions: $f\left( \varepsilon \right) \rightarrow G\left( \varepsilon
\right) $ and $g\left( \varepsilon \right) \rightarrow \phi \left(
\varepsilon ,\mathbf{r}\right) ,$ we obtain the Newton form for the NMTO
which most clearly exhibits the step-down property (\ref{d18}): 
\begin{eqnarray}
\chi ^{\left( N\right) }\left( \mathbf{r}\right) 
&=&\sum\nolimits_{M=N}^{0}\phi \left( \left[ M..N\right] ,\mathbf{r}\right) G%
\left[ 0..M\right] G\left[ 0...N\right] ^{-1}  \notag \\[0.15cm]
&=&\phi _{N}\left( \mathbf{r}\right) +\phi \left( \left[ N-1,N\right] ,%
\mathbf{r}\right) \left( E^{\left( N\right) }-\varepsilon _{N}\right) +..
\label{d24} \\
&&\quad \quad \quad \quad ..+\phi \left( \left[ 0...N\right] ,\mathbf{r}%
\right) \left( E^{\left( 1\right) }-\varepsilon _{1}\right) ..\left(
E^{\left( N\right) }-\varepsilon _{N}\right) ,  \notag
\end{eqnarray}
since, from (\ref{e2}) and (\ref{d16}), 
\begin{eqnarray}
\left( \mathcal{H}-\varepsilon _{N}\right) \phi _{N}\left( \mathbf{r}\right)
&=&-\delta _{N,0}\delta \left( \mathbf{r}\right) K_{0},  \notag \\[0.15cm]
\left( \mathcal{H}-\varepsilon _{N}\right) \phi \left( \left[ M...N\right] ,%
\mathbf{r}\right)  &=&\phi \left( \left[ M..N-1\right] ,\mathbf{r}\right) .
\label{c4}
\end{eqnarray}
We thus realize that the energy matrices in the Newton series for the NMTO
set are the matrices for stepping down to the sets of lower order. For some
purposes, the 'reversed' series, obtained from (\ref{aa22}) with $f\left(
\varepsilon \right) \rightarrow \phi \left( \varepsilon ,\mathbf{r}\right)
G\left( \varepsilon \right) $ and $g\left( \varepsilon \right) \rightarrow
G\left( \varepsilon \right) $: 
\begin{eqnarray}
\chi ^{\left( N\right) }\left( \mathbf{r}\right) 
&=&\sum\nolimits_{M=0}^{N}\phi \left( \left[ 0..M\right] ,\mathbf{r}\right) G%
\left[ M..N\right] G\left[ 0...N\right] ^{-1}  \notag \\[0.15cm]
&=&\phi _{0}\left( \mathbf{r}\right) +\phi \left( \left[ 01\right] ,\mathbf{r%
}\right) \left( E^{\left( N\right) }-\varepsilon _{0}\right) +..  \label{d25}
\\
&&\quad \quad \quad \quad ..+\phi \left( \left[ 0...N\right] ,\mathbf{r}%
\right) \left( E^{\left( 1\right) }-\varepsilon _{N-1}\right) ..\left(
E^{\left( N\right) }-\varepsilon _{0}\right) ,  \notag
\end{eqnarray}
is more convenient. This expression clearly exhibits the correspondence with
the Newton polynomial approximation (\ref{e33}) to the energy dependence of $%
\phi \left( \varepsilon ,\mathbf{r}\right) .$ Conceptually, a
divided-difference series is more desirable than the Lagrange series,
because the Lagrange weights (\ref{d22}) 'fluctuate wildly' as a function of 
$n,$ taken in the order of monotonically increasing energies.

For a condensed mesh, (\ref{d24}) and (\ref{d25}) obviously reduce to
one-and-the-same matrix-equivalent of the Taylor series for $\phi \left(
\varepsilon ,\mathbf{r}\right) :$%
\begin{eqnarray*}
\chi ^{\left( N\right) }\left( \mathbf{r}\right) \; &\rightarrow &\;\phi
\left( \mathbf{r}\right) +\dot{\phi}\left( \mathbf{r}\right) \left(
E^{\left( N\right) }-\varepsilon _{\nu }\right) +.. \\
&&\quad \quad \quad \quad \quad \quad \quad \quad ..+\frac{1}{N!}\overset{%
\left( N\right) }{\phi }\left( \mathbf{r}\right) \left( E^{\left( 1\right)
}-\varepsilon _{\nu }\right) ..\left( E^{\left( N\right) }-\varepsilon _{\nu
}\right) ,
\end{eqnarray*}
and (\ref{c4}) becomes: 
\begin{equation*}
\left( \mathcal{H}-\varepsilon _{\nu }\right) \phi \left( \mathbf{r}\right)
=-\delta _{N,0}\delta \left( \mathbf{r}\right) K,\quad \;\left( \mathcal{H}%
-\varepsilon _{\nu }\right) \frac{\overset{\left( N-M\right) }{\phi \left( 
\mathbf{r}\right) }}{\left( N-M\right) !}=\frac{\overset{\left( N-M-1\right) 
}{\phi \left( \mathbf{r}\right) }}{\left( N-M-1\right) !}.
\end{equation*}

Readers used to the LMTO-ASA method, where --according to (\ref{a10})-- the
KKR matrix is basically the two-center TB Hamiltonian, may not like the
thought of having to differentiate its inverse, the Green matrix, with
respect to energy. (The computer seems to work well with the formalism based
on the Green matrix). Such readers might therefore prefer an NMTO formalism
in terms of kink matrices. For a discrete mesh many ugly relations exist,
but the one relation which is conceptually pleasing is the following: 
\begin{eqnarray}
&&0=  \label{d27} \\
&&K_{0}+K\left[ 01\right] \left( E^{\left( N\right) }-\varepsilon
_{0}\right) +..+K\left[ 0..N\right] \left( E^{\left( 1\right) }-\varepsilon
_{N\mathrm{-1}}\right) ..\left( E^{\left( N\right) }-\varepsilon _{0}\right)
,  \notag
\end{eqnarray}
because it looks like the matrix form of the secular KKR equation: $\left|
K\left( \varepsilon \right) \right| =0.$ This relation may be obtained by
taking the $N$th divided difference of the equation: $K\left( \varepsilon
\right) G\left( \varepsilon \right) \equiv 1,$ using the binomial expression
(\ref{aa22}) for a product like in (\ref{d25}), but with $K\left(
\varepsilon \right) $ substituted for $\phi \left( \varepsilon ,\mathbf{r}%
\right) ,$ and multiplying the result from the right by $G\left[ 0...N\right]
^{-1}.$ To find the transfer matrices from (\ref{d27}), we may solve for $%
E^{\left( N\right) }-\varepsilon _{0}$ and do recursion starting from $N$=1.
The results are$:$%
\begin{eqnarray}
E^{\left( 1\right) }-\varepsilon _{0} &=&-K\left[ 01\right] ^{-1}K_{0}\quad
\rightarrow \quad -\dot{K}^{-1}K\,,  \notag \\[0.1cm]
E^{\left( 2\right) }-\varepsilon _{0} &=&-\left( K\left[ 01\right] +K\left[
012\right] \left( E^{\left( 1\right) }-\varepsilon _{1}\right) \right)
^{-1}K_{0}  \label{d26} \\
&\rightarrow &-\left( \dot{K}-\ddot{K}\dot{K}^{-1}K/2\right) ^{-1}K,  \notag
\end{eqnarray}
a.s.o. These low-$N$ expressions are reasonably simple. For $N$=1, the
discrete form is seen to be identical with (\ref{d15}) and, for a condensed
mesh, it reduces to the well-known expression for the 3rd-generation LMTO.
We conclude that the energy matrices, $E^{\left( M\right) },$ are
well-behaved functions of the kink matrix and its divided differences, up to
and including $M$th order. With $M$ increasing, the corresponding
expressions for $E^{\left( M\right) }$ however become more and more
complicated. The simplest expression for $E^{\left( M\right) }$ is therefore
(\ref{d17}), the one which uses $G$-language.

\subsection{Variational NMTO method}

\label{var}The NMTO set has been defined through (\ref{c1}) and (\ref{c2})
in such a way that its leading errors are proportional to $\left(
\varepsilon -\varepsilon _{0}\right) ..\left( \varepsilon -\varepsilon
_{N}\right) $. By virtue of the variational principle, solution of the
generalized eigenvalue problem (\ref{a3}) with this basis set will therefore
provide \emph{one-electron energies,} $\varepsilon _{i},$ with a leading
error $\propto \left( \varepsilon _{i}-\varepsilon _{0}\right) ^{2}..\left(
\varepsilon _{i}-\varepsilon _{N}\right) ^{2}.$ The error of the \emph{wave
function} will of course still be of order $\left( \varepsilon
_{i}-\varepsilon _{0}\right) ..\left( \varepsilon _{i}-\varepsilon
_{N}\right) ,$ but that is usually all right because, as mentioned at the
beginning of the present section, the MTO scheme is based on the
factorization: $\gamma \left( \varepsilon ,\mathbf{r}\right) =\phi \left(
\varepsilon ,\mathbf{r}\right) G\left( \varepsilon \right) ,$ where $\phi
\left( \varepsilon ,\mathbf{r}\right) $ has a \emph{smooth} energy
dependence and $G\left( \varepsilon \right) $ provides the poles at the
one-electron energies.

\subsubsection{Hamiltonian and overlap matrices.}

For a variational calculation, we need expressions for the NMTO overlap and
Hamiltonian matrices, $\left\langle \chi ^{\left( N\right) }\mid \chi
^{\left( N\right) }\right\rangle $ and $\left\langle \chi ^{\left( N\right)
}\left| \mathcal{H}\right| \chi ^{\left( N\right) }\right\rangle $. From (%
\ref{c1}), the $N$th divided difference of the contracted Green function (%
\ref{e6}) is: 
\begin{equation}
\gamma ^{\left( N\right) }\left( \left[ 0..N\right] ,\mathbf{r}\right) =\chi
^{\left( N\right) }\left( \mathbf{r}\right) G\left[ 0..N\right]
=\sum_{n=0}^{N}\frac{\phi _{n}\left( \mathbf{r}\right) \,G_{n}}{%
\prod_{m=0,\neq n}^{N}\left( \varepsilon _{n}-\varepsilon _{m}\right) }
\label{d5}
\end{equation}
and using now (\ref{b55}), we obtain for the integral over the product of
the $M$th and $N$th divided differences of contracted Green functions: 
\begin{eqnarray}
&&\left\langle \gamma \left[ 0...M\right] \mid \gamma \left[ 0....N\right]
\right\rangle =\sum_{n=0}^{N}\sum_{n^{\prime }=0}^{M}\frac{-G\left[
n,n^{\prime }\right] }{\prod\limits_{m=0,\neq n}^{N}\left( \varepsilon
_{n}-\varepsilon _{m}\right) \prod\limits_{m^{\prime }=0,\neq n^{\prime
}}^{M}\left( \varepsilon _{n^{\prime }}-\varepsilon _{m^{\prime }}\right) } 
\notag \\[0.15cm]
&=&-G\left[ \left[ 0...M\right] ..N\right] \;\rightarrow \,-\frac{\overset{%
\left( M+N+1\right) }{G}}{\left( M+N+1\right) !}.  \label{d14}
\end{eqnarray}
This is simply the negative of the $\left( M+N+1\right) $st \emph{Hermite
divided difference }(\ref{h5}) of the Green matrix, as proved in Eq. (\ref
{h11}) in the Appendix!

Note that the meaning of a matrix equation like (\ref{b55}) is: 
\begin{eqnarray*}
\left\langle \gamma _{RL}\left( \varepsilon _{n}\right) \mid \gamma
_{R^{\prime }L^{\prime }}\left( \varepsilon _{n^{\prime }}\right)
\right\rangle &=&-G_{RL,R^{\prime }L^{\prime }}\left[ n,n^{\prime }\right] \\
&=&-G_{RL,R^{\prime }L^{\prime }}\left[ n^{\prime },n\right] =\left\langle
\gamma _{RL}\left( \varepsilon _{n^{\prime }}\right) \mid \gamma _{R^{\prime
}L^{\prime }}\left( \varepsilon _{n}\right) \right\rangle .
\end{eqnarray*}
In \emph{matrix} notation, that is: $\left\langle \gamma _{n}\mid \gamma
_{n^{\prime }}\right\rangle =\left\langle \gamma _{n^{\prime }}\mid \gamma
_{n}\right\rangle ,$ and not: $\left\langle \gamma _{n}\mid \gamma
_{n^{\prime }}\right\rangle =\left\langle \gamma _{n^{\prime }}\mid \gamma
_{n}\right\rangle ^{\ast }.$ Even without the symmetry of the matrix $G\left[
n,n^{\prime }\right] $ with respect to the exchange of $n$ and $n^{\prime },$
it is of course always true that 
\begin{equation*}
\left\langle \gamma _{RL}\left( \varepsilon _{n}\right) \mid \gamma
_{R^{\prime }L^{\prime }}\left( \varepsilon _{n^{\prime }}\right)
\right\rangle =\left\langle \gamma _{R^{\prime }L^{\prime }}\left(
\varepsilon _{n^{\prime }}\right) \mid \gamma _{RL}\left( \varepsilon
_{n}\right) \right\rangle ^{\ast },
\end{equation*}
i.e. that a matrix like $\left\langle \gamma _{n}\mid \gamma _{n^{\prime
}}\right\rangle $ is Hermitian: $\left\langle \gamma _{n}\mid \gamma
_{n^{\prime }}\right\rangle \;=\;\left\langle \gamma _{n^{\prime }}\mid
\gamma _{n}\right\rangle ^{\dagger }.$ The point is, that $n$ is an argument
-- not an index -- of a matrix. Similarly, $N$ and $M$ are not matrix
indices in (\ref{d14}). Since the first expression (\ref{d14}) is symmetric
under exchange of $N$ and $M,$ because $G\left[ n,n^{\prime }\right] $ is
symmetric, we may choose $M\leq N,$ and this has in fact been done in the
second expression.

From (\ref{d16}) and (\ref{d14}), we now see that the Hamiltonian matrix
between the $N$th divided differences of contracted Green functions becomes: 
\begin{eqnarray}
\left\langle \gamma \left[ 0...N\right] \left| \mathcal{H}-\varepsilon
_{N}\right| \gamma \left[ 0...N\right] \right\rangle  &=&\left\langle \gamma %
\left[ 0...N\right] \mid \gamma \left[ 0..N-1\right] \right\rangle   \notag
\\
&=&-G\left[ \left[ 0..N-1\right] N\right] \;\rightarrow \,-\frac{\overset{%
\left( 2N\right) }{G}}{\left( 2N\right) !}.  \label{d13}
\end{eqnarray}
Hence, we have arrived at the important results: The NMTO overlap matrix may
be expressed in terms of the $N$th-order divided difference and the $(2N+1)$%
st Hermite divided difference of the Green matrix as: 
\begin{equation}
\left\langle \chi ^{\left( N\right) }\mid \chi ^{\left( N\right)
}\right\rangle \;=\;-G\left[ 0...N\right] ^{-1}G\left[ \left[ 0...N\right] %
\right] \,G\left[ 0...N\right] ^{-1},  \label{d12}
\end{equation}
where the --even simpler-- result for a condensed mesh was quoted in the
Overview (\ref{a23}). The Hermite derivative $G\left[ \left[ 0,...,N\right] %
\right] $ is thus negative definite. The NMTO Hamiltonian matrix may be
expressed analogously, in terms of a 2$N$th-order Hermite divided
difference: 
\begin{equation}
\left\langle \chi ^{\left( N\right) }\left| \mathcal{H}-\varepsilon
_{N}\right| \chi ^{\left( N\right) }\right\rangle =-G\left[ 0...N\right]
^{-1}G\left[ \left[ 0..N-1\right] N\right] \,G\left[ 0...N\right] ^{-1}.
\label{d11}
\end{equation}
Here again, the result given in (\ref{a23}) for a condensed mesh is even
simpler. The NMTO Green function is 
\begin{eqnarray*}
&&\left\langle \chi ^{\left( N\right) }\left| z-\mathcal{H}\right| \chi
^{\left( N\right) }\right\rangle ^{-1}= \\
&&G\left[ 0...N\right] \left\{ G\left[ \left[ 0..N-1\right] N\right] -\left(
z-\varepsilon _{N}\right) G\left[ \left[ 0...N\right] \right] \right\} ^{-1}G%
\left[ 0...N\right] 
\end{eqnarray*}

Expressions (\ref{d12}) and (\ref{d11}) for the NMTO overlap and Hamiltonian
matrices are not only simple and beautiful, but they also offer sweet coding
and speedy computation. For a crystal, and transforming to $\mathbf{k}$%
-representation, one may even use the representation of contracted Green
functions where the overlap and Hamiltonian matrices --according to (\ref
{d14}) and (\ref{d13})-- are merely $-G\left[ \left[ 0...N\right] \right] $
and $-G\left[ \left[ 0..N-1\right] N\right] .$ In Section \ref{linear} we
shall see that an energy-dependent linear transformation of the kinked
partial waves does \emph{not} change the Hilbert space spanned by an
energy-independent NMTO set --but only the individual basis functions.
Therefore, we might also use kinked partial waves $\phi ^{\alpha }\left(
\varepsilon ,\mathbf{r}\right) $ and the Green matrix $G^{\alpha }\left(
\varepsilon \right) $ with phase-shift normalization.

In summary: The variational NMTO scheme requires computation of the kink
matrix and its first energy derivative at the $N+1$ mesh points. It delivers
energies and wave functions which are correct to order $2N+1$ and $N$,
respectively. This lower accuracy of the wave functions is appropriate
because the kinked partial waves are rather smooth functions of energy. For
the computation of the $\dot{\partial}_{n}$'s entering $\dot{K}_{n}\equiv
a\left( \dot{B}_{n}-\dot{\partial}_{n}\right) ,$ radial
normalization-integrals should be used.

As an example, for the LMTO method, the Hamiltonian and overlap matrices are
respectively: 
\begin{eqnarray}
&&\left\langle \chi ^{\left( 1\right) }\left| \mathcal{H}-\varepsilon
_{1}\right| \chi ^{\left( 1\right) }\right\rangle \;=\;-G\left[ 01\right]
^{-1}\,G\left[ \left[ 0\right] 1\right] \,G\left[ 01\right] ^{-1}  \notag \\
&=&\;\left( \varepsilon _{0}-\varepsilon _{1}\right) \left(
G_{0}-G_{1}\right) ^{-1}\left( -\dot{G}_{0}+G\left[ 01\right] \right) \left(
G_{0}-G_{1}\right) ^{-1}  \label{a17} \\[0.1cm]
&\rightarrow &\;-\dot{G}^{-1}\frac{\ddot{G}}{2!}\dot{G}^{-1}\;=\;-K+K\dot{K}%
^{-1}\frac{\ddot{K}}{2!}\dot{K}^{-1}K,  \notag
\end{eqnarray}
and 
\begin{eqnarray}
&&\left\langle \chi ^{\left( 1\right) }\mid \chi ^{\left( 1\right)
}\right\rangle \;=\;-G\left[ 01\right] ^{-1}\,G\left[ \left[ 01\right] %
\right] \,G\left[ 01\right] ^{-1}  \notag \\
&=&\;\left( G_{0}-G_{1}\right) ^{-1}\left( -\dot{G}_{0}+2G\left[ 01\right] -%
\dot{G}_{1}\right) \left( G_{0}-G_{1}\right) ^{-1}  \label{a18} \\[0.1cm]
&\rightarrow &\;-\dot{G}^{-1}\frac{\dddot{G}}{3!}\dot{G}^{-1}\;=\;\dot{K}-K%
\dot{K}^{-1}\frac{\ddot{K}}{2!}-\frac{\ddot{K}}{2!}\dot{K}^{-1}K+K\dot{K}%
^{-1}\frac{\dddot{K}}{3!}\dot{K}^{-1}K.  \notag
\end{eqnarray}
The result for a condensed mesh in terms of the kink matrix and its first
three energy derivatives is seen to be almost identical to the one (\ref{a13}%
), which in previous LMTO generations required the ASA. To get exactly to (%
\ref{a13}), one needs to transform to the LMTO set: $\hat{\chi}^{\left(
1\right) }\left( \mathbf{r}\right) \equiv \chi ^{\left( 1\right) }\left( 
\mathbf{r}\right) \dot{K}^{-1/2},$ which in fact corresponds to a L\"{o}wdin
orthonormalization of the 0th-order set. We shall return to this matter in
Sect. \ref{ASA}. From the above relations we realize that --even for a
condensed mesh and $N$ as low as 1--$\;G$-language is far simpler than $K$%
-language.

\subsubsection{Orthonormal NMTOs.}

In many cases one would like to work with a representation of \emph{%
orthonormal} NMTOs, which preserves the $RL$-character of each NMTO. In
order to arrive at this, we should -- in the language of L\"{o}wdin --
perform a \emph{symmetrical orthonormalization} of the NMTO set. According
to (\ref{d12}) such a representation is obtained by the following
transformation: 
\begin{equation}
\check{\chi}^{\left( N\right) }\left( \mathbf{r}\right) =\chi ^{\left(
N\right) }\left( \mathbf{r}\right) \,G\left[ 0...N\right] \sqrt{-G\left[ %
\left[ 0...N\right] \right] }^{-1},  \label{o28}
\end{equation}
because it yields: 
\begin{equation*}
\left\langle \check{\chi}^{\left( N\right) }\mid \check{\chi}^{\left(
N\right) }\right\rangle =-\sqrt{-G\left[ \left[ 0...N\right] \right] }%
^{-1\dagger }G\left[ \left[ 0...N\right] \right] \sqrt{-G\left[ \left[ 0...N%
\right] \right] }^{-1}=1.
\end{equation*}
Note that this means: $-G\left[ \left[ 0..N\right] \right] =\sqrt{-G\left[ %
\left[ 0..N\right] \right] }^{\dagger }\sqrt{-G\left[ \left[ 0..N\right] %
\right] }.$ In this orthonormal representation, the Hamiltonian matrix
becomes 
\begin{eqnarray}
\left\langle \check{\chi}^{\left( N\right) }\left| \mathcal{H}-\varepsilon
_{N}\right| \check{\chi}^{\left( N\right) }\right\rangle  &=&-\sqrt{-G\left[ %
\left[ 0...N\right] \right] }^{-1\,\dagger }\times   \label{o29} \\
&&G\left[ \left[ 0..N-1\right] N\right] \,\sqrt{-G\left[ \left[ 0...N\right] %
\right] }^{-1}.  \notag
\end{eqnarray}
To find an efficient way to \emph{compute} the \emph{square root} of the
Hermitian, positive definite matrix $-G\left[ \left[ 0...N\right] \right] $
may be a problem. Of course one may diagonalize the matrix, take the square
root of the eigenvalues, and then back-transform, but this is time
consuming. Cholesky decomposition is a better alternative, but that usually
amounts to staying in the original representation. L\"{o}wdin
orthogonalization works if the set is nearly orthogonal, because then the
overlap matrix is nearly diagonal, and L\"{o}wdin's solution was to
normalize the matrix such that it becomes 1 along the diagonal and then
expand in the off-diagonal part, $O:$ 
\begin{equation}
\sqrt{1+O}^{-1}\;=\;1-\frac{1}{2}O+\frac{3}{8}O^{2}-...  \label{o32}
\end{equation}
This should work for the NMTO overlap matrix (\ref{d12}) when the NMTOs are
nearly orthogonal, but it hardly works for $-G\left[ \left[ 0...N\right] %
\right] .$ There is therefore no advantage in pulling out the factor $G\left[
0...N\right] ,$ on the contrary. The other way around: In order to take the
square root of $-G\left[ \left[ 0...N\right] \right] ,$ we should find a
transformation, $T,$ such that $T^{\dagger }G\left[ \left[ 0...N\right] %
\right] T$ is nearly diagonal, and then perform the L\"{o}wdin
orthonormalization on the latter matrix. We shall return to this problem in
Sect. \ref{ortho}.

\input{fig5.tex}

\subsubsection{One-orbital model: switching behavior of $H^{\left( N\right)
},$ $L_{n}^{\left( N\right) },$ and the variational energy.}

Our development of the NMTO formalism has been focused on its matrix aspects
and, through the introduction of energy \emph{matrices} and by pointing to
the correspondence with classical Lagrange and Newton interpolation of the
energy-dependent kinked partial waves, we have tried to make the reader
accept the seemingly uncomfortable fact, that the quantities of interest do
arise by energy differentiations of a Green matrix.

Let us now illustrate the Green-function aspects by considering the $1\times
1$ Green matrix (\ref{d4}) for \emph{one,} \emph{normalized orbital:} $%
\check{\chi}^{\left( N\right) }\left( \mathbf{r}\right) =\Psi _{j}\left( 
\mathbf{r}\right) \;$with $\left\langle \left| \check{\chi}^{\left( N\right)
}\right| ^{2}\right\rangle =1.$ Note that in this model, $j$ runs over the
one-electron energies, which is a different set --with much larger spacing--
than the energy mesh whose points are denoted $n$ and $m.$ For a crystal,
and using Bloch-symmetrized NMTOs and Green matrices, $\check{\chi}^{\left(
N\right) }\left( \mathbf{k,r}\right) $ and $\check{G}\left( \varepsilon ,%
\mathbf{k}\right) ,$ this would be an $s$-band model with $j$ being the
radial quantum number. We want the NMTO to describe the $i$-band and
therefore choose the mesh between $\varepsilon _{i-1}\left( \mathbf{k}%
\right) $ and $\varepsilon _{i+1}\left( \mathbf{k}\right) .$ In the
following we shall drop the Bloch vector and not necessarily consider a
crystal.

We first demonstrate how $\check{E}^{\left( N\right) }\equiv H^{\left(
N\right) }$ --in this case a $1\times 1$ Hamiltonian (see Sect.\ref{ortho}%
)-- expressed in terms of ratios of energy derivatives of a Green function,
with its singular behavior, produces correct results for the one-electron
energy and how, when the mesh is swept over a large energy interval, $%
H^{\left( N\right) }$ switches between bands with different radial quantum
numbers$.$ From (\ref{d17}) and (\ref{d20}) we get: 
\begin{equation*}
H^{\left( N\right) }-\varepsilon _{N}=\frac{\check{G}\left[ 0..N-1\right] }{%
\check{G}\left[ 0...N\right] }=\left( \varepsilon _{i}-\varepsilon
_{N}\right) \frac{1+\sum\limits_{j\neq i}\prod_{m=0}^{N-1}\frac{\varepsilon
_{i}-\varepsilon _{m}}{\varepsilon _{j}-\varepsilon _{m}}}{%
1+\sum\limits_{j\neq i}\prod_{m=0}^{N}\frac{\varepsilon _{i}-\varepsilon _{m}%
}{\varepsilon _{j}-\varepsilon _{m}}}.
\end{equation*}
Hence, for the model and an energy mesh with $N+1$ points, $H^{\left(
N\right) }$ equals$\;\varepsilon _{i}$ to order $N,$ with an error
proportional to $\left( \varepsilon _{i}-\varepsilon _{0}\right) ..\left(
\varepsilon _{i}-\varepsilon _{N}\right) ,$ which for a condensed mesh
becomes $\left( \varepsilon _{i}-\varepsilon _{\nu }\right) ^{N+1}.$ In Fig. 
\ref{Fig5} we show $H^{\left( N\right) }\left( \varepsilon _{\nu }\right) $
for $N=1$ to 6,$\;$computed from the above expression for a four-level model
with $\varepsilon _{j}=$ 0, 1, 2, and 3, and a condensed mesh. We see that $%
H^{\left( N\right) }\left( \varepsilon _{\nu }\right) $ behaves as it
should: It switches from one level to the next, with the plateau around each
level flattening out as $N$ increases. For $N$ odd, the switching-curve is
step-like and, for $N$ even, the switching is via $-\infty \rightarrow
+\infty .$ This comes from the ability of the denominator in the expression
for $H^{\left( N\right) }$ to be zero when $N+1$ is odd. An energy-\emph{in}%
dependent orbital, as considered in the present model, can of course only
describe \emph{one} band. With the NMTO defined for a mesh condensed onto a
chosen energy $\varepsilon _{\nu },$ we want to describe the band near $%
\varepsilon _{\nu }$ as well as possible --also if the distance to the next
band is small-- and with a result which over a large region is insensitive
to the choice of $\varepsilon _{\nu }.$ In a \emph{multi-orbital}
calculation, we should fold down those channels which are switching in the
energy range of interest into the screened spherical waves. This will remove
schizophrenic members of the NMTO set and prevent the possible occurrence of
ghost bands.

In the one-orbital model, the estimate of a true, normalized wave function, $%
\check{\phi}\left( \varepsilon _{i},\mathbf{r}\right) ,$ is the $N$th-order
muffin-tin orbital: $\check{\chi}^{\left( N\right) }\left( \mathbf{r}\right)
=\sum_{n}^{N}\check{\phi}_{n}\left( \mathbf{r}\right) L_{n}^{\left( N\right)
}.$ If we now use (\ref{d4}) and (\ref{d20}) to evaluate expression (\ref{d1}%
) for the Lagrange weights, we find: 
\begin{equation*}
L_{n}^{\left( N\right) }\;=\;\frac{\sum\limits_{j}\frac{1}{\varepsilon
_{j}-\varepsilon _{n}}}{\sum\limits_{j}\frac{1}{\varepsilon _{j}-\varepsilon
_{n}}\prod_{m=0,\neq n}^{N}\frac{\varepsilon _{n}-\varepsilon _{m}}{%
\varepsilon _{j}-\varepsilon _{m}}}\;=\;l_{n}^{\left( N\right) }\left(
\varepsilon _{i}\right) \frac{1+\sum\limits_{j\neq i}\frac{\varepsilon
_{i}-\varepsilon _{n}}{\varepsilon _{j}-\varepsilon _{n}}}{%
1+\sum\limits_{j\neq i}\prod_{m=0}^{N}\frac{\varepsilon _{i}-\varepsilon _{m}%
}{\varepsilon _{j}-\varepsilon _{m}}},
\end{equation*}
where $l_{n}^{\left( N\right) }\left( \varepsilon \right) $ is the Lagrange
polynomial (\ref{e32}) of degree $N$. We have therefore reached the
conclusion that --in our orthonormal model, and to leading order-- the wave
function is the energy-dependent MTO, $\check{\phi}\left( \varepsilon ,%
\mathbf{r}\right) ,$ Lagrange interpolated over the ($N$+1)-point mesh.

Since the error of an NMTO set is of order $N$+1, use of the \emph{%
variational} principle will reduce the error of the one-electron \emph{%
energies,} $\varepsilon _{i},$ from that of the \emph{highest} transfer
matrix, $H^{\left( N\right) }-\varepsilon _{N},$ to order 2($N$+1). The
variational energies are thus correct to order 2$N$+1. For a condensed mesh,
this also follows trivially from (\ref{d12})-(\ref{d11}), which show that
the variational energy, with respect to $\varepsilon _{\nu },$ is: 
\begin{equation*}
\frac{\left\langle \chi ^{\left( N\right) }\left| \mathcal{H}-\varepsilon
_{\nu }\right| \chi ^{\left( N\right) }\right\rangle }{\left\langle \chi
^{\left( N\right) }\mid \chi ^{\left( N\right) }\right\rangle }=\frac{%
\overset{\left( 2N\right) }{G}}{\left( 2N\right) !}\left/ \frac{\overset{%
\left( 2N+1\right) }{G}}{\left( 2N+1\right) !}\right. =H^{\left( 2N+1\right)
}-\varepsilon _{\nu }.
\end{equation*}
The odd-ordered switching curves $H^{\left( 1\right) }\left( \varepsilon
_{\nu }\right) ,$ $H^{\left( 3\right) }\left( \varepsilon _{\nu }\right) ,$
and $H^{\left( 5\right) }\left( \varepsilon _{\nu }\right) $ shown in the
left-hand panel of Fig. \ref{Fig5} are thus the variational estimates
resulting from the use of respectively the 0th, 1st, and 2nd-order NMTO,
that is, the MTO, the LMTO, and the QMTO. These curves are well behaved.

The expression for the variational energy in the one-band model can be
evaluated exactly, also for a discrete mesh, and yields a transparent
result. We use the double-mesh procedure explained in the Appendix after (%
\ref{h5}), and let the differences $\epsilon _{n}\equiv \varepsilon
_{n+N+1}-\varepsilon _{n}$ shrink to zero. From (\ref{d20}) we then get: 
\begin{eqnarray}
\check{G}\left[ \left[ 0...N\right] \right] &=&-\sum_{j}\frac{1}{%
\prod_{m=0}^{N}\left( \varepsilon _{j}-\varepsilon _{m}\right) ^{2}},
\label{c5} \\
\check{G}\left[ \left[ 0..N-1\right] N\right] &=&-\sum_{j}\frac{1}{\left(
\varepsilon _{j}-\varepsilon _{N}\right) \prod_{m=0}^{N-1}\left( \varepsilon
_{j}-\varepsilon _{m}\right) ^{2}},  \notag
\end{eqnarray}
and for the variational energy (\ref{o29}): 
\begin{equation}
\left\langle \check{\chi}^{\left( N\right) }\left| \mathcal{H}-\varepsilon
_{N}\right| \check{\chi}^{\left( N\right) }\right\rangle =\left( \varepsilon
_{i}-\varepsilon _{N}\right) \frac{1+\sum\limits_{j\neq i}\frac{\varepsilon
_{i}-\varepsilon _{N}}{\varepsilon _{j}-\varepsilon _{N}}\prod_{m=0}^{N-1}%
\left( \frac{\varepsilon _{i}-\varepsilon _{m}}{\varepsilon _{j}-\varepsilon
_{m}}\right) ^{2}}{1+\sum\limits_{j\neq i}\prod_{m=0}^{N}\left( \frac{%
\varepsilon _{i}-\varepsilon _{m}}{\varepsilon _{j}-\varepsilon _{m}}\right)
^{2}},  \notag
\end{equation}
which of course agrees with the variational principle.

\input{fig6.tex}

\subsubsection{Treating semi-core and excited states: GaAs.}

An accurate\textit{\ }description of the cohesive properties of GaAs
requires a good band-structure calculation of the five Ga $3d^{10}$
semi-core, the As $4s^{2}$-band, and the three As $4p^{2}$ Ga $4sp^{3}$
valence bands. If also the four lowest conduction bands must be described,
one is faced with the problem of computing a band structure containing
extremely narrow as well as wide bands over a 20 eV-region. To do this 
\textit{ab initio} with a \emph{minimal }Ga $spd$ As $sp$ basis set (13
orbitals per GaAs), has hitherto not been possible.

With 1st and 2nd-generation LMTO-ASA methods one would normally use $Rl$%
-dependent $\varepsilon _{\nu }$'s and employ a 36-orbital-per-GaAs basis,
consisting of the $spd$ LMTOs centered on the Ga, the As, and the
interstitial sites in the zincblende structure. The conduction-band errors
arising from the choice $\kappa ^{2}$=0 are so large that the combined
correction is needed. Downfolding works for the $p$ and $d$ orbitals on the
two interstitial spheres, but not for the interstitial $s$ and the As $d$
orbitals. With the 3rd-generation LMTO method, downfolding works much
better, but the energy window is now screening dependent, and the use of $Rl$%
-dependent $\varepsilon _{\nu }$'s is avoided because it messes up the
formalism.

In Fig. \ref{Fig6} we show --in full lines-- the exact (up to 7eV) LDA band
structure calculated by the screened KKR method, i.e.\textit{\ }by the
3rd-generation LMTO method using \emph{many} energy panels and the Ga $spd$
As $sp$ basis. The five Ga $3d^{10}$ semi-core bands\ are at --15 eV, the As 
$4s^{2}$-band is around --12 eV, and the three As $4p^{2}$ Ga $4sp^{3}$
valence bands extend from --7 to 0 eV. Above the gap, there are the four As $%
4p^{4}$ Ga $4sp^{3}$ conduction bands. The dotted lines give results of
3rd-generation LMTO variational calculations with a condensed mesh and an $%
\varepsilon _{\nu }$ in the middle of the three valence bands. In the
left-hand
\input{fig7.tex}

\noindent
figure, the screening-sphere radii for the active Ga $spd$ and As $%
sp$ channels were chosen at the Ga and As default values, respectively $0.82t
$ and $0.78t,$ where $t$ is half the nearest-neighbor distance. We see that
the entire valence-band structure is distorted by hybridization with Ga $d$
ghost bands. The dotted bands in the right-hand figure result after changing
the Ga $d$ screening-sphere radius to $0.35t,$ which is close to the actual
radius of the Ga $3d$ core. Now, the band structure looks reasonable: The
valence bands near $\varepsilon _{\nu }$ are perfect, but the Ga $3d$ bands
are nearly 0.5 eV to high \cite{PolyT}.

That the variational LMTO method with a minimal basis and a single $%
\varepsilon _{\nu }$ cannot describe all occupied states of GaAs with
sufficient accuracy, becomes even more obvious from the left-hand side of
Fig. \ref{Fig7}, where we show --as functions of $\varepsilon _{\nu }$-- the
average errors of the five Ga $3d$ bands, those of the As $4s$ band, and
those of the three valence bands$.$ The error $\propto \left( \varepsilon
_{i}\left( \mathbf{k}\right) -\varepsilon _{\nu }\right) ^{4}$ of the
variational energy is clearly visible for the narrow Ga $3d$ and As $4s$
bands. With $\varepsilon _{\nu }$'s in a narrow range around --11 eV, the
variational error in the sum of the one-electron energies gets down to about
250 meV per GaAs. On the right-hand side, we show the same quantities, but
obtained with the QMTO method. Now the errors $\propto \left( \varepsilon
_{i}\left( \mathbf{k}\right) -\varepsilon _{\nu }\right) ^{6}$ are
acceptable, and there is a comfortable range of $\varepsilon _{\nu }$'s
around --10 eV where the error in the sum of the one-electron energies does
not exceed 25 meV per GaAs. The screening-sphere radii chosen in these
calculations \cite{PolyT} were: $0.93t,\,1.05t,$ and $0.35t$ for
respectively Ga $s,\,p,$ and $d,$ and $0.89t$ and $1.00t$ for respectively
As $s$ and $p.$

\input{fig8.tex}

In Fig. \ref{Fig8} we show the same kind of results, but this time obtained
with the discrete (Lagrange) LMTO and QMTO methods. The size of the basis
set, the screening-sphere radii, etc., were as in Fig. \ref{Fig7}. For the
LMTO method, $\varepsilon _{0}$ was fixed at the position of the Ga $3d$
bands and the figure shows the result of varying the position $\varepsilon
_{1}$ of the other mesh point. The quadratic dependence on $\varepsilon _{1}$
of the variational energy-error $\propto \left( \varepsilon _{i}\left( 
\mathbf{k}\right) -\varepsilon _{0}\right) ^{2}\left( \varepsilon _{i}\left( 
\mathbf{k}\right) -\varepsilon _{1}\right) ^{2}$ is clearly recognized.
Compared with the results of the tangent LMTO method shown in the previous
figure, those of the chord-LMTO are far superior: With $\varepsilon _{1}$'s
around --5 eV, the variational error in the sum of the one-electron energies
gets down to about 30 meV per GaAs, and yet, for $N$ given, the method
employing a discrete mesh is computationally simpler than the one employing
a condensed mesh. On the right-hand side of the figure, we show the QMTO
results as functions of $\varepsilon _{2},$ with $\varepsilon _{0}$ fixed at
the Ga $3d$ position, and $\varepsilon _{1}$ at the As $4s$ position$.$ Here
again, the quadratic dependence on $\varepsilon _{2}$ of the variational
energy-error $\propto \left( \varepsilon _{i}\left( \mathbf{k}\right)
-\varepsilon _{0}\right) ^{2}\left( \varepsilon _{i}\left( \mathbf{k}\right)
-\varepsilon _{1}\right) ^{2}\left( \varepsilon _{i}\left( \mathbf{k}\right)
-\varepsilon _{2}\right) ^{2}$ may be seen. We realize, that with this
discrete QMTO method, meV-accuracy for the sum of the one-electron energies
can be reached.

Finally, in Fig \ref{Fig9} we show the GaAs band structure in a wide ($40$
eV) range around the gap. Further conduction bands now appear\ above 7 eV
and we needed to employ a basis consisting of the Ga $spd$ As $spdf$ 2E $s$
QMTOs. $\varepsilon _{0}$ was chosen at the Ga $3d$ position, $\varepsilon
_{1}$ near the gap, and $\varepsilon _{2}$ 10 eV above the gap. The results
of this discrete QMTO calculation shown by the dotted curves agree superbly
with those of a multi-panel LMTO (=KKR) calculation shown in full line \cite
{PolyT}. This proves the power of the 3rd-generation NMTO method.

\input{fig9.tex}

\subsubsection{Massive downfolding: CaCuO$_{2}.$}

An increasingly important field of research is the electronic structure of
real materials with strongly correlated conduction electrons. Within a given
class of materials, fine-tuning of the interesting properties will require
detailed knowledge of the single-electron part --the orbitals, hopping
integrals and basic on-site terms-- of the correlated Hamiltonian. In the
previous review \cite{MRS} of the 3rd-generation 0th- and 1st-order
differential MTO method, we demonstrated for the idealized high-temperature
superconductor, CaCuO$_{2}$ with dimpled CuO$_{2}$ planes, how one could
extract low-energy, few-band Hamiltonians by massive downfolding; in the
extreme limit: Downfolding to \emph{one} Cu $d_{x^{2}-y^{2}}$ orbital per Cu
site \cite{Tan,Ladder}. Let us now reconsider this example in the light of
the new NMTO methods.

In Fig. \ref{Fig10} the full lines in all four parts show the (same) full
LDA band structure in a $\pm 3$ eV region around the Fermi level, which for
the doping levels of interests would be near the energy $-0.8$ eV of the
so-called extended saddle-point at \textbf{X}. The conduction band has
mostly O-Cu anti-bonding $pd\sigma $-character (O $p_{x}$ -- Cu $%
d_{x^{2}-y^{2}}$) with the bonding partner lying 10 eV lower in energy. The
bottom of the conduction band is seen to cross and hybridize with a
multitude of O-Cu $pd\pi $-bands lying below --1.2 eV. The top of the
conduction band hybridizes strongly with a broad O-Ca bonding $pd\pi $ (O $%
p_{x}$ --Ca $d_{xy})$ band near \textbf{A}. In this situation, one clearly
does not want to use the rather ill-defined and very long-ranged Wannier
orbital for describing the low-energy electronic structure. Rather, one
wants an orbital which describes the band (including its dependence on other
relevant low-energy excitations such as spin-fluctuations and phonons) in
the $\pm $200meV range around $\varepsilon _{F},$ that is an NMTO with \emph{%
all} channels, except Cu $d_{x^{2}-y^{2}},$ downfolded and with as short a
range as possible. The four dotted bands shown in each of the sub-figures
result from such calculations \cite{PolyT}. In all cases, the
screening-sphere radius of Cu $d_{x^{2}-y^{2}}$ was taken to be $0.62t.$ The
upper figures illustrate a problem with the 3rd-generation tangent LMTO
method: If $\varepsilon _{\nu }$ is taken where we want it to be, at the $%
-0.8$ eV saddle-point deep down in the anti-bonding $pd\sigma $-band, then
the method develops a \emph{schizophrenia} near the top of the band, above 1
eV and near \textbf{M}, which is apparently sufficiently far away from $%
\varepsilon _{\nu }$ that the LMTO 'might consider' describing the bonding
rather than the anti-bonding state. 
\input{fig10.tex}

\noindent
The resulting orbital has very long
range due to the high Fourier components caused by the schizophrenia and, as
a result, we are forced to take $\varepsilon _{\nu }$ at a higher energy
than we actually want. With $\varepsilon _{\nu }$=--0.3 eV, we still get
long range as seen in the upper left-hand figure, and in order to cure that
problem we need to go to $\varepsilon _{\nu }$=+0.3 eV, but then the
description of the bottom of the anti-bonding band, the extended
saddle-point in particular, has substantially deteriorated. In the lower
left-hand figure we have now switched from the tangent to the chord LMTO,
and that is seen to help considerably. Finally, the lower right-hand figure
presents what might be called an 'overkill': We have used the discrete CMTO (%
$N$=3) method, and the agreement with the exact result is superb.

\subsubsection{Using integrals of divided differences of MTOs.}

In all previous derivations of the variational LMTO method, the LMTO was
expressed as a matrix Taylor series (\ref{a1}) and the Hamiltonian and
overlap matrices (\ref{a4}) were worked out using expressions (\ref{a10})
for $\left\langle \phi \mid \dot{\phi}\right\rangle $ and $\left\langle \dot{%
\phi}\mid \dot{\phi}\right\rangle .$

The same may be done for the general, discrete NMTO method, although the
number of terms in the resulting series increases quadratically with $N$.
For this, we first use a divided-difference form --such as (\ref{d25})-- for
the NMTO and then need expressions for the overlap integrals, $\left\langle
\phi \left[ 0..N\right] \mid \phi \left[ 0..M\right] \right\rangle ,$ and
Hamiltonians, $\left\langle \phi \left[ 0..N\right] \left| \mathcal{H}%
\right| \phi \left[ 0..M\right] \right\rangle ,$ between divided differences
of kinked partial waves. Since expressions (\ref{b54}) and (\ref{b55}) are
formally equivalent, we find that, analogous to (\ref{d14}), 
\begin{eqnarray}
&&\left\langle \phi \left[ 0..M\right] \mid \phi \left[ 0...N\right]
\right\rangle \;=\;\left\langle \phi \left[ 0...N\right] \mid \phi \left[
0..M\right] \right\rangle \;=\;K\left[ \left[ 0..M\right] .N\right]
\label{e70} \\
&\rightarrow &\;\left\langle \frac{\overset{\left( M\right) }{\phi }}{M!}%
\mid \frac{\overset{\left( N\right) }{\phi }}{N!}\right\rangle
\;=\;\left\langle \frac{\overset{\left( N\right) }{\phi }}{N!}\mid \frac{%
\overset{\left( M\right) }{\phi }}{M!}\right\rangle \;=\;\frac{\overset{%
\left( M+N+1\right) }{K}}{\left( M+N+1\right) !},  \notag
\end{eqnarray}
where we have assumed $M\leq N.$ From this result for $M=N,$ it follows that
the odd-ordered Hermite divided differences of the kink matrix are positive
definite. For a contracted mesh, this overlap matrix is seen to depend only
on $M+N.$

For the matrix elements of the Hamiltonian we must use: 
\begin{eqnarray}
&&\left\langle \phi \left[ 0..M\right] \left| \mathcal{H}-\varepsilon
_{n}\right| \phi \left[ 0...N\right] \right\rangle =\left\langle \phi \left[
0..M\right] \mid \phi \left[ 0..n-1,n+1..N\right] \right\rangle  \notag \\%
[0.15cm]
&=&\left\{ 
\begin{array}{l}
K\left[ \left[ 0..n-1,n+1..\min \left( M,N\right) \right] n..\max \left(
M,N\right) \right] \\ 
K\left[ \left[ 0..\min \left( M,N\right) \right] ..n-1,n+1..\max \left(
M,N\right) \right]
\end{array}
\right.  \label{e71} \\
&\rightarrow &\;\left\langle \frac{\overset{\left( M\right) }{\phi }}{M!}%
\left| \mathcal{H}-\varepsilon _{\nu }\right| \frac{\overset{\left( N\right) 
}{\phi }}{N!}\right\rangle \;=\;\left\langle \frac{\overset{\left( M\right) 
}{\phi }}{M!}\mid \frac{\overset{\left( N-1\right) }{\phi }}{\left(
N-1\right) !}\right\rangle \;=\;\frac{\overset{\left( M+N\right) }{K}}{%
\left( M+N\right) !}\,.  \notag
\end{eqnarray}
where the upper and lower results on the second line correspond to $%
n\lesseqgtr \min \left( M,N\right) $. Here again, for a condensed mesh the
Hamiltonian matrix depends only on $M+N.$

The resulting expressions for $\left\langle \chi ^{\left( N\right) }\mid
\chi ^{\left( N\right) }\right\rangle $ and $\left\langle \chi ^{\left(
N\right) }\left| \mathcal{H}-\varepsilon _{n}\right| \chi ^{\left( N\right)
}\right\rangle $ contain the above-mentioned integrals \emph{times }products
of $\left( E^{\left( N-M+1\right) }-\varepsilon _{M-1}\right) $-matrices.
These expressions are by far not as explicit as equations (\ref{d12}) and (%
\ref{d11}), and they are more complicated for a discrete than for a
condensed mesh. We shall now consider a more useful application of (\ref{e70}%
)-(\ref{e71}).

\subsubsection{Charge density and total energy: Si phase diagram.}

The wave function obtained from a variational calculation is: $\Psi
_{i}\left( \mathbf{r}\right) =\chi \left( \mathbf{r}\right) c_{i}\,,$ where
we have dropped the superscript $\left( N\right) $ on the NMTO. The
eigen(column)vector, $c_{i},$ of the generalized eigenvalue equation (\ref
{a3}) should be normalized according to: $c_{i}^{\dagger }\left\langle \chi
\mid \chi \right\rangle c_{i^{\prime }}=\delta _{ii^{\prime }},$ or
--regarding $c_{RL,i}$ as a matrix-- according to: $c^{\dagger }\left\langle
\chi \mid \chi \right\rangle c=1.$ The charge density is now given by (\ref
{a15}), which to a very good approximation is (\ref{b56}) with the
energy-dependent wave functions in expressions (\ref{b57})-(\ref{b58})
substituted by their matrix Lagrange or Newton series. The computer code
would use the Lagrange form: 
\begin{equation*}
\rho \left( \mathbf{r}\right) =\chi \left( \mathbf{r}\right) cc^{\dagger
}\chi \left( \mathbf{r}\right) ^{\dagger }=\sum_{nn^{\prime }}\phi
_{n}\left( \mathbf{r}\right) \,L_{n}cc^{\dagger }L_{n^{\prime }}^{\dagger
}\,\phi _{n^{\prime }}\left( \mathbf{r}\right) ^{\dagger },
\end{equation*}
so that in this case, the density-of-states matrix $\Gamma \left(
\varepsilon \right) $ in (\ref{b59}) should be substituted by: 
\begin{equation}
\Gamma _{nn^{\prime }}\equiv L_{n}\left( \sum_{i}^{occ}c_{i}c_{i}^{\dagger
}\right) L_{n^{\prime }}^{\dagger }.  \label{d31}
\end{equation}
Equations (\ref{b57})-(\ref{b58}) then become: 
\begin{eqnarray}
\rho ^{\psi }\left( \mathbf{r}\right) &\equiv &\sum_{RR^{\prime
}}\sum_{LL^{\prime }}\sum_{nn^{\prime }}\psi _{RL,n}\left( \mathbf{r}%
_{R}\right) \,\Gamma _{RL,n;R^{\prime }L^{\prime },n^{\prime }}\,\psi
_{R^{\prime }L^{\prime },n^{\prime }}\left( \mathbf{r}_{R^{\prime }}\right)
^{\ast },  \label{d30} \\
\rho _{R}^{\varphi }\left( \mathbf{r}\right) &=&\sum_{LL^{\prime
}}Y_{L}\left( \mathbf{\hat{r}}\right) Y_{L^{\prime }}^{\ast }\left( \mathbf{%
\hat{r}}\right)\sum_{nn^{\prime}}
\varphi _{Rl,n}\left( r\right) \,\Gamma _{RL,n;RL^{\prime
},n^{\prime }}\,\varphi _{Rl^{\prime },n^{\prime }}\left( r\right) ,  \notag
\\
\rho _{R}^{\varphi ^{\circ }}\left( \mathbf{r}\right) &=&\sum_{LL^{\prime
}}Y_{L}\left( \mathbf{\hat{r}}\right) Y_{L^{\prime }}^{\ast }\left( \mathbf{%
\hat{r}}\right)\sum_{nn^{\prime}}
\varphi _{Rl,n}^{\circ }\left( r\right) \,\Gamma
_{RL,n;RL^{\prime },n^{\prime }}\,\varphi _{Rl^{\prime },n^{\prime }}^{\circ
}\left( r\right) .  \notag
\end{eqnarray}
If one feels that, with the variational NMTO method, the KKR equations have
been solved with sufficient accuracy, then one may even use (\ref{b57})-(\ref
{b59}) as they stand, and interpolate the energy dependences of the wave
functions using the \emph{classical} Lagrange or Newton methods (\ref{e32})
and (\ref{e33}).

In order to solve Poisson's equation and to compute the Coulomb- and
exchange-correlation integrals for the total energy and forces, we need to
fit the charge density by suitable functions. The properties of $\rho (%
\mathbf{r})$ to which we have most easy access are its spherical-harmonics
expansions around the various sites. For the fitting we therefore choose
atom-centered NMTO-like functions which have the following advantages: (1)
they are the unitary functions for \emph{continuous fitting} at non-touching 
$a$-spheres, (2) they are localized, (3) we know the result of operating on
them with $\nabla ^{2},$ and (4) the integral of any product of two such
functions is the energy derivative of a kink matrix (\ref{e70})-(\ref{e71}).

\input{fig11.tex}

Our fitting procedure \cite{Ras} can be outlined as follows: We first place
a set of screening spheres around each atomic site. This defines our
screened Hankel functions (\ref{b8a}) and divides space into non overlapping
intra-sphere parts and an interstitial part. It is not necessary to place
screening spheres at interstitial sites, even though the resulting
interstitial can be very large. In the intra-sphere region we use a
spherical-harmonics expansion of the charge density, with the components $%
\rho _{RL}(r)$ known on a radial mesh. As the screening spheres are
relatively small this summation can be truncated at $l$=3 or 4. In the
interstitial we expand in the screened Hankel functions, $n_{RL}^{a}\left(
\varepsilon ,\mathbf{r}_{R}\right) ,$ normalized as in (\ref{b24}) and with
3 different, negative energies, of which the lowest is about 4 times the
work function, that is: 
\begin{eqnarray}
\rho (\mathbf{r}) &\approx &\sum_{n=0}^{2}\sum_{RL}n_{RL}^{a}\left(
\varepsilon _{n},\mathbf{r}_{R}\right) \,\lambda _{RL;n}=\sum_{RL}\breve{n}%
_{RL}^{a}\left( \mathbf{r}_{R}\right) \mu _{RL}+  \label{d32} \\
&&\quad \quad \sum_{RL}\left( n_{RL}^{a}\left( \mathbf{r}_{R}\right) \rho
_{RL}\left( a\right) +n_{RL}^{a}\left( \left[ 01\right] ,\mathbf{r}%
_{R}\right) \sum_{R^{\prime }L^{\prime }}X_{RL,R^{\prime }L^{\prime }}\,\rho
_{R^{\prime }L^{\prime }}\right)   \notag
\end{eqnarray}
for all $r_{R}\geq a_{RL}.$ With three energies, we can in principle fit
continuously with continuous 1st and 2nd derivatives. However, in practice
it is difficult to compute the 2nd radial derivatives of the high-$l$
components of the charge density. We therefore determine the matrix $X$ in
such a way that the fitting is continuous and \emph{once} differentiable,
that is: $X=B^{a}\left[ 01\right] ^{-1}\left( \partial \left\{ \rho \left(
a\right) \right\} -B_{0}^{a}\right) .$ The functions $\breve{n}%
_{RL}^{a}\left( \mathbf{r}_{R}\right) $ in (\ref{d32}) are those linear
combinations of the three $n^{a}\left( \varepsilon _{n},\mathbf{r}\right) $%
's whose value and radial slope vanish in \emph{all} channels at the
screening spheres. These functions therefore peak in the interstitial region
and their coefficients $\mu _{RL}$ are determined by a least squares fit in
the region interstitial to the MT-spheres, by sampling the full charge (\ref
{d30}), as well as the expansion (\ref{d32}) at a set of pseudo-random
points. Once the expansion is obtained, it is very easy to solve Poisson's
equation. In the intra-sphere part this is done numerically and in the
interstitial analytically by virtue of the screened Hankel functions solving
the wave equation. The same expansion procedure can be applied to the
exchange-correlation energy density $\epsilon (\mathbf{r})$ and potential $%
\mu (\mathbf{r})$. This gives a full potential. The total energy $E_{tot}$
is also easy to evaluate. The interstitial part of the integrals reduces
simply to a summation over Hermite divided differences of the slope matrix.

We have applied this procedure to look at the total energy of various
possible structures for silicon \cite{SiPhase}. For each structure we
perform a standard self consistent LMTO-ASA calculation. In the last
iteration an expansion of the full charge density is made and $E_{tot}$
evaluated correctly. The result is shown in Fig. \ref{Fig11} where, for
comparison, we show the full-potential LMTO result from Ref. \cite{MethFP}.

\input{fig13.tex}

\subsubsection{Overlapping MT-potential: Si without empty spheres.}

The phase diagram of Si just shown was calculated using LMTOs defined for
MT-potentials with empty spheres. We now consider the possibility offered by
Eq. (\ref{ovl}) of allowing the atom-centered sphere a substantial overlap
--like the 50\% radial overlap shown in Figs. \ref{Fig2}-\ref{Fig4a}-- and,
hence, of getting rid of the empty spheres.

The first question is: How to construct such a potential? Our answer is \cite
{Catia} that the potential should be constructed such as to minimize the
mean squared deviation of the valence-band energies from the ones for the
full potential. From this condition, it then follows that the overlapping
MT-potential, $\sum_{R}v\left( r_{R}\right) ,$ should be the least-squares
approximation to the full-potential, $V\left( \mathbf{r}\right) $, weighted
with the valence charge density. This yields a set of coupled equations for
the \emph{shape,} $f\left( r\right) \equiv v\left( r\right) -g,$ and the 
\emph{zero,} $g,$ of the MT-potential. The equation which arises from
requiring stationarity with respect to $\delta g$ is of course: $\int \left(
V-\sum v\right) \rho d^{3}r=0,$ and it means that the error in the sum of
the valence-band energies should vanish to leading order. The other
equations, which arise by requiring stationarity with respect to $\delta
f\left( r\right) ,$ are coupled integral equations, which are complicated
due to the presence of the charge-density weighting. Taking the charge
density to be constant in space, corresponds to minimizing the mean squared
energy-deviation for the \emph{entire} spectrum, rather than merely for the
valence band. Now, in our present implementation, we only took the spatial
behavior of the charge density into account in the $\delta g$-equation. The
resulting potentials for diamond-structured Si were shown in Figs. 10 and 11
of Ref. \cite{MRS}. We have recently succeeded in obtaining the overlapping
MT potential from the full potential obtained from the charge density (\ref
{d32}) \cite{Catia2}, but in the present paper we shall only show results
obtained by taking the full potential to be the Si+E ASA potential --like in
Ref. \cite{MRS}.

\input{fig12.tex}

Fig. \ref{Fig13} shows three different results for the rms error of the
valence-band energy as a function of the linear overlap, $\omega \equiv
\left( s/t\right) -1.$ For the overlap increasing up to about 30\%, the rms
error falls in all cases, simply because the overlapping MT-potential
becomes an increasingly better approximation to the full potential. Without
any overlap correction, the kinetic-energy error (\ref{ovl}), which is of
second order in the potential overlap, initially rises proportional to $%
v\left( s\right) ^{2}\omega ^{4}$ \cite{MRS}, and this is seen to limit the
maximum overlap to about 30\%. We may, however, use the LMTO equivalent \cite
{Catia2} of Eq. (\ref{ovl}) to correct each band energy, $\varepsilon
_{i}\left( \mathbf{k}\right) ,$ and the results are shown by the two other
curves. The dashed curve --marked 'present technique'-- uses the $\delta g$%
-equation as given above, whereas the 'ideal' curve was obtained by
adjusting $g$ --iteratively, because $g$ enters the $\delta f\left( r\right) 
$ equations-- to have the mean error of the valence-band energy vanish
exactly. It \emph{is} possible to improve upon the 'present technique'
without knowing the valence-band energy \textit{a priori,} and we are
currently including charge-density weighting in the $\delta f\left( r\right) 
$-equations. This makes the curve flatten out --like the one marked 'ideal' 
\cite{Catia2}.

The solid curves in Fig. \ref{Fig12} show the Si band structure obtained
with the 60\% overlapping MT-potential, including the LMTO overlap
correction, and determining $g$ to yield vanishing mean error of the valence
band. The dotted curve is the 'exact' result as obtained with a
(3rd-generation) LMTO calculation for the Si+E potential. The errors seen in
the valence band are certainly no larger than 30 meV, but those in the
conduction band are larger.

\section{Energy-dependent linear transformations}

\label{linear}If one considers Fig. \ref{Fig1}, it might seem as if the
energy-window over which an NMTO set yields good approximations to the wave
functions will be wider if one starts out from energy-dependent linear
combinations of kinked partial waves: 
\begin{equation}
\hat{\phi}\left( \varepsilon ,\mathbf{r}\right) \;\equiv \;\phi \left(
\varepsilon ,\mathbf{r}\right) \hat{T}\left( \varepsilon \right) ,
\label{e34}
\end{equation}
which have smoother energy dependencies. Normalized kinked partial waves and
L\"{o}wdin orthonormalized kinked partial waves are examples of cases where
the divergences of the kinked partial waves at the energies, $\varepsilon
_{RL}^{a},$ where a node passes through the screening radius, are avoided.
The transformation given by the -- in general non-Hermitian -- matrix $\hat{T%
}\left( \varepsilon \right) $ mixes kinked partial waves with the \emph{same
energy} and \emph{different} $RL$'s \emph{linearly.} Although the Hilbert
spaces spanned by the energy-dependent sets, $\phi \left( \varepsilon ,%
\mathbf{r}\right) $ and $\hat{\phi}\left( \varepsilon ,\mathbf{r}\right) ,$
are identical, it is not obvious that those spanned by the respective
polynomial approximations, $\chi ^{\left( N\right) }\left( \mathbf{r}\right) 
$ and $\hat{\chi}^{\left( N\right) }\left( \mathbf{r}\right) ,$ are also
identical, particularly not if one bears only Fig. \ref{Fig1} in mind.

Depending on the transformation, the resulting $\hat{\phi}\left( \varepsilon
,\mathbf{r}\right) $ may completely have lost its original $RL$-character.
Since the linear combination, $\hat{\phi}\left( \varepsilon ,\mathbf{r}%
\right) ,$ of kinked partial waves has \emph{active} radial functions on 
\emph{other} sites, as well as at its own site for other $L$'s, it is \emph{%
not} a kinked partial wave in the usual sense, that is, one which could have
been obtained by a screening transformation. Remember, that for
3rd-generation kinked partial waves, a screening transformation is not
linear. In the following, we shall assume that the screening radii have been
chosen at the previous step, in the screening calculation for the structure
matrix, and perhaps by subsequent re-screening of the $G_{n}$'s using (\ref
{b44}).

A further motivation for considering transformed kinked partial waves is
that they might provide the freedom to obtain energy matrices (\ref{d17})
which are \emph{Hermitian. }This would simplify the finite-difference
expressions (\ref{d24}) and (\ref{d25}) for the NMTO so that they take the
simpler form (\ref{a1}) which then --like in (\ref{a2})-- could be
diagonalized to leading order by the eigenvectors of $\hat{E}^{\left(
N\right) }.$ From expression (\ref{d23}) for the transfer matrix, we realize
that the condition that a transformed $\hat{E}^{\left( M\right) }$ be a
Hamiltonian matrix, is that we can find a transformation with the property
that 
\begin{equation}
\left\langle \hat{\chi}^{\left( M\right) }\mid \hat{\chi}^{\left( M-1\right)
}\right\rangle \;=\;1.  \label{e3}
\end{equation}
This formalism could therefore also be the basis for obtaining an \emph{%
orthonormal} NMTO set.

Let us finally express the important equations (\ref{e0})-(\ref{b55}) in
terms of the transformed kinked partial waves: 
\begin{equation}
\left( \mathcal{H}-\varepsilon \right) \hat{\phi}\left( \varepsilon ,\mathbf{%
r}\right) =-\delta \left( \mathbf{r}\right) \hat{K}\left( \varepsilon
\right) ,\mathrm{\;}\left( \mathcal{H}-\varepsilon \right) \hat{\phi}\left(
\varepsilon ,\mathbf{r}\right) \hat{G}\left( \varepsilon \right) =-\delta
\left( \mathbf{r}\right) ,  \label{e4}
\end{equation}
where we have defined the \emph{non-Hermitian} matrices 
\begin{equation}
\hat{K}\left( \varepsilon \right) \equiv K\left( \varepsilon \right) \hat{T}%
\left( \varepsilon \right) ,\mathrm{\quad \quad }\hat{G}\left( \varepsilon
\right) \equiv \hat{K}\left( \varepsilon \right) ^{-1}=\hat{T}\left(
\varepsilon \right) ^{-1}G\left( \varepsilon \right) .  \label{e5}
\end{equation}
Note that these definitions do \emph{not} correspond to similarity
transformations. The kink matrix, $K\left( \varepsilon \right) ,$ and
thereby its inverse, $G\left( \varepsilon \right) ,$ were originally defined
in such a way that they are Hermitian, but they are \emph{inherently 'skew',}
because (\ref{e4}) tells us that it is the 'one-sided' contraction of the
Green function, 
\begin{equation}
\gamma \left( \varepsilon ,\mathbf{r}\right) =\phi \left( \varepsilon ,%
\mathbf{r}\right) G\left( \varepsilon \right) =\hat{\phi}\left( \varepsilon ,%
\mathbf{r}\right) \hat{G}\left( \varepsilon \right) ,  \label{e7}
\end{equation}
which is \emph{invariant.} For the same reason, the integrals of the
products of two contracted Green functions, with possibly different
energies, form an overlap matrix, 
\begin{equation}
\hat{G}\left( \varepsilon \right) ^{\dagger }\left\langle \hat{\phi}\left(
\varepsilon \right) \mid \hat{\phi}\left( \varepsilon ^{\prime }\right)
\right\rangle \hat{G}\left( \varepsilon ^{\prime }\right) =-\frac{G\left(
\varepsilon \right) -G\left( \varepsilon ^{\prime }\right) }{\varepsilon
-\varepsilon ^{\prime }},  \label{e35h}
\end{equation}
which is independent of $\hat{T}\left( \varepsilon \right) .$

Adding to the discussion following (\ref{d14}) about the meaning of the 
\emph{matrix} equation $\left\langle \phi _{n}\mid \phi _{n^{\prime
}}\right\rangle =\left\langle \phi _{n^{\prime }}\mid \phi _{n}\right\rangle
,$ note that this equation does not hold in a general representation: $%
\left\langle \hat{\phi}_{n}\mid \hat{\phi}_{n^{\prime }}\right\rangle =\hat{T%
}_{n}^{\dagger }\hat{T}_{n^{\prime }}^{\dagger -1}\left\langle \hat{\phi}%
_{n^{\prime }}\mid \hat{\phi}_{n}\right\rangle \hat{T}_{n}^{-1}\hat{T}%
_{n^{\prime }}\neq \left\langle \hat{\phi}_{n^{\prime }}\mid \hat{\phi}%
_{n}\right\rangle ,$ unless$\;\hat{T}_{n}=\hat{T}_{n^{\prime }}.$ But it is
of course always true that$\;\left\langle \hat{\phi}_{n}\mid \hat{\phi}%
_{n^{\prime }}\right\rangle =\left\langle \hat{\phi}_{n^{\prime }}\mid \hat{%
\phi}_{n}\right\rangle ^{\dagger }.$

We now come to derive NMTOs from the transformed kinked partial waves (\ref
{e34}). Since the arguments around expression (\ref{c1}) concerned the
contracted Green function, which according to (\ref{e7}) is invariant, (\ref
{c1}) is unchanged but should be rewritten in the form: 
\begin{equation}
\hat{\chi}^{\left( N\right) }\left( \varepsilon ,\mathbf{r}\right) \hat{G}%
\left( \varepsilon \right) =\hat{\phi}\left( \varepsilon ,\mathbf{r}\right) 
\hat{G}\left( \varepsilon \right) -\sum_{n=0}^{N}\hat{\phi}_{n}\left( 
\mathbf{r}\right) \,\hat{G}_{n}\,A_{n}^{\left( N\right) }\left( \varepsilon
\right) .  \label{a38}
\end{equation}
As a consequence, (\ref{c2}) should be substituted by: 
\begin{equation}
\hat{\chi}^{\left( N\right) }\left( \mathbf{r}\right) =\frac{\Delta ^{N}\hat{%
\phi}\left( \mathbf{r}\right) \hat{G}}{\Delta \left[ 0..N\right] }\left( 
\frac{\Delta ^{N}\hat{G}}{\Delta \left[ 0..N\right] }\right) ^{-1}=\frac{%
\Delta ^{N}\phi \left( \mathbf{r}\right) G}{\Delta \left[ 0..N\right] }%
\left( \frac{\Delta ^{N}\hat{G}}{\Delta \left[ 0..N\right] }\right) ^{-1}.
\label{a39}
\end{equation}

The last equation (\ref{a39}) shows that the polynomial approximation to the
transformed energy-dependent NMTO, $\hat{\chi}^{\left( N\right) }\left(
\varepsilon ,\mathbf{r}\right) =\chi ^{\left( N\right) }\left( \varepsilon ,%
\mathbf{r}\right) \hat{T}\left( \varepsilon \right) ,$ is 
\begin{equation}
\hat{\chi}^{\left( N\right) }\left( \mathbf{r}\right) =\chi ^{\left(
N\right) }\left( \mathbf{r}\right) \,G\left[ 0...N\right] \,\hat{G}\left[
0...N\right] ^{-1},  \label{a44}
\end{equation}
which is a \emph{linear} transformation. Hence, regardless of the
energy-dependent transformation $\hat{T}\left( \varepsilon \right) $ of the
kinked partial waves, \emph{all} NMTO sets span the \emph{same} Hilbert
space and all energy-windows are therefore identical. This disproves the
above-mentioned naive conclusion drawn from Fig. \ref{Fig1}. Since $G\left(
\varepsilon \right) =\hat{T}\left( \varepsilon \right) \hat{G}\left(
\varepsilon \right) ,$ we may express the NMTO transformation (\ref{a44}) as
a Newton series (\ref{d25}) for $\hat{T}\left( \varepsilon \right) :$%
\begin{eqnarray}
G\left[ 0...N\right] \,\hat{G}\left[ 0...N\right] ^{-1} &=&\left( \hat{T}%
\hat{G}\right) \left[ 0...N\right] \,\hat{G}\left[ 0...N\right] ^{-1}
\label{a45} \\
&=&\sum\nolimits_{M=0}^{N}\hat{T}\left[ 0..M\right] \,\hat{G}\left[ M..N%
\right] \,\hat{G}\left[ 0...N\right] ^{-1}  \notag \\
&=&\hat{T}_{0}+..+\hat{T}\left[ 0...N\right] \left( \hat{E}^{\left( 1\right)
}-\varepsilon _{N-1}\right) ..\left( \hat{E}^{\left( N\right) }-\varepsilon
_{0}\right) .  \notag
\end{eqnarray}
Since the contracted Green function is invariant, so are equations (\ref{d14}%
) and (\ref{d13}) which relate the overlap and Hamiltonian integrals of such
functions to Hermite divided differences of $G\left( \varepsilon \right) .$
For the NMTO overlap and Hamiltonian matrices, we therefore obtain (\ref{d12}%
) and (\ref{d11}), with the prefactor substituted by $\hat{G}\left[ 0..N%
\right] ^{-1\dagger },$ the postfactor substituted by $\hat{G}\left[ 0..N%
\right] ^{-1},$ and the Hermite divided differences of $G\left( \varepsilon
\right) $ unaltered.

The first equation (\ref{a39}) shows that the expressions derived previously
for the NMTOs,\emph{\ excluding }those for \emph{integrals} over NMTOs, may
be taken over, after these expressions have been subject to the following
substitutions: 
\begin{equation}
\begin{array}{ccccc}
\phi \left( \varepsilon ,\mathbf{r}\right) \rightarrow \hat{\phi}\left(
\varepsilon ,\mathbf{r}\right) , & \quad  & K\left( \varepsilon \right)
\rightarrow \hat{K}\left( \varepsilon \right) , & \quad  & L_{n}^{\left(
N\right) }\rightarrow \hat{L}_{n}^{\left( N\right) }, \\ 
\chi \left( \varepsilon ,\mathbf{r}\right) \rightarrow \hat{\chi}\left(
\varepsilon ,\mathbf{r}\right) , & \quad  & G\left( \varepsilon \right)
\rightarrow \hat{G}\left( \varepsilon \right) , & \quad  & E^{\left(
M\right) }\rightarrow \hat{E}^{\left( M\right) }.
\end{array}
\label{a40}
\end{equation}
Remember, that the substitutions for $K\left( \varepsilon \right) $ and $%
G\left( \varepsilon \right) $ do not correspond to a similarity
transformation.

As long as we only consider $\hat{T}\left( \varepsilon \right) $%
-transformations which are independent of $N$, the step-down relation (\ref
{d18}) holds for the transformed NMTOs and for its transfer matrices,
because the derivation merely made use of (\ref{e0}), which transforms into (%
\ref{e4}). This shows that $\hat{E}^{\left( 0\right) }-\varepsilon _{0}$
equals $-\hat{K}_{0}=-K_{0}\hat{T}_{0},$ as expected, but that: $%
\left\langle \hat{\chi}^{\left( 0\right) }\mid \hat{\chi}^{\left( -1\right)
}\right\rangle =1$ does not hold. The hatted version of (\ref{d23})
therefore only holds for $N\geq 1.$ For $N=0:$%
\begin{equation}
\left\langle \hat{\chi}^{\left( 0\right) }\left| \mathcal{H}-\varepsilon
_{0}\right| \hat{\chi}^{\left( 0\right) }\right\rangle =-\hat{T}%
_{0}^{\dagger }K_{0}\hat{T}_{0}=\hat{T}_{0}^{\dagger }\left( \hat{E}^{\left(
0\right) }-\varepsilon _{0}\right) \equiv \hat{H}^{\left( 0\right)
}-\varepsilon _{0}.  \label{a47}
\end{equation}
The expressions for the transformed NMTO in terms of divided differences of 
\emph{transformed} kinked partial waves are the hatted versions of (\ref{d24}%
) and (\ref{d25}). One should remember that the divided difference, $\hat{%
\phi}\left( \left[ 0..M\right] ,\mathbf{r}\right) ,$ is a linear combination
of the $M+1$ functions $\phi _{0}\left( \mathbf{r}\right) \hat{T}%
_{0},..,\phi _{M}\left( \mathbf{r}\right) \hat{T}_{M},$ and hence, a linear
combination of the $M+1$ divided differences: $\phi _{0}\left( \mathbf{r}%
\right) ,\,..,\phi \left( \left[ 0..M\right] ,\mathbf{r}\right) $. This is
the generalization of the property: $d\phi \left( \varepsilon ,\mathbf{r}%
\right) \hat{T}\left( \varepsilon \right) \left/ d\varepsilon \right|
_{\varepsilon _{\nu }}=\dot{\phi}\left( \mathbf{r}\right) \hat{T}+\phi
\left( \mathbf{r}\right) \overset{.}{\hat{T}},$ used in the 2nd-generation
LMTO formalism. Explicitly: 
\begin{eqnarray}
&&\hat{\phi}\left( \left[ 0...M\right] ,\mathbf{r}\right) =\sum_{n=0}^{M}%
\frac{\phi _{n}\left( \mathbf{r}\right) \hat{T}_{n}}{\prod_{m=0,\neq
n}^{M}\left( \varepsilon _{n}-\varepsilon _{m}\right) }  \label{a41} \\
&=&\sum_{m=0}^{M}\phi \left( \left[ m..M\right] ,\mathbf{r}\right) \hat{T}%
\left[ 0..m\right] =\phi \left( \left[ 0...M\right] ,\mathbf{r}\right) \hat{T%
}_{0}+..+\phi _{M}\left( \mathbf{r}\right) \hat{T}\left[ 0...M\right] . 
\notag
\end{eqnarray}
The transformed versions of the results (\ref{e70}), (\ref{e71}) are
complicated, unless $\hat{T}\left( \varepsilon \right) $ is independent of $%
\varepsilon .$ In that case, the right-hand sides just have $K\left(
\varepsilon \right) $ substituted by $\hat{T}^{\dagger }K\left( \varepsilon
\right) \hat{T}\equiv \bar{K}\left( \varepsilon \right) .$

Usually $\left\langle \hat{\phi}\left[ 0..M\right] \mid \hat{\phi}\left[ 0..N%
\right] \right\rangle \neq \left\langle \hat{\phi}\left[ 0..N\right] \mid 
\hat{\phi}\left[ 0..M\right] \right\rangle ,$ unless $\hat{T}\left(
\varepsilon \right) =\hat{T},$\ or\ the\ matrix\ is\ diagonal; $\left\langle 
\hat{\phi}\left[ 0..M\right] \mid \hat{\phi}\left[ 0..N\right] \right\rangle
=\left\langle \hat{\phi}\left[ 0..N\right] \mid \hat{\phi}\left[ 0..M\right]
\right\rangle ^{\dagger }$ of course always holds.

\section{Hamiltonian energy matrices and orthonormal sets}

\label{ortho}Having seen that an energy-dependent, linear transformation (%
\ref{e34}) of the MTO set does \emph{not} change the Hilbert space spanned
by the set of energy-independent NMTOs, but merely the individual basis
functions, we now turn to the objective of finding a representation in which
the energy matrices $\hat{E}^{\left( M\right) }$ --but not necessarily the
Green matrix $\hat{G}\left( \varepsilon \right) $-- are \emph{Hermitian.}
The energy matrices will then be the two-center Hamiltonians entering
expressions like (\ref{a1}) for the orbitals. From (\ref{d23}), we obviously
want: 
\begin{equation}
\hat{E}^{\left( M\right) }-\varepsilon _{M}=\left\langle \hat{\chi}^{\left(
M\right) }\left| \mathcal{H}-\varepsilon _{M}\right| \hat{\chi}^{\left(
M\right) }\right\rangle \equiv \hat{H}^{\left( M\right) }-\varepsilon _{M}
\label{e1}
\end{equation}
for$\;1\leq M\leq N,$ and since this condition leads to the \emph{%
near-orthonormality condition }(\ref{e3}), it guides the way to make \emph{%
one} of the NMTO sets --let us call it the $L$th--$\,$\emph{orthonormal.}

In order to solve the $N$ \emph{near-orthonormality} conditions for the
Hamiltonian matrices, we first insert the transformed version of expression (%
\ref{d21}) for the inverse of the $M$th divided difference of the Green
matrix in terms of the transfer matrices and $\hat{H}^{\left( 0\right)
}-\varepsilon _{0},$ defined by (\ref{a47}), 
\begin{equation}
-\hat{G}\left[ 0...M\right] ^{-1}=\hat{T}_{0}^{-1\dagger }\left( \hat{H}%
^{\left( 0\right) }-\varepsilon _{0}\right) \left( \hat{H}^{\left( 1\right)
}-\varepsilon _{1}\right) ..\left( \hat{H}^{\left( M\right) }-\varepsilon
_{M}\right) ,  \label{e11}
\end{equation}
into the transformed version of expression (\ref{d11}) for the Hamiltonian
in terms of the $2M$th Hermite divided difference of the original Green
matrix $G\left( \varepsilon \right) .$ We then use (\ref{e1}) and notice
that one factor $\hat{H}^{\left( M\right) }-\varepsilon _{M}$ cancels out so
that the equation may be solved for this highest transfer matrix: 
\begin{equation*}
\hat{H}^{\left( M\right) }-\varepsilon _{M}=\left[ 
\begin{array}{l}
\left( \hat{H}^{\left( M-1\right) }-\varepsilon _{M-1}\right) ..\left( 
\hat{H}^{\left( 1\right) }-\varepsilon _{1}\right) \left( \hat{H}^{\left(
0\right) }-\varepsilon _{0}\right)  \\ 
\quad \times \hat{T}_{0}^{-1}\left( -G\left[ \left[ 0..M-1\right] M\right]
\right) \hat{T}_{0}^{-1\dagger } \\ 
\quad \quad \times \left( \hat{H}^{\left( 0\right) }-\varepsilon _{0}\right)
\left( \hat{H}^{\left( 1\right) }-\varepsilon _{1}\right) ..\left( \hat{H}%
^{\left( M-1\right) }-\varepsilon _{M-1}\right) 
\end{array}
\right] ^{-1}
\end{equation*}
for $M\geq 1$. Solving recursively for the transfer matrices, and including (%
\ref{a47}) at the top, we obtain the following results: 
\begin{eqnarray}
\hat{H}^{\left( 0\right) }-\varepsilon _{0} &=&-\hat{T}_{0}^{\dagger }G\left[
\left[ \;\right] 0\right] ^{-1}\hat{T}_{0}  \notag \\
\hat{H}^{\left( 1\right) }-\varepsilon _{1} &=&-\hat{T}_{0}^{-1}G\left[ 
\left[ \;\right] 0\right] \,G\left[ \left[ 0\right] 1\right] ^{-1}G\left[ 
\left[ \;\right] 0\right] \hat{T}_{0}^{-1\dagger }  \notag \\
\hat{H}^{\left( 2\right) }-\varepsilon _{2} &=&-\hat{T}_{0}^{\dagger }G\left[
\left[ \;\right] 0\right] ^{-1}G\left[ \left[ 0\right] 1\right] \,G\left[ 
\left[ 01\right] 2\right] ^{-1}G\left[ \left[ 0\right] 1\right] \,G\left[ 
\left[ \;\right] 0\right] ^{-1}\hat{T}_{0}  \notag \\[0.15cm]
\hat{H}^{\left( M\right) }-\varepsilon _{M} &=&-\hat{T}_{0}^{\left(
-1\right) ^{M}\left( \dagger \right) ^{M+1}}G\left[ \left[ \;\right] 0\right]
^{\left( -1\right) ^{M+1}}...\,G\left[ \left[ 0..M-1\right] M\right] ^{-1} 
\notag \\
&&\quad \quad \quad \quad \quad \quad \quad ...G\left[ \left[ \;\right] 0%
\right] ^{\left( -1\right) ^{M+1}}\hat{T}_{0}^{\left( -1\right) ^{M}\left(
\dagger \right) ^{M}},  \label{e8}
\end{eqnarray}
where for reasons of systematics we have used the notation (\ref{h2}): 
\begin{equation*}
G\left[ \left[ \;\right] 0\right] =G\left[ 0\right] =G_{0}=K_{0}^{-1},
\end{equation*}
explained in the Appendix.

The divided differences (\ref{e11}) of the transformed Green matrix are
needed for specification of the transformation via (\ref{e5}), the orbitals
via (\ref{a44}), or the transformed kinked partial waves via (\ref{e7}), and
are seen to be given by: 
\begin{eqnarray}
\hat{G}\left[ 0\right] ^{-1} &=&\;G\left[ \left[ \;\right] 0\right] ^{-1}%
\hat{T}_{0}  \notag \\
\hat{G}\left[ 01\right] ^{-1} &=&-G\left[ \left[ 0\right] 1\right] ^{-1}G%
\left[ \left[ \;\right] 0\right] \hat{T}_{0}^{-1\dagger }  \label{e10} \\
\hat{G}\left[ 012\right] ^{-1} &=&\;G\left[ \left[ 01\right] 2\right] ^{-1}G%
\left[ \left[ 0\right] 1\right] \,G\left[ \left[ \;\right] 0\right] ^{-1}%
\hat{T}_{0}  \notag \\
\hat{G}\left[ 0...M\right] ^{-1} &=&\left( -\right) ^{M}G\left[ \left[ 0..M-1%
\right] M\right] ^{-1}...G\left[ \left[ \;\right] 0\right] ^{\left(
-1\right) ^{M+1}}\hat{T}_{0}^{\left( -1\right) ^{M}\left( \dagger \right)
^{M}}.  \notag
\end{eqnarray}

Since we originally had the $N+1$ matrices $\hat{T}_{0}...\hat{T}_{N}$ at
our disposal and have used $N$ to satisfy the near-orthonormality
conditions, we have one, $\hat{T}_{0},$ left. This --and thereby implicitly
also the other $\hat{T}_{n}$'s-- may now be chosen equal to a matrix, $%
\check{T}_{0},$ which makes the $L$th set\emph{\ orthonormal. }Note that
whereas the transformation $\hat{T}\left( \varepsilon \right) $ did not
depend on the order of any basis set, the transformation $\check{T}\left(
\varepsilon \right) $ does; it depends on $L$.

Let us first discuss whether the transformation (\ref{o28}) to an
orthonormalized NMTO set may at all be arrived at by an energy-dependent
linear transformation of the kinked partial waves: According to (\ref{a44}),
othonormality of the $L$th set happens for any transformation $\check{T}%
\left( \varepsilon \right) $ which satisfies: $\left( \check{T}^{-1}G\right) %
\left[ 0...L\right] =\left( -G\left[ \left[ 0...L\right] \right] \right)
^{1/2},$ where $G\left[ \left[ 0..L\right] \right] $ is the $(2L+1)$st
Hermite divided difference (\ref{h5}) of the original Green matrix. Hence,
this is a linear equation between the $L+1$ values of the matrix $\check{T}%
\left( \varepsilon \right) ^{-1}$ at the first $L+1$ mesh points, and it is
therefore plausible that it may be used to fix $\check{T}_{0}.$

The better way of writing this equation is, like for the Hamiltonian matrix,
to insert (\ref{e11}) for $\hat{G}\left[ 0..L\right] ^{-1}$ into the
transformed version of expression (\ref{d12}) for the overlap matrix. As a
result: 
\begin{eqnarray}
&&\left\langle \hat{\chi}^{\left( L\right) }\mid \hat{\chi}^{\left( L\right)
}\right\rangle =\left( \hat{H}^{\left( L\right) }-\varepsilon _{L}\right)
..\left( \hat{H}^{\left( 1\right) }-\varepsilon _{1}\right) \left( \hat{H}%
^{\left( 0\right) }-\varepsilon _{0}\right) \times  \label{e9} \\
&&\hat{T}_{0}^{-1}\left( -G\left[ \left[ 0..L\right] \right] \right) \hat{T}%
_{0}^{-1\dagger }\left( \hat{H}^{\left( 0\right) }-\varepsilon _{0}\right)
\left( \hat{H}^{\left( 1\right) }-\varepsilon _{1}\right) ..\left( \hat{H}%
^{\left( L\right) }-\varepsilon _{L}\right)  \notag \\[0.2cm]
&=&-\hat{T}_{0}^{\left( -1\right) ^{L}\left( \dagger \right) ^{L+1}}G\left[ %
\left[ \;\right] 0\right] ^{\left( -1\right) ^{L+1}}..G\left[ \left[ 0..L%
\right] \right] ..G\left[ \left[ \;\right] 0\right] ^{\left( -1\right)
^{L+1}}\hat{T}_{0}^{\left( -1\right) ^{L}\left( \dagger \right) ^{L}}. 
\notag
\end{eqnarray}
We see that the equation $\left\langle \hat{\chi}^{\left( L\right) }\mid 
\hat{\chi}^{\left( L\right) }\right\rangle =1$, in contrast to the equation: 
$\left\langle \hat{\chi}^{\left( M\right) }\left| \mathcal{H}-\varepsilon
_{M}\right| \hat{\chi}^{\left( M\right) }\right\rangle =\hat{H}^{\left(
M\right) }-\varepsilon _{M}$, is \emph{quadratic }in \emph{all} Hamiltonians$%
,$ and therefore can only be solved by taking the square root of a matrix.

Hence, our strategy is to choose a $\hat{T}_{0},$ which makes the \emph{%
non-orthonormality,} 
\begin{equation}
\left\langle \hat{\chi}^{\left( L\right) }\mid \hat{\chi}^{\left( L\right)
}\right\rangle -1\;\equiv \;\hat{O}^{\left( L\right) },  \label{e13}
\end{equation}
so small, that we may use an expansion like (\ref{o32}) to find $\check{T}%
_{0}$ and the corresponding Hamiltonians $\check{H}^{\left( M\right) }.$ Of
these, $\check{H}^{\left( L\right) }$ equals the variational Hamiltonian (%
\ref{o29}) with $N$ substituted by $L,$ and its eigenvalues are therefore
correct to order $2L+1.$ Expression (\ref{e9}) now tells us that: 
\begin{equation*}
\hat{T}_{0}^{\left( -1\right) ^{L+1}\left( \dagger \right)
^{L+1}}\left\langle \hat{\chi}^{\left( L\right) }\mid \hat{\chi}^{\left(
L\right) }\right\rangle \hat{T}_{0}^{\left( -1\right) ^{L+1}\left( \dagger
\right) ^{L}}=\check{T}_{0}^{\left( -1\right) ^{L+1}\left( \dagger \right)
^{L+1}}\check{T}_{0}^{\left( -1\right) ^{L+1}\left( \dagger \right) ^{L}},
\end{equation*}
which may be solved to yield: 
\begin{equation}
\check{T}_{0}=\hat{T}_{0}\sqrt{1+\hat{O}^{\left( L\right) }}^{\left(
-1\right) ^{L+1}}=\hat{T}_{0}\left\{ 
\begin{array}{l}
1+\frac{1}{2}\hat{O}^{\left( L\right) }-\frac{1}{8}\left( \hat{O}^{\left(
L\right) }\right) ^{2}+.. \\ 
1-\frac{1}{2}\hat{O}^{\left( L\right) }+\frac{3}{8}\left( \hat{O}^{\left(
L\right) }\right) ^{2}-..
\end{array}
\right.  \label{e16}
\end{equation}
Here, the upper result is for $L$ odd and the lower for $L$ even. Since $%
\hat{O}^{\left( L\right) }$ will be chosen small, and for $L>1$ is usually
of order $\left( \varepsilon _{i}-\varepsilon _{1}\right) \left( \varepsilon
_{i}-\varepsilon _{0}\right) ,$ as we shall argue in (\ref{c8}) and (\ref
{e38}), this transformation preserves the $RL$-character of each NMTO. The
Hamiltonian matrix (\ref{e8}) is seen to transform like the overlap matrix (%
\ref{e9}) with $M$ substituted for $L$ and, as a consequence, 
\begin{eqnarray}
&&\check{H}^{\left( M\right) }-\varepsilon _{M}=  \notag \\[0.15cm]
&&\sqrt{1+\hat{O}^{\left( L\right) }}^{\left( -1\right) ^{L-M+1}}\left( \hat{%
H}^{\left( M\right) }-\varepsilon _{M}\right) \sqrt{1+\hat{O}^{\left(
L\right) }}^{\left( -1\right) ^{L-M+1}}.  \label{e17}
\end{eqnarray}
Similarly, from (\ref{e10}): 
\begin{equation}
\check{G}\left[ 0...M\right] ^{-1}\;=\;\hat{G}\left[ 0...M\right] ^{-1}\sqrt{%
1+\hat{O}^{\left( L\right) }}^{\left( -1\right) ^{L-M+1}}.  \label{e22}
\end{equation}

A procedure for computing $\left[ 1+O\right] ^{\pm \frac{1}{2}},$ which is
more robust than the matrix Taylor series (\ref{e16}), is included in our
codes \cite{SQRT}.

\subsubsection{Choosing $\hat{T}_{0}.$}

Since the near-orthonormality conditions (\ref{e3}) merely fix the \emph{%
geometrical average} $\left\langle \hat{\chi}^{\left( M\right) }\mid \hat{%
\chi}^{\left( M-1\right) }\right\rangle $ of successive sets, the nearly
orthonormal scheme (\ref{e8})-(\ref{e9}) only makes sense if the
transformation $\hat{T}_{0}$ of the kinked partial waves at $\varepsilon
_{0} $ is chosen in such a way that the non-orthonormality $\hat{O}^{\left(
0\right) }$ is small compared with the unit matrix. The nearly-orthonormal
scheme alone, does not make the orthonormalization integrals $\left\langle 
\hat{\chi}^{\left( M\right) }\mid \hat{\chi}^{\left( M\right) }\right\rangle 
$ converge towards the unit matrix, but make them behave like: 
\begin{equation*}
\left\langle \hat{\chi}^{\left( M\right) }\mid \hat{\chi}^{\left( M\right)
}\right\rangle \;\sim \;\left\langle \hat{\chi}^{\left( 0\right) }\mid \hat{%
\chi}^{\left( 0\right) }\right\rangle ^{\left( -1\right) ^{M}}.
\end{equation*}
This alternates with fluctuations depending on the size of $\left\langle 
\hat{\chi}^{\left( 0\right) }\mid \hat{\chi}^{\left( 0\right) }\right\rangle
.$

The first thing to do is therefore to \emph{renormalize} the MTOs in such a
way that $\hat{T}_{0}^{a\dagger }\left\langle \left| \phi _{RL}^{a}\right|
^{2}\right\rangle \hat{T}_{0}^{a}=1,$ instead of (\ref{b26}). Hence, the
first choice is: 
\begin{equation}
\hat{T}_{0}^{a}\;=\;\left( \dot{k}_{0}^{a}\right) ^{-\frac{1}{2}}
\label{e19}
\end{equation}
where $\dot{k}_{0}^{a}$ is the energy-\emph{in}dependent \emph{diagonal}
matrix with elements 
\begin{equation}
\left\langle \left| \phi _{RL}^{a}\left( \varepsilon _{0}\right) \right|
^{2}\right\rangle \;=\;\dot{K}_{RL,RL}^{a}\left( \varepsilon _{0}\right)
\;\equiv \;\dot{k}_{RL,RL}^{a}\left( \varepsilon _{0}\right) .  \label{e18}
\end{equation}
Another choice is to start with a \emph{L\"{o}wdin orthonormalized}
0th-order set: 
\begin{equation}
\hat{T}_{0}^{a}\;=\;\left( \dot{k}_{0}^{a}\right) ^{-\frac{1}{2}}\sqrt{%
1+O^{a}}^{-1}  \label{e20}
\end{equation}
where $O^{a}$ is the non-orthonormality of the 0th-order, renormalized MTO
set: 
\begin{equation}
O^{a}\;\equiv \;\left( \dot{k}_{0}^{a}\right) ^{-\frac{1}{2}}\dot{K}%
_{0}^{a}\left( \dot{k}_{0}^{a}\right) ^{-\frac{1}{2}}-1.  \label{e21}
\end{equation}
This choice therefore corresponds to taking $L=0.$

\subsubsection{Test case: GaAs.}

We have tested this orthonormalization method for GaAs using the minimal Ga $%
spd$ As $sp$ basis set and going all the way up to $L=3,$ that is, to a CMTO
basis with the properties that $\check{H}^{\left( 3\right) }=\left\langle 
\check{\chi}^{\left( 3\right) }\left| \mathcal{H}\right| \check{\chi}%
^{\left( 3\right) }\right\rangle $ and $\left\langle \check{\chi}^{\left(
3\right) }\mid \check{\chi}^{\left( 3\right) }\right\rangle =1,$ so that $%
\check{H}^{\left( 3\right) }$ is a 7th-order Hamiltonian. $\check{H}^{\left(
2\right) }$ and $\check{H}^{\left( 1\right) }$ are of lower order, however,
and neither of the three Hamiltonians commute.

We diagonalized $\check{H}^{\left( L\right) }$ for $L=1,2,3$ and compared
with the band structures obtained with the corresponding non-orthonormal
variational method discussed in Sect. \ref{var}. Both starting choices (\ref
{e19}) and (\ref{e20}) were tried, and both gave fast convergence of the
square-root expansions. The first choice which only requires evaluation of a
square root at the last stage (\ref{e17}) but whose non-orthonormality $\hat{%
O}^{\left( L\right) }$ is larger, was found to be the fastest \cite{PolyT}.

\subsubsection{ Aleph-representation.}

The renormalization (\ref{e19}) is of the same nature as --but simpler than
(due to lack of energy dependence)-- the one performed in Subsection \ref
{hard}, where we went from phase-shift normalization to screening-sphere
normalization. That diagonal transformation was given by (\ref{b24}) for the
screened spherical waves, by (\ref{b25}) and (\ref{b26}) for the 0th-order
MTOs, and by (\ref{b34}) for the KKR matrix. Since we distinguished between
those two normalizations by using respectively Greek and Latin superscripts
for the screening, e.g.\textit{\ }$\alpha $ and $a,$ and since it is
irrelevant, whether one arrives at a nearly orthonormal representation from
quantities normalized one-or-another way, it is logical to label quantities
having the integral normalization (\ref{e19}) by \emph{Hebraic}
superscripts, e.g.\textit{\ }$\aleph $ as corresponding to the same
screening as $\alpha $ and $a.$ Although not diagonal, and therefore
influencing the shape of the kinked partial waves, also the L\"{o}wdin
orthonormalization (\ref{e20}) is an energy-\emph{in}dependent \emph{%
similarity} transformation, and so is any of the following transformations: 
\begin{equation}
\begin{array}{lll}
\phi ^{\aleph }\left( \varepsilon ,\mathbf{r}\right) \;\equiv \;\phi
^{a}\left( \varepsilon ,\mathbf{r}\right) \hat{T}_{0}^{a} & \quad  & \chi
^{\aleph \left( N\right) }\left( \varepsilon ,\mathbf{r}\right) \;\equiv
\;\chi ^{a\left( N\right) }\left( \varepsilon ,\mathbf{r}\right) \hat{T}%
_{0}^{a} \\ 
&  &  \\ 
K^{\aleph }\left( \varepsilon \right) \;\equiv \;\hat{T}_{0}^{a\,\dagger
}\,K^{a}\left( \varepsilon \right) \,\hat{T}_{0}^{a} & \quad  & G^{\aleph
}\left( \varepsilon \right) \;\equiv \;\hat{T}_{0}^{a\,-1}\,G^{a}\left(
\varepsilon \right) \,\hat{T}_{0}^{a\,-1\,\dagger }
\end{array}
\label{e23}
\end{equation}
with $\hat{T}_{0}^{a}$ arbitrary. From the latter energy-independent
similarity transformation of $G\left( \varepsilon \right) ,$ the \emph{non}%
-Hermitian matrices $L_{n}^{\left( N\right) }$ and $E^{\left( N\right) },$
which are given in terms of $G\left( \varepsilon \right) $ by respectively (%
\ref{d1}) and (\ref{d17}), are seen to transform like: 
\begin{equation}
\begin{array}{ccc}
L_{n}^{\aleph \left( N\right) }\;=\;\hat{T}_{0}^{a\,-1}L_{n}^{a\left(
N\right) }\,\hat{T}_{0}^{a} & \quad \mathrm{and\quad } & E^{\aleph \left(
M\right) }\;=\;\hat{T}_{0}^{a\,-1}E^{a\left( M\right) }\,\hat{T}_{0}^{a}
\end{array}
.  \label{e31}
\end{equation}
This --(\ref{e23})-(\ref{e31})-- has all concerned an energy-independent
similarity transformation of \emph{un}-hatted quantities.

In order to ensure that the \emph{hatted} quantities are independent of
which representation --$a$ or $\aleph $-- we start out from, e.g. 
\begin{equation*}
\hat{\phi}^{\aleph }\left( \varepsilon ,\mathbf{r}\right) =\hat{\phi}%
^{a}\left( \varepsilon ,\mathbf{r}\right) =\phi ^{a}\left( \varepsilon ,%
\mathbf{r}\right) \hat{T}^{a}\left( \varepsilon \right) =\phi ^{\aleph
}\left( \varepsilon ,\mathbf{r}\right) \hat{T}^{\aleph }\left( \varepsilon
\right)
\end{equation*}
and 
\begin{equation*}
\hat{G}^{\aleph }\left( \varepsilon \right) =\hat{G}^{a}\left( \varepsilon
\right) =\hat{T}^{a}\left( \varepsilon \right) ^{-1}G^{a}\left( \varepsilon
\right) \hat{T}^{a}\left( \varepsilon \right) ^{-1\dagger }=\hat{T}^{\aleph
}\left( \varepsilon \right) ^{-1}G^{\aleph }\left( \varepsilon \right) \hat{T%
}^{\aleph }\left( \varepsilon \right) ^{-1\,\dagger }
\end{equation*}
where, from the latter, it follows that 
\begin{equation*}
\begin{array}{ccc}
\hat{L}_{n}^{\aleph \left( N\right) }\;=\;\hat{L}_{n}^{a\left( N\right) } & 
\quad \mathrm{and\quad } & \hat{E}_{n}^{\aleph \left( M\right) }\;=\;\hat{E}%
_{n}^{a\left( M\right) }
\end{array}
,
\end{equation*}
it suffices to satisfy the relation: 
\begin{equation}
\begin{array}{ccccc}
\hat{T}^{\aleph }\left( \varepsilon \right) \;\equiv \;\hat{T}_{0}^{a\,-1}%
\hat{T}^{a}\left( \varepsilon \right) \,,\, & \; & \mathrm{which\;leads\;to:}
& \; & \hat{T}_{0}^{\aleph }=1.
\end{array}
\label{e25}
\end{equation}
In conclusion, under the substitution $a\rightarrow \aleph ,$ all previous
equations remain valid, and the factors $\hat{T}_{0}^{\aleph }$ may be
deleted.

The virtue of this notation is that, once we have decided upon the
normalization and the screening, we can drop the superscripts; and this is
what we shall do: From now on, and throughout the remainder of this paper,
un-hatted quantities, i.e.\textit{\ }the kinked partial waves, the kink and
the Green matrices, and the Lagrange and energy matrices, are all supposed
to have the integral (ortho)normalization (\ref{e19}) or (\ref{e20}), that
is, they are all in the Aleph-representation. All equations derived
previously are then unchanged, and $\hat{T}_{0}$ may be dropped.

\subsubsection{Accuracies of Hamiltonians.}

The accuracies of the Hamiltonians depend on the sizes of the corresponding
non-orthonormalities. Specifically, since the residual error of the
one-electron energy after use of the variational principle (\ref{a3}) for
the set $\hat{\chi}^{\left( M\right) }\left( \mathbf{r}\right) $, 
\begin{equation*}
\hat{H}^{\left( M\right) }v_{i}=\,\varepsilon _{i}v_{i}+\left( \varepsilon
_{i}-\varepsilon _{M}\right) \hat{O}^{\left( M\right) }v_{i}\,,
\end{equation*}
is proportional to $\left( \varepsilon _{i}-\varepsilon _{0}\right)
^{2}..\left( \varepsilon _{i}-\varepsilon _{M}\right) ^{2},$ neglect of the
non-orthonormality, leads to the error: 
\begin{equation}
\delta \hat{\varepsilon}_{i}^{\,\left( M\right) }\;=\;\left( \varepsilon
_{i}-\varepsilon _{M}\right) \hat{O}_{ii}^{\left( M\right) }\;+\;\mathcal{O}%
\left\{ \left( \varepsilon _{i}-\varepsilon _{0}\right) ^{2}..\left(
\varepsilon _{i}-\varepsilon _{M}\right) ^{2}\right\} ,  \label{c7}
\end{equation}
where $\hat{O}_{ii}^{\left( M\right) }\equiv $ $v_{i}^{\dagger }\hat{O}%
^{\left( M\right) }v_{i}$ and $\mathcal{O}$ means at the order of. The goal
should thus be to reduce the non-orthonormality to: 
\begin{equation*}
\hat{O}_{ii}^{\left( M\right) }\;=\;\mathcal{O}\left\{ \left( \varepsilon
_{i}-\varepsilon _{0}\right) ^{2}..\left( \varepsilon _{i}-\varepsilon
_{M-1}\right) ^{2}\left( \varepsilon _{i}-\varepsilon _{M}\right) \right\}
\end{equation*}
because in that case, the error from non-orthonormality will be of the same
order as that of the residual error. This can usually only achieved for $%
M=L. $

The order of the non-orthonormality may be found by use of the difference
function: 
\begin{eqnarray*}
&&\hat{\chi}^{\left( M\right) }\left( \mathbf{r}\right) -\hat{\chi}^{\left(
M-1\right) }\left( \mathbf{r}\right) \;=\;\hat{\phi}\left( \left[ 01\right] ,%
\mathbf{r}\right) \left( \hat{H}^{\left( M\right) }-\hat{H}^{\left(
M-1\right) }\right) +\;\hat{\phi}\left( \left[ 012\right] ,\mathbf{r}\right)
\\
&&\quad \quad \quad \quad \quad \quad \times \left\{ 
\begin{array}{l}
\left( \hat{H}^{\left( M-1\right) }-\varepsilon _{1}\right) \left( \hat{H}%
^{\left( M\right) }-\varepsilon _{0}\right) \\ 
\quad \quad \quad -\left( \hat{H}^{\left( M-2\right) }-\varepsilon
_{1}\right) \left( \hat{H}^{\left( M-1\right) }-\varepsilon _{0}\right)
\end{array}
\right\} \;+\;..\,,
\end{eqnarray*}
obtained from (\ref{d25}) and where we should take $\hat{H}^{\left( m\right)
}\equiv 0$ if $m<1.$ As a result: 
\begin{eqnarray}
\hat{O}^{\left( M\right) } &=&\left\langle \hat{\chi}^{\left( M\right) }\mid 
\hat{\chi}^{\left( M\right) }-\hat{\chi}^{\left( M-1\right) }\right\rangle 
\notag \\[0.1cm]
&=&\left\langle \hat{\phi}_{0}\mid \hat{\phi}\left[ 01\right] \right\rangle
\left( \hat{H}^{\left( M\right) }-\hat{H}^{\left( M-1\right) }\right)
\label{c8} \\
&&+\left( \hat{H}^{\left( M\right) }-\varepsilon _{0}\right) \left\langle 
\hat{\phi}\left[ 01\right] \mid \hat{\phi}\left[ 01\right] \right\rangle
\left( \hat{H}^{\left( M-1\right) }-\varepsilon _{1}\right) \left( \hat{H}%
^{\left( M\right) }-\varepsilon _{0}\right)  \notag \\
&&+\left\langle \hat{\phi}_{0}\mid \hat{\phi}\left[ 012\right] \right\rangle
\left\{ 
\begin{array}{l}
\left( \hat{H}^{\left( M-1\right) }-\varepsilon _{1}\right) \left( \hat{H}%
^{\left( M\right) }-\varepsilon _{0}\right) \\ 
-\left( \hat{H}^{\left( M-2\right) }-\varepsilon _{1}\right) \left( \hat{H}%
^{\left( M-1\right) }-\varepsilon _{0}\right)
\end{array}
\right\} +..  \notag
\end{eqnarray}
which is usually of order $\left( \hat{H}^{\left( M-1\right) }-\varepsilon
_{1}\right) \left( \hat{H}^{\left( M\right) }-\varepsilon _{0}\right) $ when 
$M>1.$

To evaluate integrals like $\left\langle \hat{\phi}_{0}\mid \hat{\phi}\left[
01\right] \right\rangle $ we must transform to the original representation
using (\ref{a41}) and then use (\ref{e70}). In this way we get: 
\begin{equation}
\left\langle \hat{\phi}_{0}\mid \hat{\phi}\left[ 01\right] \right\rangle
=\left\langle \phi _{0}\mid \phi \left[ 01\right] \right\rangle
+\left\langle \phi _{0}\mid \phi _{1}\right\rangle \hat{T}\left[ 01\right] =K%
\left[ \left[ 0\right] 1\right] +K\left[ 01\right] \hat{T}\left[ 01\right] .
\label{c9}
\end{equation}
Remember, that we are using the Aleph-normalization (\ref{e23}), because
this influences the right-hand sides. For a condensed mesh, (\ref{c9})
reduces to: 
\begin{equation*}
\left\langle \hat{\phi}\mid \overset{.}{\hat{\phi}}\right\rangle
\;=\;\left\langle \phi \mid \dot{\phi}\right\rangle +\,\overset{.}{\hat{T}}%
\;=\;\frac{\ddot{K}}{2!}\,+\,\overset{.}{\hat{T}}.
\end{equation*}
We shall conclude this study of the accuracy of the Hamiltonians in Eq. (\ref
{e37}) below.

\section{Connecting back to the ASA formalism}

\label{ASA}What remains to be demonstrated is that the NMTO sets, $\chi
^{\left( N\right) }\left( \mathbf{r}\right) ,$ $\hat{\chi}^{\left( N\right)
}\left( \mathbf{r}\right) ,$ and $\check{\chi}^{\left( N\right) }\left( 
\mathbf{r}\right) ,$ of which the two former are based on L\"{o}%
wdin-orthonormalized kinked partial waves at the first mesh point (\ref{e20}%
), and the last corresponds to the $L$=1-set being orthonormal, are the
generalizations to overlapping MT-potentials, arbitrary $N,$ and discrete
meshes of the well-known LMTO-ASA sets given in the Overview by respectively
(\ref{a1}), (\ref{a5}), and (\ref{a6}).

Since in the present paper we have not made use of the ASA, but merely a
MT-potential --plus redefinition of the partial waves followed by a
L\"{o}wdin-orthonormalization-- we merely need to show that the formalism
developed above reduces to the one given in the Overview for the case $N$=1
and a condensed mesh. In order to bridge the gap between the new and old
formalisms, a bit more will be done though.

\subsubsection{$N=0,\;L=0.$}

For the 0th-order set we have: 
\begin{equation*}
\chi ^{\left( 0\right) }\left( \mathbf{r}\right) =\hat{\chi}^{\left(
0\right) }\left( \mathbf{r}\right) =\phi _{0}\left( \mathbf{r}\right) =\hat{%
\phi}_{0}\left( \mathbf{r}\right) .
\end{equation*}
All un-hatted quantities in the present section will correspond to using
kinked partial waves, transformed to be orthonormal at this first mesh
point, $\varepsilon _{0}.$ That is: All un-hatted quantities are in the
Aleph-representation (\ref{e23})-(\ref{e25}) with $\hat{T}_{0}^{a}$ given by
(\ref{e20}). In this representation all previously derived relations hold,
and in addition: 
\begin{equation}
\hat{T}_{0}=1\quad \quad \mathrm{and\quad \quad }\dot{K}_{0}=1.  \label{e35}
\end{equation}
Relating back to the Overview, this means that instead of the ASA-relation (%
\ref{a9}), we have (\ref{e23}) with $\hat{T}_{0}^{a}$ given by (\ref{e20}).
The latter is the proper definition of $\dot{K}_{0}^{a\,-1/2},$ now that $%
\dot{K}_{0}^{a}=\left\langle \phi _{0}^{a}\mid \phi _{0}^{a}\right\rangle $
is no longer diagonal. We now see that the un-hatted quantities used in the
Overview were, in fact, in the Aleph representation.

The overlap and Hamiltonian matrices for the 0th-order set are thus: 
\begin{eqnarray}
\left\langle \chi ^{\left( 0\right) }\mid \chi ^{\left( 0\right)
}\right\rangle &=&\left\langle \phi _{0}\mid \phi _{0}\right\rangle
=\left\langle \hat{\chi}^{\left( 0\right) }\mid \hat{\chi}^{\left( 0\right)
}\right\rangle =\left\langle \hat{\phi}_{0}\mid \hat{\phi}_{0}\right\rangle
=1  \notag \\
\left\langle \chi ^{\left( 0\right) }\left| \mathcal{H}-\varepsilon
_{0}\right| \chi ^{\left( 0\right) }\right\rangle &=&\left\langle \hat{\chi}%
^{\left( 0\right) }\left| \mathcal{H}-\varepsilon _{0}\right| \hat{\chi}%
^{\left( 0\right) }\right\rangle =H^{\left( 0\right) }-\varepsilon
_{0}=-K_{0},  \label{e27}
\end{eqnarray}
and with the 0th-order set being orthonormal, the Hamiltonian is
variational. Hence, $H^{\left( 0\right) }=\hat{H}^{\left( 0\right) }$ is the 
\emph{first}-order, two-center, TB Hamiltonian of the 3rd-generation scheme.

\subsubsection{$N=1,\;L=0.$}

For the LMTO set we have: 
\begin{equation*}
\chi ^{\left( 1\right) }\left( \mathbf{r}\right) =\phi _{0}\left( \mathbf{r}%
\right) +\phi \left( \left[ 01\right] ,\mathbf{r}\right) \left( E^{\left(
1\right) }-\varepsilon _{0}\right) \;\rightarrow \;\phi \left( \mathbf{r}%
\right) +\dot{\phi}\left( \mathbf{r}\right) \left( H^{\left( 0\right)
}-\varepsilon _{\nu }\right) ,
\end{equation*}
where $E^{\left( 1\right) }$ --as given by (\ref{d26})-- is seen to become
the Hermitian, \emph{first-}order Hamiltonian $H^{\left( 0\right) }$ given
by (\ref{e27}) if the mesh condenses. This proves (\ref{a1}).

The Hamiltonian and overlap matrices were given in respectively (\ref{a17})
and (\ref{a18}), and using now $\dot{K}=1$ together with (\ref{e70})$,$ we
see that for a condensed mesh 
\begin{eqnarray*}
\left\langle \chi ^{\left( 1\right) }\left| \mathcal{H}-\varepsilon
_{1}\right| \chi ^{\left( 1\right) }\right\rangle \; &\rightarrow &\;-\dot{G}%
^{-1}\frac{\ddot{G}}{2!}\dot{G}^{-1}\;=\;-K+K\frac{\ddot{K}}{2!}K \\[0.15cm]
&=&H^{\left( 0\right) }-\varepsilon _{\nu }+\left( H^{\left( 0\right)
}-\varepsilon _{\nu }\right) \left\langle \phi \mid \dot{\phi}\right\rangle
\left( H^{\left( 0\right) }-\varepsilon _{\nu }\right)
\end{eqnarray*}
and 
\begin{eqnarray*}
\left\langle \chi ^{\left( 1\right) }\mid \chi ^{\left( 1\right)
}\right\rangle \; &\rightarrow &\;-\dot{G}^{-1}\frac{\dddot{G}}{3!}\dot{G}%
^{-1}\;=\;1-K\frac{\ddot{K}}{2!}-\frac{\ddot{K}}{2!}K+K\frac{\dddot{K}}{3!}K
\\[0.15cm]
&=&1+\left( H^{\left( 0\right) }-\varepsilon _{\nu }\right) \left\langle 
\dot{\phi}\mid \phi \right\rangle +\left\langle \phi \mid \dot{\phi}%
\right\rangle \left( H^{\left( 0\right) }-\varepsilon _{\nu }\right) \\
&&\quad \quad \quad \quad \quad \quad \quad +\left( H^{\left( 0\right)
}-\varepsilon _{\nu }\right) \left\langle \dot{\phi}\mid \dot{\phi}%
\right\rangle \left( H^{\left( 0\right) }-\varepsilon _{\nu }\right) ,
\end{eqnarray*}
which are exactly (\ref{a4}). Merely $\left\langle \phi \mid \dot{\phi}%
\right\rangle $ is not a \emph{diagonal} matrix of radial integrals like in
the ASA.

The nearly orthonormal LMTO set is: 
\begin{equation*}
\hat{\chi}^{\left( 1\right) }\left( \mathbf{r}\right) \;=\;\hat{\phi}%
_{0}\left( \mathbf{r}\right) \;+\;\hat{\phi}\left( \left[ 01\right] ,\mathbf{%
r}\right) \left( \hat{H}^{\left( 1\right) }-\varepsilon _{0}\right) ,
\end{equation*}
and the two conditions: $\left\langle \hat{\chi}^{\left( 0\right) }\mid \hat{%
\chi}^{\left( 0\right) }\right\rangle =1=\left\langle \hat{\chi}^{\left(
1\right) }\mid \hat{\chi}^{\left( 0\right) }\right\rangle ,$ therefore lead
to: 
\begin{equation*}
\left\langle \hat{\phi}\left[ 01\right] \mid \hat{\phi}_{0}\right\rangle
=0=\left\langle \hat{\phi}_{0}\mid \hat{\phi}\left[ 01\right] \right\rangle
,\quad \mathrm{and}\quad \left\langle \hat{\phi}_{1}\mid \hat{\phi}%
_{0}\right\rangle =1=\left\langle \hat{\phi}_{0}\mid \hat{\phi}%
_{1}\right\rangle .
\end{equation*}
Of these matrix equations, the first means that \emph{any} $\hat{\phi}%
_{RL}\left( \left[ 01\right] ,\mathbf{r}\right) $ is \emph{orthogonal} to 
\emph{any} $\hat{\phi}_{R^{\prime }L^{\prime }}\left( \varepsilon _{0},%
\mathbf{r}\right) .$ As a consequence, the \emph{leading term} of the
non-orthonormality (\ref{c8}) \emph{vanishes.} The non-orthonormality of
this LMTO set is then: 
\begin{equation}
\hat{O}^{\left( 1\right) }\;=\;\left( \hat{H}^{\left( 1\right) }-\varepsilon
_{0}\right) \left\langle \hat{\phi}\left[ 01\right] \mid \hat{\phi}\left[ 01%
\right] \right\rangle \left( \hat{H}^{\left( 1\right) }-\varepsilon
_{0}\right) ,  \label{e36}
\end{equation}
which by use of (\ref{c7}) shows that the errors of the $\hat{H}^{\left(
1\right) }$-eigenvalues are: 
\begin{equation}
\delta \hat{\varepsilon}_{i}^{\,\left( 1\right) }\approx \left\langle \hat{%
\phi}\left[ 01\right] \mid \hat{\phi}\left[ 01\right] \right\rangle
_{ii}\,\left( \varepsilon _{i}-\varepsilon _{1}\right) \left( \varepsilon
_{i}-\varepsilon _{0}\right) ^{2}.  \label{c16}
\end{equation}
This is one order better than the error $\propto \left( \varepsilon
_{i}-\varepsilon _{0}\right) ^{2}$ obtained by diagonalization of $H^{\left(
0\right) },$ but one order worse than the error $\propto \left( \varepsilon
_{i}-\varepsilon _{1}\right) ^{2}\left( \varepsilon _{i}-\varepsilon
_{0}\right) ^{2}$ obtained variationally using the LMTO set. Hence, $\hat{H}%
^{\left( 1\right) }$ is a \emph{second}-order Hamiltonian. From (\ref{e8}): 
\begin{eqnarray*}
&&\hat{H}^{\left( 1\right) }-\varepsilon _{1}\;=\;-G_{0}\,G\left[ \left[ 0%
\right] 1\right] ^{-1}G_{0}\;\rightarrow \;-G\left[ \frac{\ddot{G}}{2!}%
\right] ^{-1}G\;= \\
&&\left( 1-K\frac{\ddot{K}}{2!}\right) ^{-1}\left( -K\right) \;=\;\left[
1+\left( H^{\left( 0\right) }-\varepsilon _{\nu }\right) \left\langle \dot{%
\phi}\mid \phi \right\rangle \right] ^{-1}\left( H^{\left( 0\right)
}-\varepsilon _{\nu }\right) ,
\end{eqnarray*}
which for a condensed mesh is exactly (\ref{a5}).

For the transformation (\ref{a44}) from the $\chi $ to the $\hat{\chi}$-set,
we get by use of (\ref{e10}): 
\begin{eqnarray*}
&&G\left[ 01\right] \hat{G}\left[ 01\right] ^{-1}\;=\;-G\left[ 01\right] G%
\left[ \left[ 0\right] 1\right] ^{-1}G_{0} \\[0.15cm]
&\rightarrow &\;-\dot{G}\left[ \frac{\ddot{G}}{2!}\right] ^{-1}G\;=\;G^{2}%
\left[ \frac{\ddot{G}}{2!}\right] ^{-1}G\;=\;\left[ 1+\left\langle \dot{\phi}%
\mid \phi \right\rangle \left( H^{\left( 0\right) }-\varepsilon _{\nu
}\right) \right] ^{-1}
\end{eqnarray*}
which --since from (\ref{e70}): $\left\langle \dot{\phi}\mid \phi
\right\rangle =\left\langle \phi \mid \dot{\phi}\right\rangle $-- is exactly
(\ref{a5}).

The transformation (\ref{a41}) of the kinked partial waves is most easily
found by using the orthogonality of $\hat{\phi}_{0}\left( \mathbf{r}\right) $
and $\hat{\phi}\left( \left[ 01\right] ,\mathbf{r}\right) $ together with (%
\ref{c9}). For a condensed mesh, the result is simple: 
\begin{equation*}
\overset{.}{\hat{\phi}}\left( \mathbf{r}\right) \;=\;\dot{\phi}\left( 
\mathbf{r}\right) +\phi \left( \mathbf{r}\right) \overset{.}{\hat{T}\;}=\;%
\dot{\phi}\left( \mathbf{r}\right) -\phi \left( \mathbf{r}\right)
\left\langle \phi \mid \dot{\phi}\right\rangle \;=\;\dot{\phi}\left( \mathbf{%
r}\right) -\phi \left( \mathbf{r}\right) \frac{\ddot{K}}{2!},
\end{equation*}
and well known --see Eqs. (\ref{a5}) and (\ref{a10}). For a \emph{discrete
mesh,} things look more complicated in $K$-language: From (\ref{c9}), 
\begin{equation}
\hat{T}\left[ 01\right] \;=\;-K\left[ 01\right] ^{-1}K\left[ \left[ 0\right]
1\right] \;=\;-K\left[ 01\right] ^{-1}\frac{1-K\left[ 01\right] }{%
\varepsilon _{0}-\varepsilon _{1}}\,,  \notag
\end{equation}
where the 2nd equation has been obtained by use of (\ref{h2}): $F\left[ %
\left[ 0\right] 1\right] =\frac{\dot{F}_{0}-F\left[ 01\right] }{\varepsilon
_{0}-\varepsilon _{1}},$ together with: $\dot{K}_{0}=1.$ For (\ref{a41}) we
thus obtain: 
\begin{eqnarray}
\hat{\phi}\left( \left[ 01\right] ,\mathbf{r}\right) \; &=&\;\phi \left( %
\left[ 01\right] ,\mathbf{r}\right) \;+\;\phi _{1}\left( \mathbf{r}\right) 
\hat{T}\left[ 01\right]  \notag \\
&=&\;\phi \left( \left[ 01\right] ,\mathbf{r}\right) \left( 1\;+\;\left(
\varepsilon _{1}-\varepsilon _{0}\right) \hat{T}\left[ 01\right] \right)
\;+\;\phi _{0}\left( \mathbf{r}\right) \hat{T}\left[ 01\right]  \notag \\
&=&\;\phi \left( \left[ 01\right] ,\mathbf{r}\right) K\left[ 01\right]
^{-1}\;+\;\phi _{0}\left( \mathbf{r}\right) \hat{T}\left[ 01\right]  \notag
\\
&=&\;\left\{ \phi \left( \left[ 01\right] ,\mathbf{r}\right) \;+\;\phi
_{0}\left( \mathbf{r}\right) \hat{T}\left[ 01\right] K\left[ 01\right]
\right\} K\left[ 01\right] ^{-1}  \notag \\
&=&\;\left\{ \phi \left( \left[ 01\right] ,\mathbf{r}\right) \;-\;\phi
_{0}\left( \mathbf{r}\right) K\left[ \left[ 0\right] 1\right] \right\} K%
\left[ 01\right] ^{-1}  \label{e26}
\end{eqnarray}
where from (\ref{e70}): $K\left[ \left[ 0\right] 1\right] =\left\langle \phi
_{0}\mid \phi \left[ 01\right] \right\rangle $ is the equivalent to the
usual radial integral and the new factor $K\left[ 01\right] $ in the
transformation is caused by the presence of $\phi _{1}\left( \mathbf{r}%
\right) $ rather than $\phi _{0}\left( \mathbf{r}\right) $ on the right-hand
side of the top line in (\ref{e26}).

In order to complete the identification of the nearly-orthonormal LMTO
representation for a discrete mesh with the ASA version (\ref{a5}) and (\ref
{a10}), we need an explicit expression for the third parameter, which is the
matrix entering the non-orthonormality (\ref{e36}). With the help of (\ref
{e26}), and remembering that $\hat{\phi}_{0}\left( \mathbf{r}\right) $ and $%
\hat{\phi}\left( \left[ 01\right] ,\mathbf{r}\right) $ are orthogonal, we
get: 
\begin{eqnarray*}
\left\langle \hat{\phi}\left[ 01\right] \mid \hat{\phi}\left[ 01\right]
\right\rangle \; &=&\;K\left[ 01\right] ^{-1}\left\langle \phi \left[ 01%
\right] \mid \hat{\phi}\left[ 01\right] \right\rangle \\
&=&\;K\left[ 01\right] ^{-1}\left( \left\langle \phi \left[ 01\right] \mid
\phi \left[ 01\right] \right\rangle -K\left[ \left[ 0\right] 1\right]
^{2}\right) K\left[ 01\right] ^{-1} \\
&=&\;K\left[ 01\right] ^{-1}\left( K\left[ \left[ 01\right] \right] -K\left[ %
\left[ 0\right] 1\right] ^{2}\right) K\left[ 01\right] ^{-1} \\
&\rightarrow &\;\left\langle \overset{.}{\hat{\phi}}\mid \overset{.}{\hat{%
\phi}}\right\rangle \;=\;\frac{\dddot{K}}{3!}-\left[ \frac{\ddot{K}}{2!}%
\right] ^{2},
\end{eqnarray*}
where, in the third equation, we have used (\ref{e70}).

\subsubsection{TB parametrization}

For tight-binding parametrizations of many bands over a relatively wide
energy range, it is usually important to have as few parameters as possible.
Our experience \cite{TanusriTB,MRS} for the occupied and lowest excited
bands of semiconductors and transition metals is that the off-diagonal
elements of $\left\langle \phi _{0}\mid \phi _{1}\right\rangle =K\left[ 01%
\right] ,\quad \left\langle \phi _{0}\mid \phi \left[ 01\right]
\right\rangle ,$ and $\left\langle \hat{\phi}\left[ 01\right] \mid \hat{\phi}%
\left[ 01\right] \right\rangle $ may be neglected. This is in the spirit of
the ASA. We therefore need to tabulate only those few diagonal elements,
together with the single TB matrix $H^{\left( 0\right) }.$ These quantities
may then be used to construct for instance the Hamiltonian and overlap
matrices $\left\langle \chi ^{\left( 1\right) }\left| \mathcal{H}%
-\varepsilon _{1}\right| \chi ^{\left( 1\right) }\right\rangle $ and $%
\left\langle \chi ^{\left( 1\right) }\mid \chi ^{\left( 1\right)
}\right\rangle .$ This is like in the ASA, but now, we neither need this
approximation nor a condensed mesh.

\subsubsection{$N>1,\;L=0.$}

The nearly-orthonormal QMTO set is: 
\begin{equation*}
\hat{\chi}^{\left( 2\right) }\left( \mathbf{r}\right) \;=\;\hat{\phi}%
_{0}\left( \mathbf{r}\right) +\left\{ \hat{\phi}\left( \left[ 01\right] ,%
\mathbf{r}\right) +\hat{\phi}\left( \left[ 012\right] ,\mathbf{r}\right)
\left( \hat{H}^{\left( 1\right) }-\varepsilon _{1}\right) \right\} \left( 
\hat{H}^{\left( 2\right) }-\varepsilon _{0}\right)
\end{equation*}
with the non-orthonormality: 
\begin{eqnarray*}
&&\hat{O}^{\left( 2\right) }=\left\langle \hat{\chi}^{\left( 2\right) }\mid 
\hat{\chi}^{\left( 2\right) }-\hat{\chi}^{\left( 1\right) }\right\rangle
=\left\langle \hat{\phi}_{0}\mid \hat{\phi}\left[ 012\right] \right\rangle
\left( \hat{H}^{\left( 1\right) }-\varepsilon _{1}\right) \left( \hat{H}%
^{\left( 2\right) }-\varepsilon _{0}\right) + \\
&&\quad \quad \quad \quad \quad \quad \quad \quad \quad +\left( \hat{H}%
^{\left( 2\right) }-\varepsilon _{0}\right) \left\langle \hat{\phi}\left[ 10%
\right] \mid \hat{\phi}\left[ 01\right] \right\rangle \left( \hat{H}^{\left(
2\right) }-\hat{H}^{\left( 1\right) }\right) +..\,.
\end{eqnarray*}
This --together with (\ref{c7})-- shows that the eigenvalue errors of $\hat{H%
}^{\left( 2\right) }$ are: 
\begin{equation*}
\delta \hat{\varepsilon}_{i}^{\,\left( 2\right) }\approx \left\langle \hat{%
\phi}_{0}\mid \hat{\phi}\left[ 012\right] \right\rangle _{ii}\left(
\varepsilon _{i}-\varepsilon _{2}\right) \left( \varepsilon _{i}-\varepsilon
_{1}\right) \left( \varepsilon _{i}-\varepsilon _{0}\right) ,
\end{equation*}
which means, that $\hat{H}^{\left( 2\right) }$ is a \emph{second-}order
Hamiltonian like $\hat{H}^{\left( 1\right) },$ but different from it$.$ In
general, for $N>1,$ the leading non-orthonormality is: 
\begin{equation}
\hat{O}^{\left( N\right) }\approx \left\langle \hat{\phi}_{0}\mid \hat{\phi}%
\left[ 012\right] \right\rangle \left( \hat{H}^{\left( N-1\right)
}-\varepsilon _{1}\right) \left( \hat{H}^{\left( N\right) }-\varepsilon
_{0}\right) ,  \label{e38}
\end{equation}
as seen from (\ref{c8}). This means that $\hat{H}^{\left( N\right) }$
remains a 2nd-order Hamiltonian when $N>1,$ and that its eigenvalue errors
are: 
\begin{equation}
\delta \hat{\varepsilon}_{i}^{\,\left( N\right) }\approx \left\langle \hat{%
\phi}_{0}\mid \hat{\phi}\left[ 012\right] \right\rangle _{ii}\left(
\varepsilon _{i}-\varepsilon _{N}\right) \left( \varepsilon _{i}-\varepsilon
_{1}\right) \left( \varepsilon _{i}-\varepsilon _{0}\right) .  \label{e37}
\end{equation}
This is much inferior to the variational estimate obtainable with an NMTO
basis. Moreover, the same result would have been obtained had we started out
from the cheaper, renormalized scheme based on (\ref{e19}). Hence, with the
present scheme only the Hamiltonians $H^{\left( M\right) }$ with $M\sim L,$
have eigenvalues which are accurate approximations to the one-electron
energies.

\subsubsection{$N=1,\;L=1.$}

We finally use the general procedure (\ref{e13})-(\ref{e22}) to
orthonormalize the nearly-orthonormal LMTO set considered above. The small
parameter --the non-orthonormality $\hat{O}^{\left( L=1\right) }$-- is thus
given by (\ref{e36}).

The transformation from the nearly to the completely orthonormal set is
obtained from (\ref{e22}), with $L=M=1,$ as: 
\begin{equation*}
\check{\chi}^{\left( 1\right) }\left( \mathbf{r}\right) =\hat{\chi}^{\left(
1\right) }\left( \mathbf{r}\right) \hat{G}\left[ 01\right] \check{G}\left[ 01%
\right] ^{-1}=\hat{\chi}^{\left( 1\right) }\left( \mathbf{r}\right) \left[ 1+%
\hat{O}^{\left( 1\right) }\right] ^{-\frac{1}{2}},
\end{equation*}
which is the generalization to discrete meshes and (overlapping)
MT-potentials of the first equation (\ref{a6}). The resulting, orthonormal
LMTO set is: 
\begin{equation*}
\check{\chi}^{\left( 1\right) }\left( \mathbf{r}\right) \;=\;\check{\phi}%
_{0}\left( \mathbf{r}\right) \;+\;\check{\phi}\left( \left[ 01\right] ,%
\mathbf{r}\right) \left( \check{H}^{\left( 1\right) }-\varepsilon
_{0}\right) ,
\end{equation*}
with the \emph{third}-order Hamiltonian obtained from (\ref{e17}) with $%
L=M=1 $ as: 
\begin{equation*}
\check{H}^{\left( 1\right) }-\varepsilon _{1}=\left[ 1+\hat{O}^{\left(
1\right) }\right] ^{-\frac{1}{2}}\left( \hat{H}^{\left( 1\right)
}-\varepsilon _{1}\right) \left[ 1+\hat{O}^{\left( 1\right) }\right] ^{-%
\frac{1}{2}}.
\end{equation*}
This is the second ASA equation (\ref{a6}).

For the transformation of the kinked partial waves, we have from (\ref{e16}%
): 
\begin{equation*}
\check{\phi}_{0}\left( \mathbf{r}\right) =\hat{\phi}_{0}\left( \mathbf{r}%
\right) \left[ 1+\hat{O}^{\left( 1\right) }\right] ^{\frac{1}{2}}
\end{equation*}
and putting all of this together, we may obtain: 
\begin{equation*}
\check{\phi}\left( \left[ 01\right] ,\mathbf{r}\right) \;\approx \;\hat{\phi}%
\left( \left[ 01\right] ,\mathbf{r}\right) \;-\;\hat{\phi}_{0}\left( \mathbf{%
r}\right) \left( \hat{H}^{\left( 1\right) }-\varepsilon _{0}\right)
\left\langle \hat{\phi}\left[ 01\right] \mid \hat{\phi}\left[ 01\right]
\right\rangle ,
\end{equation*}
which is a new result. Finally, we may check that: 
\begin{eqnarray*}
&&\left\langle \check{\chi}^{\left( 1\right) }\mid \check{\chi}^{\left(
0\right) }\right\rangle =\left\langle \check{\phi}_{0}\mid \check{\phi}%
_{0}\right\rangle +\left( \check{H}^{\left( 1\right) }-\varepsilon
_{0}\right) \left\langle \check{\phi}\left[ 01\right] \mid \check{\phi}%
_{0}\right\rangle = \\
&&1+\hat{O}^{\left( 1\right) }-\left( \check{H}^{\left( 1\right)
}-\varepsilon _{0}\right) \left\langle \hat{\phi}\left[ 01\right] \mid \hat{%
\phi}\left[ 01\right] \right\rangle \left( \check{H}^{\left( 1\right)
}-\varepsilon _{0}\right) \left\langle \check{\phi}_{0}\mid \check{\phi}%
_{0}\right\rangle \approx 1.
\end{eqnarray*}

\section{Outlook}

Of the new developments described above, only the use of overlapping
MT-potentials and efficient computation of total energies and forces from
TB-LMTO-ASA charge densities were planned. Those parts turned out to be the
hardest and still await their completion. But on the way, we did pick up a
number of beautiful and useful instruments. Now that we have an accordion
for playing Schr\"{o}dinger, maybe Poisson can be learned as well.

\section{Acknowledgments}

It is a pleasure to thank Mark van Schilfgaarde for drawing our attention to
a weakness in the 3rd-generation tangent-LMTO scheme. This triggered the
development of the general and robust NMTO method. To make it presentable,
took longer than expected, and we are most grateful to Hugues Dreysse and
all other contributors to this book for their patience and encouragement.

\section{Appendix: Classical Polynomial Approximations}

\subsubsection{Lagrange and Newton interpolation.}

In these interpolation schemes, a function $f\left( \varepsilon \right) $ is
approximated by that \emph{polynomial} of $N$th degree, $f^{\left( N\right)
}\left( \varepsilon \right) ,$ which coincides with the function at the $N+1$%
\textit{\ }energies, $\varepsilon _{0},$\thinspace $\varepsilon
_{1},..,\,\varepsilon _{N},$ forming the \emph{mesh.} The error is
proportional to $\left( \varepsilon -\varepsilon _{0}\right) \left(
\varepsilon -\varepsilon _{1}\right) ..\left( \varepsilon -\varepsilon
_{N}\right) .$

The expression for the approximating polynomial in terms of the $N+1$ values
of the function, $f\left( \varepsilon _{n}\right) \equiv f_{n},$ with $%
n=0,1,..,N,$ is: 
\begin{equation}
f^{\left( N\right) }\left( \varepsilon \right)
=\sum_{n=0}^{N}f_{n}\,l_{n}^{\left( N\right) }\left( \varepsilon \right)
,\quad \mathrm{where}\quad l_{n}^{\left( N\right) }\left( \varepsilon
\right) \equiv \prod_{m=0,\neq n}^{N}\frac{\varepsilon -\varepsilon _{m}}{%
\varepsilon _{n}-\varepsilon _{m}}  \label{e32}
\end{equation}
is the \emph{Lagrange }polynomial of $N$th degree. It has nodes at all mesh
points, except at the \textit{n}th, where it takes the value 1. Since
Lagrange interpolation is exact for all functions $\varepsilon ^{M}$ with $%
M\leq N,$ the Lagrange polynomials satisfy the sum rules: $\varepsilon
^{M}=\sum_{n=0}^{N}\left( \varepsilon _{n}\right) ^{M}\,\,l_{n}^{\left(
N\right) }\left( \varepsilon \right) ,$ for\textrm{\ }$M=0,...,N.$

The \emph{same }approximating\emph{\ }polynomial may be expressed as a \emph{%
divided difference} --or \emph{Newton}-- series: 
\begin{eqnarray}
&&f^{\left( N\right) }\left( \varepsilon \right) =\sum_{M=0}^{N}f\left[
0,..,M\right] \prod_{n=0}^{M-1}\left( \varepsilon -\varepsilon _{n}\right)
\label{e33} \\[0.15cm]
&=&f\left[ 0\right] +f\left[ 0,1\right] \left( \varepsilon -\varepsilon
_{0}\right) +..+f\left[ 0...N\right] \left( \varepsilon -\varepsilon
_{N-1}\right) ..\left( \varepsilon -\varepsilon _{1}\right) \left(
\varepsilon -\varepsilon _{0}\right) ,  \notag
\end{eqnarray}
where the square parentheses denote divided differences as defined in the
following table: 
\begin{equation*}
\begin{array}{ccccccc}
\varepsilon _{0} &  & f_{0}\equiv f\left[ 0\right] &  &  &  &  \\ 
&  &  &  & \frac{f\left[ 0\right] -f\left[ 1\right] }{\varepsilon
_{0}-\varepsilon _{1}}\equiv f\left[ 0,1\right] &  &  \\ 
\varepsilon _{1} &  & f_{1}\equiv f\left[ 1\right] &  &  &  & \frac{f\left[
0,1\right] -f\left[ 1,2\right] }{\varepsilon _{0}-\varepsilon _{2}}\equiv f%
\left[ 0,1,2\right] \\ 
&  &  &  & \frac{f\left[ 1\right] -f\left[ 2\right] }{\varepsilon
_{1}-\varepsilon _{2}}\equiv f\left[ 1,2\right] &  &  \\ 
\varepsilon _{2} &  & f_{2}\equiv f\left[ 2\right] &  &  &  & 
\end{array}
\end{equation*}
In general, that is: 
\begin{equation}
f\left[ m,m+1,..,n,n+1\right] \equiv \frac{f\left[ m,m+1,.,n\right] -f\left[
m+1,.,n,n+1\right] }{\varepsilon _{m}-\varepsilon _{n+1}},  \label{aa4}
\end{equation}
where$\;m\leq n.$ Note that the two energies in the denominator are those
which refer to the mesh points \emph{not} common to the two divided
differences in the nominator. Also, note their order, which defines the
sign. A divided difference, $f\left[ 0...M\right] ,$ is thus a \emph{linear
combination} of $f_{0},\,f_{1},...,f_{M}.$ The divided differences entering (%
\ref{e33}) are those descending along the upper string in the table, but
other forms are possible. Besides, the order of the energies need not be
monotonic. In fact, \emph{all} divided differences of degree $M+1$ involving 
$M$ specific mesh points are identical. This means that the \emph{order }of
the \emph{arguments} in $f\left[ 0,1,.,M-1,M\right] $ is \emph{irrelevant, }%
as may be seen explicitly from expression (\ref{aa6}) below. When we have a
long string of arguments, we usually order them after increasing mesh
number, for simplicity of notation.

We may express any divided difference, $f\left[ 0..M\right] ,$ entering the
Newton\emph{\ }form (\ref{e33}) as a linear combination of the $f_{n}$'s
with $n\leq M,$ and thereby establish the relation to the Lagrange form (\ref
{e32}). To do this, we apply both Newton and Lagrange interpolation to a
function, which we take to be that $M$th degree polynomial, $f^{\left(
M\right) }\left( \varepsilon \right) ,$ which coincides with $f\left(
\varepsilon \right) $ at the first $M+1$ mesh points. This is allowed,
because $f\left[ 0..M\right] $ is independent of the $f_{n}$'s with $n>M.$
In this way, we get the identity: 
\begin{equation*}
f^{\left( M\right) }\left( \varepsilon \right) \;=\;\sum_{m=0}^{M}f\left[
0..m\right] \prod_{n=0}^{m-1}\left( \varepsilon -\varepsilon _{n}\right)
\;=\;\sum_{n=0}^{M}f_{n}\,l_{n}^{\left( M\right) }\left( \varepsilon \right)
\end{equation*}
and taking now the \emph{highest derivative,} we obtain the important
relation: 
\begin{equation}
f\left[ 0...M\right] \;=\;\sum_{n=0}^{M}\frac{f_{n}}{\prod_{m=0,\neq
n}^{M}\left( \varepsilon _{n}-\varepsilon _{m}\right) }.  \label{aa6}
\end{equation}
The inverse relation, that is the expression for $f_{n}$ in terms of divided
differences for a (sub)mesh containing $\varepsilon _{n}$, is of course just
the Newton series (\ref{e33}) evaluated at the mesh point $\varepsilon _{n}.$

In order to factorize $\left( \phi G\right) \left[ 0...N\right] $ in
expression (\ref{c2}) for the NMTO, we shall need to express the $N$th-order
divided difference of a \emph{product function,} $f\left( \varepsilon
\right) g\left( \varepsilon \right) ,$ in terms of divided differences on
the same mesh of the individual functions. Since the product is local in
energy, we start by expressing its divided difference in the Lagrange form (%
\ref{aa6}): 
\begin{equation*}
\left( fg\right) \left[ 0...N\right] =\sum_{n=0}^{N}\frac{f_{n}g_{n}}{%
\prod_{m=0,\neq n}^{N}\left( \varepsilon _{n}-\varepsilon _{m}\right) }.
\end{equation*}
For $f\left( \varepsilon \right) $ we may choose to use the divided
differences in the upper, descending string of the table. We therefore use (%
\ref{e33}) to express $f_{n}$ in terms of the divided differences on the $%
\left( 0..n\right) $-part of the mesh and thereafter reorder the summations: 
\begin{eqnarray*}
\left( fg\right) \left[ 0...N\right] &=&\sum_{n=0}^{N}\sum\limits_{M=0}^{N}%
\,f\left[ 0..M\right] \prod_{m^{\prime }=0}^{M-1}\left( \varepsilon
_{n}-\varepsilon _{m^{\prime }}\right) \frac{g_{n}}{\prod_{m=0,\neq
n}^{N}\left( \varepsilon _{n}-\varepsilon _{m}\right) } \\
&=&\sum\limits_{M=0}^{N}\,f\left[ 0..M\right] \sum_{n=0}^{N}\frac{%
\prod_{m^{\prime }=0}^{M-1}\left( \varepsilon _{n}-\varepsilon _{m^{\prime
}}\right) }{\prod_{m=0,\neq n}^{N}\left( \varepsilon _{n}-\varepsilon
_{m}\right) }g_{n}.
\end{eqnarray*}
Since $\prod_{m^{\prime }=0}^{M-1}\left( \varepsilon _{n}-\varepsilon
_{m^{\prime }}\right) =0$ for $n<M,$%
\begin{eqnarray*}
\sum_{n=0}^{N}\frac{\prod_{m^{\prime }=0}^{M-1}\left( \varepsilon
_{n}-\varepsilon _{m^{\prime }}\right) }{\prod_{m=0,\neq n}^{N}\left(
\varepsilon _{n}-\varepsilon _{m}\right) }g_{n} &=&\sum_{n=M}^{N}\frac{%
\prod_{m^{\prime }=0}^{M-1}\left( \varepsilon _{n}-\varepsilon _{m^{\prime
}}\right) }{\prod_{m=0,\neq n}^{N}\left( \varepsilon _{n}-\varepsilon
_{m}\right) }g_{n} \\
&=&\sum_{n=M}^{N}\frac{g_{n}}{\prod_{m=M,\neq n}^{N}\left( \varepsilon
_{n}-\varepsilon _{m}\right) }=g\left[ M..N\right] ,
\end{eqnarray*}
according to (\ref{aa6}). We have thus proved the \emph{binomial formula:} 
\begin{equation}
\left( fg\right) \left[ 0...N\right] =\sum_{M=0}^{N}f\left[ 0..M\right] \,g%
\left[ M..N\right] ,  \label{aa22}
\end{equation}
which expresses the $N$th divided difference of a product on the $\left(
0...N\right) $-mesh as a sum of products of divided differences on
respectively the $\left( 0..M\right) $- and $\left( M..N\right) $-parts of
the mesh, with $M$ being the only point in common. Hence, this formula is in
terms of the divided differences descending forwards along the upper string
for $f,$ and the divided differences descending backwards along the lower
string for $g,$ but this is merely one of many possibilities. For the
special case: $g\left( \varepsilon \right) =\varepsilon ,$ we get the useful
result: 
\begin{equation}
\left( \varepsilon f\right) \left[ 0...N\right] \;=\;f\left[ 0..N-1\right]
\;+\;\varepsilon _{N}f\left[ 0...N\right] .\;  \label{aa5}
\end{equation}
Since the numbering of the points is irrelevant, we could of course have
singled out \emph{any} of the $N+1$ points, not merely the last.

Newton interpolation has the conceptual advantage over Lagrange
interpolation that the 1st divided differences, $f\left[ n-1,n\right] ,$ are
the slopes of the chords connecting points $n-1$\textit{\ }and $n,$\textit{\ 
}and hence approximations to the 1st derivatives, the 2nd divided
differences, $f\left[ n-1,n,n+1\right] ,$ are 'local' approximations to $%
\frac{1}{2!}$ times the 2nd derivatives, and so on, as expressed by (\ref{c3}%
). For the mesh condensing onto the one energy, $\varepsilon _{\nu },$
Newton \emph{interpolation} becomes \emph{Taylor expansion, }which is of
course simpler. An example of this is the binomial expression for the $N$th
derivative of a product: For a discrete mesh, there are many alternatives to
(\ref{aa22}), but for a condensed mesh, there is only one expression.

\subsubsection{Hermite interpolation.}

It will turn out that the NMTO Hamiltonian and overlap matrices are best
understood and computed using the formalism of \emph{Hermite} interpolation.
Here, one seeks the polynomial of degree $M+N+1$ which fits not only the
values, $f_{n},$ at the $N+1$ points, but also the \emph{slopes}, $\dot{f}%
_{n},$ at a subset of $M+1$ points. We shall number the points in such a
way, that the $M+1$ points are the \emph{first. }This polynomial is: 
\begin{eqnarray*}
&&f^{\left( M+N+1\right) }\left( \varepsilon \right) \;= \\
&&\;\sum_{n=0}^{M}\left[ f_{n}+\left( \dot{f}_{n}-f_{n}\left( \sum_{m=0,\neq
n}^{M}\frac{2}{\varepsilon _{n}-\varepsilon _{m}}+\sum_{m=M+1}^{N}\frac{1}{%
\varepsilon _{n}-\varepsilon _{m}}\right) \right) \left( \varepsilon
-\varepsilon _{n}\right) \right] \, \\
&&\times l_{n}^{\left( M\right) }\left( \varepsilon \right) \,l_{n}^{\left(
N\right) }\left( \varepsilon \right)
\;+\;\sum_{n=M+1}^{N}f_{n}\,l_{n}^{\left( M+1\right) }\left( \varepsilon
\right) \,l_{n}^{\left( N\right) }\left( \varepsilon \right) .
\end{eqnarray*}

For those interested in \emph{why} this is so, here are the arguments: The
product of Lagrange polynomials 
\begin{equation*}
l_{n}^{\left( M\right) }\left( \varepsilon \right) \,l_{n}^{\left( N\right)
}\left( \varepsilon \right) =\prod_{m=0,\neq n}^{M}\left( \frac{\varepsilon
-\varepsilon _{m}}{\varepsilon _{n}-\varepsilon _{m}}\right)
^{2}\prod_{m=M+1}^{N}\frac{\varepsilon -\varepsilon _{m}}{\varepsilon
_{n}-\varepsilon _{m}},
\end{equation*}
with$\;0\leq n\leq M,$ is of degree $M+N.$ At a mesh point, $\varepsilon
=\varepsilon _{n^{\prime }},$ this product has value 1 when $0\leq n^{\prime
}=n\leq M,$ value 0 and slope 0 when $0\leq n^{\prime }\neq n\leq M,$ and
value 0 when $M<n^{\prime }\leq N.$ Since the slope is: 
\begin{equation*}
\left( \sum_{m=0,\neq n}^{M}\frac{2}{\varepsilon -\varepsilon _{m}}%
+\sum_{m=M+1}^{N}\frac{1}{\varepsilon -\varepsilon _{m}}\right)
\,l_{n}^{\left( M\right) }\left( \varepsilon \right) \,l_{n}^{\left(
N\right) }\left( \varepsilon \right) ,
\end{equation*}
the polynomial of degree $M+N+1:$ 
\begin{equation*}
\left( 1-\left( \varepsilon -\varepsilon _{n}\right) \left( \sum_{m=0,\neq
n}^{M}\frac{2}{\varepsilon _{n}-\varepsilon _{m}}+\sum_{m=M+1}^{N}\frac{1}{%
\varepsilon _{n}-\varepsilon _{m}}\right) \right) \,l_{n}^{\left( M\right)
}\left( \varepsilon \right) \,l_{n}^{\left( N\right) }\left( \varepsilon
\right) ,
\end{equation*}
with$\;0\leq n\leq M,$ has value $1$ and slope 0 if $\varepsilon
=\varepsilon _{n}.$ If $\varepsilon =\varepsilon _{n^{\prime }}\neq
\varepsilon _{n}$, it has value 0 and slope 0 when $0\leq n^{\prime }\leq M$
, and value 0 and some slope when $M<n^{\prime }\leq N.$ The polynomial of
degree $M+N+1:$%
\begin{equation*}
\left( \varepsilon -\varepsilon _{n}\right) l_{n}^{\left( M\right) }\left(
\varepsilon \right) \,l_{n}^{\left( N\right) }\left( \varepsilon \right) ,
\end{equation*}
with$\;0\leq n\leq M,$ vanishes at \emph{all }mesh points, has slope 1 for $%
\varepsilon =\varepsilon _{n},$ slope 0 for $\varepsilon =\varepsilon
_{n^{\prime }}\neq \varepsilon _{n}$ when $n^{\prime }$ and $0\leq n^{\prime
}\leq M,$ and some slope when $M<n^{\prime }\leq N.$ Finally, the product: 
\begin{equation*}
l_{n}^{\left( M+1\right) }\left( \varepsilon \right) \,l_{n}^{\left(
N\right) }\left( \varepsilon \right) =\prod_{m=0}^{M}\left( \frac{%
\varepsilon -\varepsilon _{m}}{\varepsilon _{n}-\varepsilon _{m}}\right)
^{2}\prod_{m=M+1,\neq n}^{N}\frac{\varepsilon -\varepsilon _{m}}{\varepsilon
_{n}-\varepsilon _{m}},
\end{equation*}
with$\;M<n\leq N,$ is a polynomial of degree $M+N+1.$ For $\varepsilon
=\varepsilon _{n^{\prime }}$ it has value 0 and slope 0 if $0\leq n^{\prime
}\leq M,$ value 0 and some slope if $M<n^{\prime }\neq n\leq N,$ and value 1
and some slope if $M<n^{\prime }=n\leq N.$

What we shall really need is, like in (\ref{aa6}), $\frac{1}{\left(
M+N+1\right) !}$ times the \emph{highest derivative} of the polynomial $%
f^{\left( M+N+1\right) }\left( \varepsilon \right) $. Calculated as the
coefficient to the highest power of $\varepsilon ,$ this \emph{Hermite
divided difference} is: 
\begin{eqnarray}
\frac{\overset{\left( M+N+1\right) }{f^{\left( M+N+1\right) }}}{\left(
M+N+1\right) !}\; &=&\;\sum_{n=0}^{M}\frac{\dot{f}_{n}-f_{n}\left(
\sum\limits_{n^{\prime }=0,\neq n}^{M}\frac{2}{\varepsilon _{n}-\varepsilon
_{n^{\prime }}}+\sum\limits_{n^{\prime }=M+1}^{N}\frac{1}{\varepsilon
_{n}-\varepsilon _{n^{\prime }}}\right) }{\prod\limits_{m=0,\neq
n}^{M}\left( \varepsilon _{n}-\varepsilon _{m}\right)
^{2}\prod\limits_{m=M+1}^{N}\left( \varepsilon _{n}-\varepsilon _{m}\right) }
\notag \\[0.1cm]
&&+\sum_{n=M+1}^{N}\frac{f_{n}}{\prod\limits_{m=0}^{M}\left( \varepsilon
_{n}-\varepsilon _{m}\right) ^{2}\prod\limits_{m=M+1,\neq n}^{N}\left(
\varepsilon _{n}-\varepsilon _{m}\right) }  \notag \\[0.25cm]
&=&\;\lim_{\epsilon \rightarrow 0}\,f\left[ 0.....M+N+1\right] \equiv f\left[
\left[ 0...M\right] ..N\right] .  \label{h5}
\end{eqnarray}
In the last line, we have indicated that the Hermite divided difference may
be considered as the divided difference for the folded and paired mesh: 
\begin{equation*}
\begin{array}{cccccccccccccccccccccccccccccc}
\varepsilon _{0} & \varepsilon _{N+1} &  &  &  & \varepsilon _{1} & 
\varepsilon _{N+2} &  &  & \cdot  & \cdot  &  &  & \cdot  & \cdot  &  &  & 
\varepsilon _{M} & \varepsilon _{M+N+1} &  &  &  & \cdot  &  &  &  & \cdot 
&  &  & \varepsilon _{N}
\end{array}
\end{equation*}
in the limit that the energy differences, $\epsilon _{n}\equiv \varepsilon
_{n+N+1}-\varepsilon _{n},$ between the pairs tend to zero. In analogy with
the notation for the divided differences, we have denoted the $\left(
M+N+1\right) $st Hermite divided difference:$\,f\left[ \left[ 0...M\right]
..N\right] ,$ which means that the mesh points listed inside \emph{two}
square parentheses have both $f_{n}$ and $\dot{f}_{n}$ associated with them,
whereas those listed inside only \emph{one} square parenthesis have merely $%
f_{n}.$ Like for the divided differences, the order of the arguments inside
a square parenthesis is irrelevant, but for long strings we usually choose
the order of increasing $n.$ For a condensed mesh, 
\begin{equation}
f\left[ \left[ 0...M\right] ..N\right] \quad \rightarrow \quad \frac{%
\overset{\left( M+N+1\right) }{f}}{\left( M+N+1\right) !}.  \label{h12}
\end{equation}
As examples of Hermite divided differences we have: 
\begin{equation}
\begin{array}{llll}
f\left[ \left[ 0\right] \right] =\dot{f}_{0} & \quad  & \quad  & f\left[ 
\left[ 0\right] 1\right] =\frac{\dot{f}_{0}-f\left[ 01\right] }{\varepsilon
_{0}-\varepsilon _{1}} \\ 
&  &  &  \\ 
f\left[ \left[ 01\right] \right] =\frac{\dot{f}_{0}-2f\left[ 0,1\right] +%
\dot{f}_{1}}{\left( \varepsilon _{0}-\varepsilon _{1}\right) ^{2}} & \quad 
& \quad  & f\left[ \left[ \;\right] 0..N\right] =f\left[ 0..N\right] 
\end{array}
\label{h2}
\end{equation}

In the NMTO formalism the Hermite divided difference (\ref{h5}) comes in the
disguise of the following double sum (\ref{d14}): 
\begin{equation}
\sum_{n=0}^{N}\sum_{n^{\prime }=0}^{M}\frac{f\left[ n,n^{\prime }\right] }{%
\prod_{m=0,\neq n}^{N}\left( \varepsilon _{n}-\varepsilon _{m}\right)
\prod_{m^{\prime }=0,\neq n^{\prime }}^{M}\left( \varepsilon _{n^{\prime
}}-\varepsilon _{m^{\prime }}\right) }\,,  \label{h6}
\end{equation}
which may, in fact, be viewed as a divided difference (\ref{aa6}) --albeit
in two dimensions-- but that brings little simplification. So let us prove
that (\ref{h5}) and (\ref{h6}) are identical: First of all, the $\dot{f}_{n}$%
-terms of the double sum (\ref{h6}) are those for which $n=n^{\prime },$ and
they obviously equal those of the single sum (\ref{h5}). Secondly, the $%
f_{n} $-terms in (\ref{h6}) are: 
\begin{eqnarray}
&&\sum_{n=0}^{N}\sum_{n^{\prime }=0,\neq n}^{M}\frac{f_{n}\left( \varepsilon
_{n}-\varepsilon _{n^{\prime }}\right) ^{-1}+f_{n^{\prime }}\left(
\varepsilon _{n^{\prime }}-\varepsilon _{n}\right) ^{-1}}{\prod_{m=0,\neq
n}^{N}\left( \varepsilon _{n}-\varepsilon _{m}\right) \prod_{m=0,\neq
n^{\prime }}^{M}\left( \varepsilon _{n^{\prime }}-\varepsilon _{m}\right) }=
\notag \\
&&\sum_{n=M+1}^{N}\frac{f_{n}}{\prod_{m=0,\neq n}^{N}\left( \varepsilon
_{n}-\varepsilon _{m}\right) }\sum_{n^{\prime }=0}^{M}\frac{\left(
\varepsilon _{n}-\varepsilon _{n^{\prime }}\right) ^{-1}}{\prod_{m=0,\neq
n^{\prime }}^{M}\left( \varepsilon _{n^{\prime }}-\varepsilon _{m}\right) }+
\label{h7} \\
&&\sum_{n=0}^{M}\frac{f_{n}}{\prod_{m=0,\neq n}^{N}\left( \varepsilon
_{n}-\varepsilon _{m}\right) }\sum_{n^{\prime }=0,\neq n}^{M}\frac{\left(
\varepsilon _{n}-\varepsilon _{n^{\prime }}\right) ^{-1}}{\prod_{m=0,\neq
n^{\prime }}^{M}\left( \varepsilon _{n^{\prime }}-\varepsilon _{m}\right) }+
\notag \\
&&\sum_{n=0}^{M}\frac{f_{n}}{\prod_{m=0,\neq n}^{M}\left( \varepsilon
_{n}-\varepsilon _{m}\right) }\sum_{n^{\prime }=0,\neq n}^{N}\frac{\left(
\varepsilon _{n}-\varepsilon _{n^{\prime }}\right) ^{-1}}{\prod_{m=0,\neq
n^{\prime }}^{N}\left( \varepsilon _{n^{\prime }}-\varepsilon _{m}\right) }.
\notag
\end{eqnarray}
Now, according to (\ref{aa6}), 
\begin{equation}
\sum_{n^{\prime }=0}^{M}\frac{\frac{1}{\varepsilon _{n}-\varepsilon
_{n^{\prime }}}}{\prod_{m=0,\neq n^{\prime }}^{M}\left( \varepsilon
_{n^{\prime }}-\varepsilon _{m}\right) }=\frac{1}{\varepsilon
_{n}-\varepsilon }\left[ 0...M\right]  \label{h8}
\end{equation}
is the $M$th divided difference of the single-pole function $1/\left(
\varepsilon _{n}-\varepsilon \right) ,$ provided that $n$ is \emph{not }on
the mesh $0...M.$ For the sum where $n$ \emph{is }on the mesh --but the $%
n^{\prime }\mathrm{=}n$-term is excluded-- we have: 
\begin{eqnarray}
\sum_{n^{\prime }=0,\neq n}^{M}\frac{\frac{1}{\varepsilon _{n}-\varepsilon
_{n^{\prime }}}}{\prod_{m=0,\neq n^{\prime }}^{M}\left( \varepsilon
_{n^{\prime }}-\varepsilon _{m}\right) } &=&\sum\limits_{n^{\prime }=0,\neq
n}^{M}\frac{\frac{-1}{\left( \varepsilon _{n}-\varepsilon _{n^{\prime
}}\right) ^{2}}}{\prod_{m=0,\neq n,\neq n^{\prime }}^{M}\left( \varepsilon
_{n^{\prime }}-\varepsilon _{m}\right) }  \notag \\[0.1cm]
&=&\frac{-1}{\left( \varepsilon _{n}-\varepsilon \right) ^{2}}\left[
0..n-1,n+1..M\right] .  \label{h9}
\end{eqnarray}
This result also holds if $M$ is named $N,$ and is therefore relevant for
both of the last terms in (\ref{h7}). We then need simpler expressions for
the divided differences of the single- and double-pole functions. Guided by
the results: 
\begin{equation*}
\frac{1}{M!}\frac{d^{M}}{d\varepsilon ^{M}}\frac{1}{\varepsilon
_{i}-\varepsilon }=\frac{1}{\left( \varepsilon _{i}-\varepsilon \right)
^{M+1}},\mathrm{\quad \quad }\frac{1}{M!}\frac{d^{M}}{d\varepsilon ^{M}}%
\frac{1}{\left( \varepsilon _{i}-\varepsilon \right) ^{2}}=\frac{M+1}{\left(
\varepsilon _{i}-\varepsilon \right) ^{M+2}},
\end{equation*}
for the derivatives, we postulate that for a discrete mesh, 
\begin{equation}
\frac{1}{\varepsilon _{i}-\varepsilon }\left[ 0...M\right] =\frac{1}{%
\prod_{m=0}^{M}\left( \varepsilon _{i}-\varepsilon _{m}\right) },\mathrm{%
\quad }\frac{1}{\left( \varepsilon _{i}-\varepsilon \right) ^{2}}\left[ 0...M%
\right] =\frac{\sum_{n=0}^{M}\frac{1}{\varepsilon _{i}-\varepsilon _{n}}}{%
\prod_{m=0}^{M}\left( \varepsilon _{i}-\varepsilon _{m}\right) }.
\label{h10}
\end{equation}
For $M$=0, these expressions obviously reduce to the correct results, $%
\left( \varepsilon _{i}-\varepsilon _{0}\right) ^{-1}$ and $\left(
\varepsilon _{i}-\varepsilon _{0}\right) ^{-2}.$ For $M>0,$ our conjectures
inserted on the right-hand side of (\ref{aa4}) and subsequent use of (\ref
{aa6}) yield: 
\begin{eqnarray*}
\frac{\frac{1}{\varepsilon _{i}-\varepsilon }\left[ 0..M-1\right] -\frac{1}{%
\varepsilon _{i}-\varepsilon }\left[ 1..M\right] }{\varepsilon
_{0}-\varepsilon _{M}} &=&\frac{1}{\prod_{m=0}^{M}\left( \varepsilon
_{i}-\varepsilon _{m}\right) }=\frac{1}{\varepsilon _{i}-\varepsilon }\left[
0...M\right] , \\
\frac{\frac{1}{\left( \varepsilon _{i}-\varepsilon \right) ^{2}}\left[ 0..M-1%
\right] -\frac{1}{\left( \varepsilon _{i}-\varepsilon \right) ^{2}}\left[
1..M\right] }{\varepsilon _{0}-\varepsilon _{M}} &=&\frac{%
\sum\limits_{n=0}^{M}\frac{1}{\varepsilon _{i}-\varepsilon _{n}}}{%
\prod_{m=0}^{M}\left( \varepsilon _{i}-\varepsilon _{m}\right) }=\frac{1}{%
\left( \varepsilon _{i}-\varepsilon \right) ^{2}}\left[ 0...M\right] ,
\end{eqnarray*}
which are obviously correct too. Hence, equations (\ref{h10}) have been
proved.

Using finally (\ref{h10}) in (\ref{h8}) and (\ref{h9}), and right back in (%
\ref{h7}), leads to the $f_{n}$-terms in (\ref{h5}). We have therefore
demonstrated that: 
\begin{equation}
\sum_{n=0}^{N}\sum_{n^{\prime }=0}^{M}\frac{f\left[ n,n^{\prime }\right] }{%
\prod\limits_{m=0,\neq n}^{N}\left( \varepsilon _{n}-\varepsilon _{m}\right)
\prod\limits_{m^{\prime }=0,\neq n^{\prime }}^{M}\left( \varepsilon
_{n^{\prime }}-\varepsilon _{m^{\prime }}\right) }=f\left[ \left[ 0...M%
\right] ..N\right] .  \label{h11}
\end{equation}

The final expression needed for the NMTO formalism, is one for the Hermite
divided difference of the product-function $\varepsilon f\left( \varepsilon
\right) .$ For this we can use (\ref{aa5}) applied to the folded and paired
mesh. As a result: 
\begin{equation}
\left( \varepsilon f\right) \left[ \left[ 0...M\right] ..N\right] \;=\;f%
\left[ \left[ 0..M-1\right] ..N\right] \;+\;\varepsilon _{M}f\left[ \left[
0...M\right] ..N\right] .  \label{h13}
\end{equation}
Since the numbering of the points is irrelevant, we could of course have
singled out \emph{any} of the $M+1$ points, not merely the last.

\end{document}

%% file: fig1.tex
\begin{figure}
\includegraphics[width=.6\textwidth]{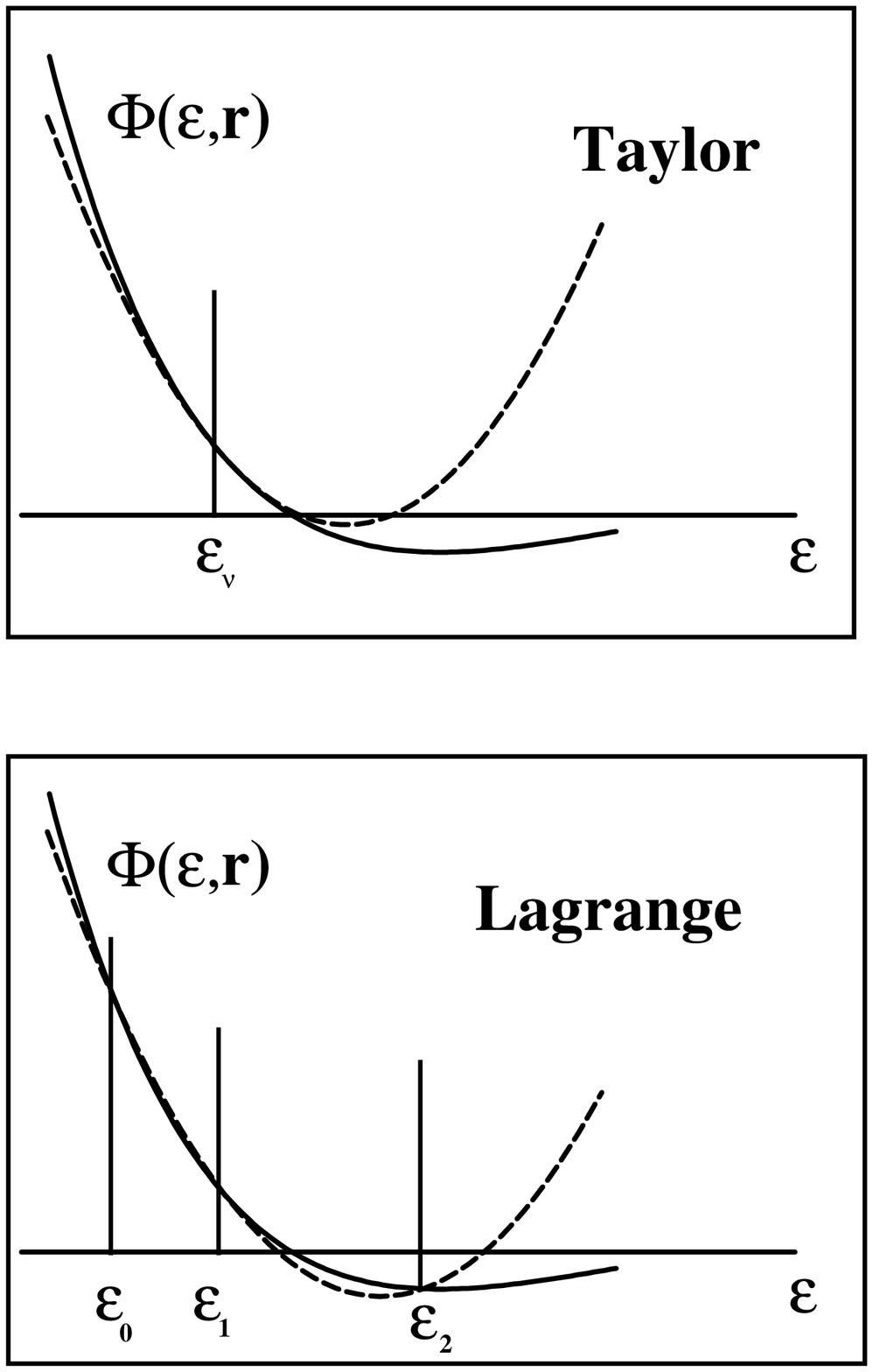}
\caption[]{Quadratic approximation to the energy dependence
of a partial wave for a condensed (Taylor) and a discrete (Lagrange)
mesh.}
\label{Fig1}
\end{figure}

%% file: fig2.tex
\begin{figure}
\includegraphics[width=.6\textwidth]{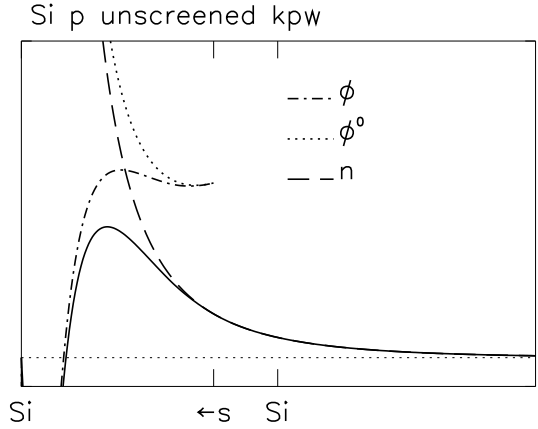}
\caption[]{Bare Si $p$ MTO according to Eq.(\ref{b2})}
\label{Fig2}
\end{figure}

%% file: fig3.tex
\begin{figure}
\includegraphics[width=.6\textwidth]{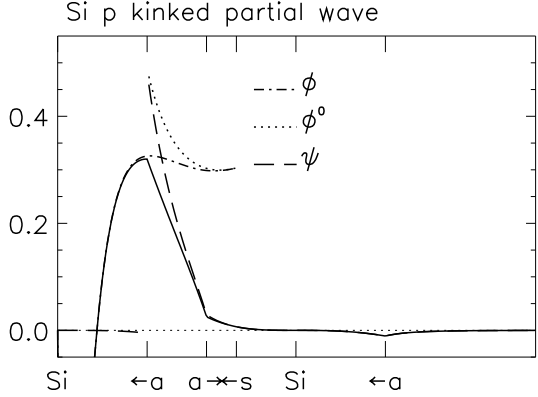}
\caption[]{Si $p_{111}$ member of a screened $spd$-set of 0th-order
MTOs (see text and Eqs.(\ref{b15}),(\ref{b23})-(\ref{b26})).}
\label{Fig3}
\end{figure}

%% file: fig4.tex
\begin{figure}[b]
\includegraphics[width=.6\textwidth]{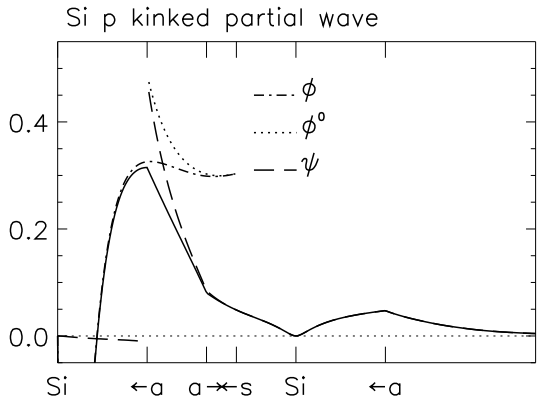}
\caption[]{
Si $p_{111}$ member of a screened minimal $sp$-set of 0th-order
MTOs (see text).}
\label{Fig4}
\end{figure}

%% file: fig4a.tex
\begin{figure}
\includegraphics[width=.6\textwidth]{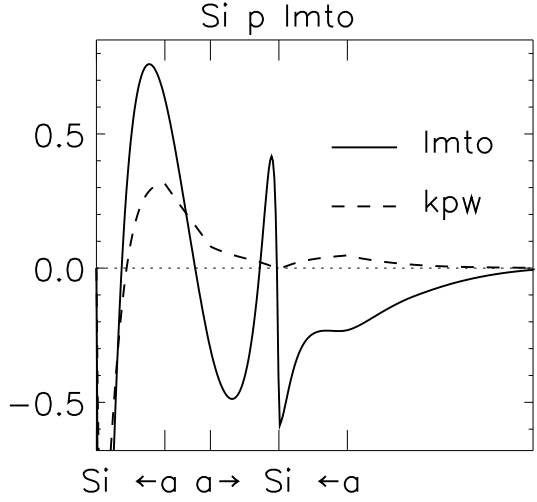}
\caption[]{
Si $p_{111}$ member of the $sp$-set of 0th (dottet) and 1th-order
MTOs (see text and Eq.(\ref{d15})).}
\label{Fig4a}
\end{figure}

%% file: fig5.tex
\begin{figure}
\vspace*{2.0cm}
\hspace*{-1cm}
\includegraphics[width=1.0\textwidth]{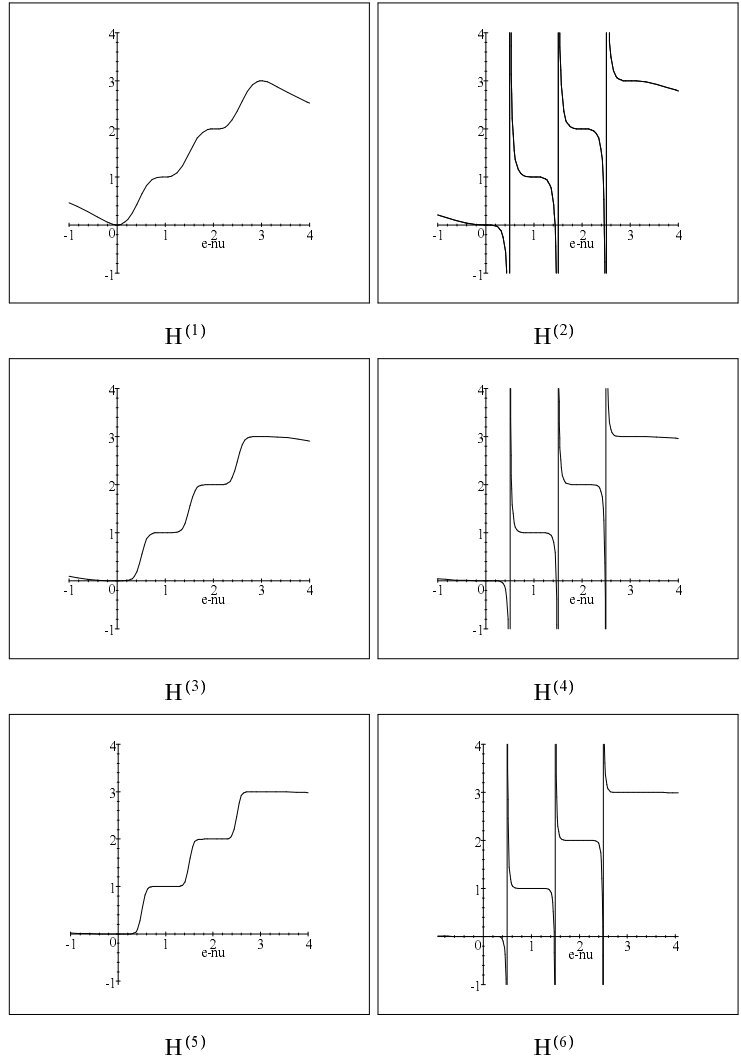}
\caption[]{
Switching behavior of $E^{\left( N\right) }\left(
\varepsilon _{\nu }\right) \equiv H^{\left( N\right) }\left(
\varepsilon
_{\nu }\right) $ for the orthonormal one-orbital model defined by Eq.
(\ref {d4}) with 4 radial levels: $\varepsilon _{j}=0,1,2,3.$}
\label{Fig5}
\end{figure}

%% file: fig6.tex
\begin{figure}
\rotatebox{-90}{
\includegraphics[height=1.0\textwidth]{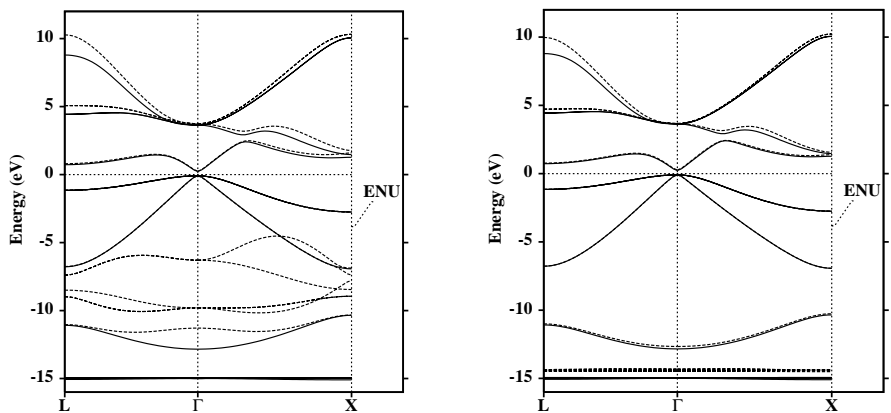}}
\caption[]{
Minimal-basis LMTO energy bands (dashed) of GaAs for two
different choices of the screening-radii compared to the exact KKR band
structure (solid). In the left-hand panel all screening radii were
$\sim
0.8t,$ while in the right-hand panel the Ga $d$ radius was reduced to
the radius of the Ga 3$d$ core \cite{PolyT}. See text.}
\label{Fig6}
\end{figure}

%% file: fig7.tex
\begin{figure}
\rotatebox{-90}{
\includegraphics[height=1.0\textwidth]{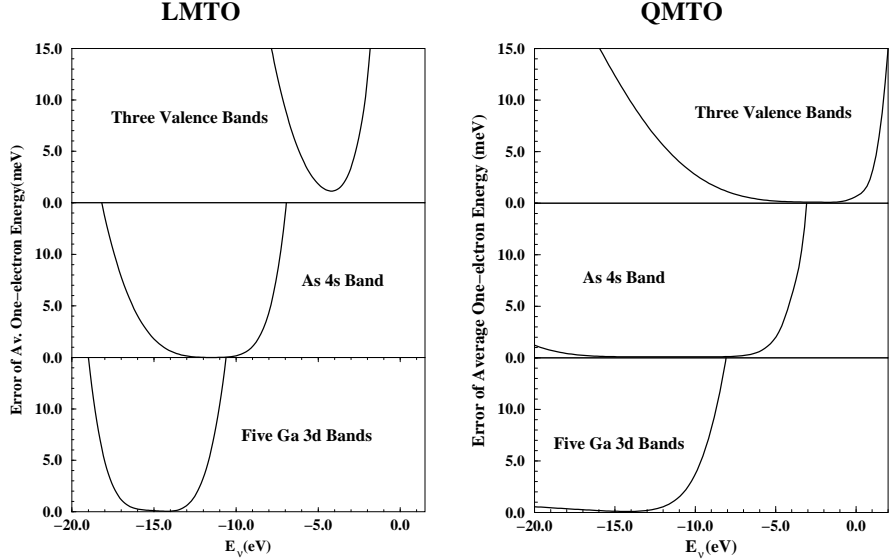}}
\caption[]{
Mean error in each of the three types of occupied
valence bands in GaAs calculated with the LMTO and QMTO methods as a
function of the expansion energy $\varepsilon _{\nu }$ for a condensed
mesh \cite{PolyT}. See Fig. \ref{Fig6} and text.}
\label{Fig7}
\end{figure}

%% file: fig8.tex
\begin{figure}
\rotatebox{-90}{
\includegraphics[height=1.0\textwidth]{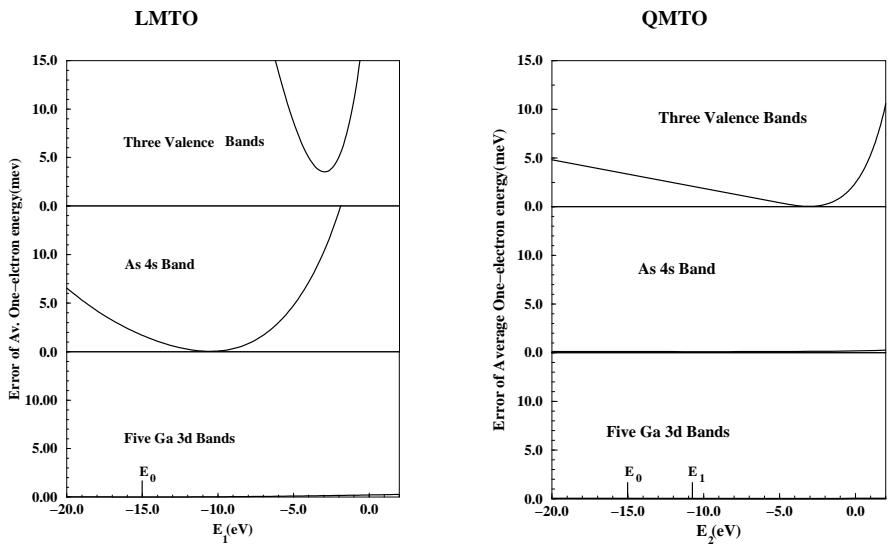}}
\caption[]{
Like Fig. \ref{Fig7}, but calculated using discrete
meshes and as functions of the position of the last energy point. The
first
energy points were fixed at the positions indicated on the abscissa
\cite {PolyT}. See text.}
\label{Fig8}
\end{figure}

%% file: fig9.tex
\begin{figure}
\rotatebox{-90}{
\includegraphics[height=0.75\textwidth]{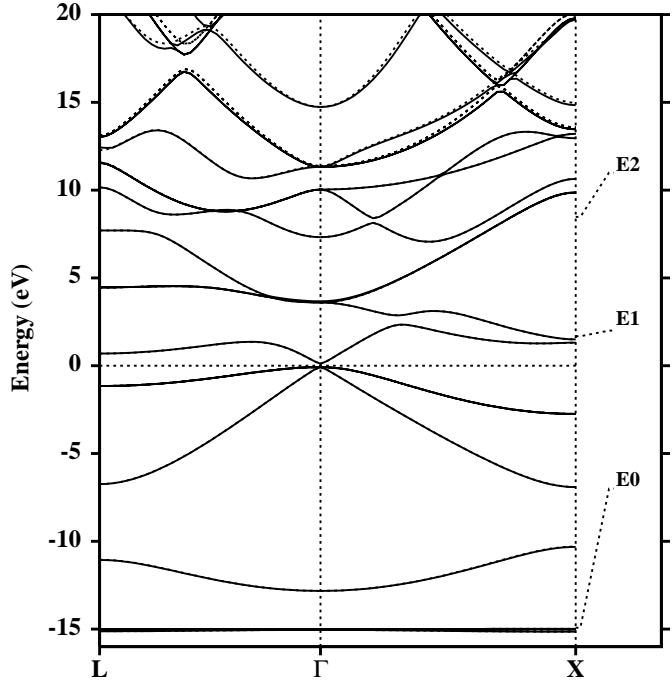}}
\caption[]{
Energy bands of GaAs calculated with the QMTO method
and the energy mesh indicated on the right-hand side (dashed) as
compared with the exact KKR result (solid) \cite{PolyT}. See text.}
\label{Fig9}
\end{figure}

%% file: fig10.tex
\begin{figure}
\includegraphics[width=1.0\textwidth]{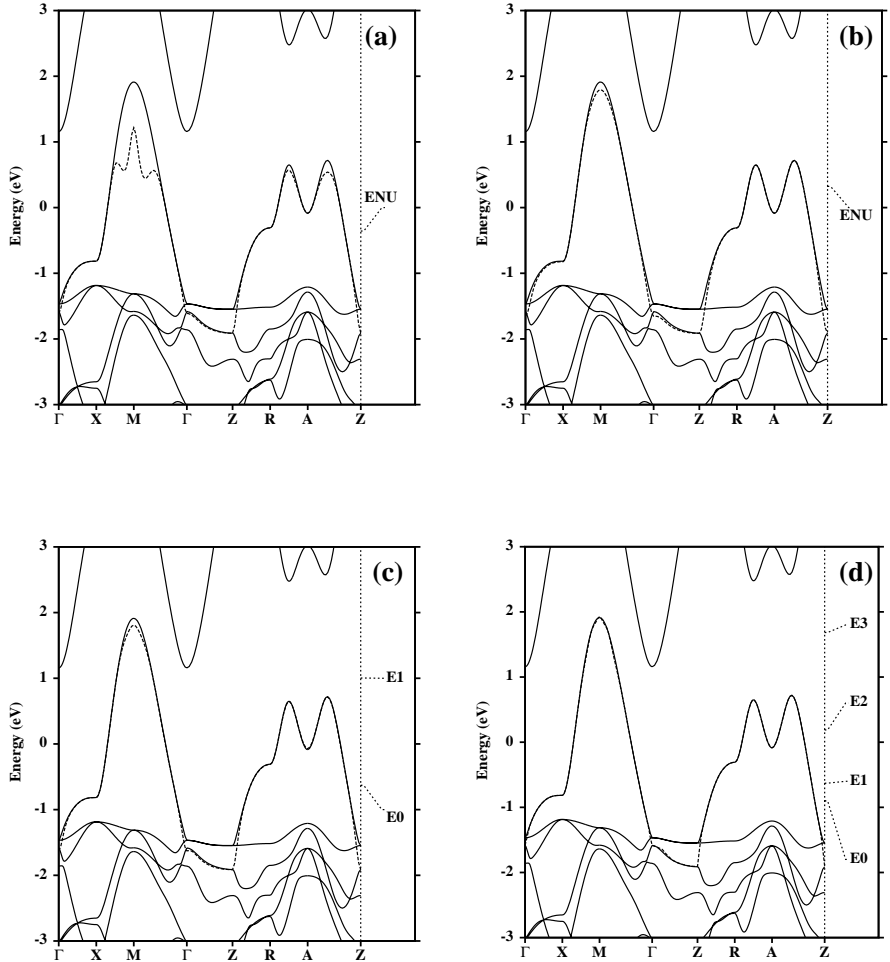}
\caption[]{
Conduction band of CaCuO$_{2}$ calculated by massive
downfolding to a single Cu $x^{2}-y^{2}$ NMTO (dotted) compared with
the full band structure (solid) \cite{PolyT}. See text.}
\label{Fig10}
\end{figure}

%% file: fig11.tex
\begin{figure}
\rotatebox{90}{
\includegraphics[height=1.0\textwidth]{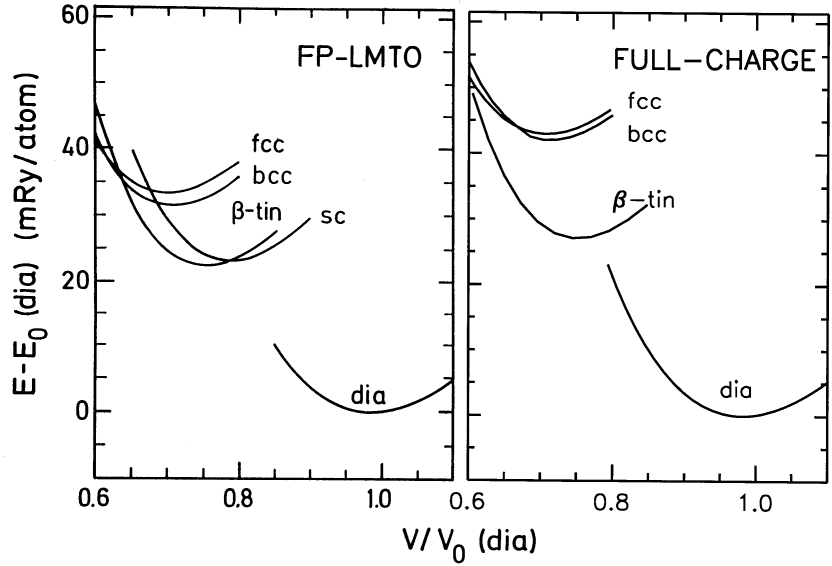}}
\caption[]{
Total energy of Si as a function of the atomic volume
for different structures calculated with the full-potential LMTO method
\cite {MethFP} and with the present full-charge scheme
\cite{SiPhase,Ras,Ras2}. See text.}
\label{Fig11}
\end{figure}

%% file: fig13.tex
\begin{figure}
\includegraphics[width=1.0\textwidth]{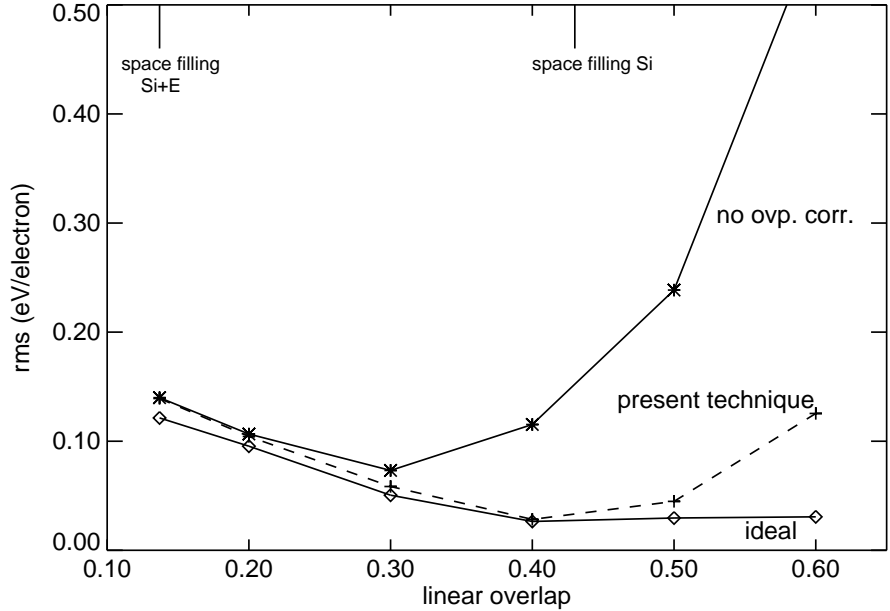}
\caption[]{
Rms error of the valence-band energies in
diamond-structured Si as a function of the overlap in the atom-centered
MT-potential \cite{Catia,Catia2}. See text.}
\label{Fig13}
\end{figure}

%% file: fig12.tex
\begin{figure}
\includegraphics[width=.75\textwidth]{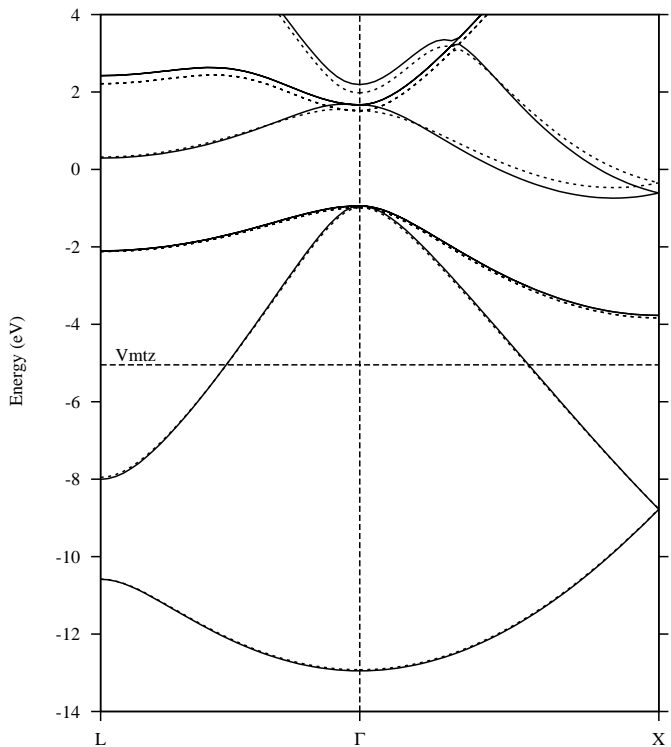}
\caption[]{
Band structure of Si calculated with the
3rd-generation LMTO method for the self-consistent Si+E MT-potential
(dashed) and for the Si-centered, 60\%-overlapping MT-approximation to
it (solid). The latter calculation included the correction for the
kinetic-energy error Eq. (\ref{ovl}) in the LMTO Hamiltonian, and the
value
of the MT-zero was adjusted in such a way that the average energy of
the
valence band was correct. Hence, the solid band structure corresponds
to the
last point on the curve marked 'ideal' in Fig. \ref{Fig13} \cite
{Catia,Catia2}. See text.}
\label{Fig12}
\end{figure}